\documentclass[a4paper,11pt]{article}
\pdfoutput=1 
\usepackage{jcappub}
\usepackage{amsmath}
\usepackage{graphicx}
\usepackage{latexsym}
\usepackage{xspace}
\usepackage{color}
\usepackage{hyperref} 
\usepackage{bm}
\usepackage{relsize}
\usepackage{tabularx}
\usepackage{multirow}
\usepackage{amssymb}
\usepackage[table]{xcolor}
\usepackage{slashbox}

\makeatletter
\gdef\@fpheader{}
\g@addto@macro\bfseries{\boldmath}
\makeatother

\newcommand{\pisto}{\pi_\mathrm{sto}}
\newcommand{\pilog}{\pi_\mathrm{log}}
\newcommand{\DKL}{D_{\mathrm{KL}}}
\newcommand{\Mmd}{M_{\mathrm{md}}}
\newcommand{\figpilogsto}{
\vspace{-0.1cm}
\hspace{4cm}
$\pilog$\hspace{6.3cm} 
$\pisto$
\vspace{-0.1cm}
}

\newcommand{\ie}{{i.e.~}}

\newcommand{\dd}{\mathrm{d}}
\newcommand{\ee}{e}

\newcommand{\sss}[1]{{\scriptscriptstyle{#1}}}

\newcommand{\uPl}{\mathrm{Pl}}

\newcommand{\umin}{\mathrm{min}}
\newcommand{\umax}{\mathrm{max}}
\newcommand{\uend}{\mathrm{end}}

\newcommand{\ureh}{\mathrm{reh}}
\newcommand{\uereh}{\mathrm{ereh}}
\newcommand{\urad}{\mathrm{rad}}

\newcommand{\uS}{\mathrm{S}}

\newcommand{\usssS}{\sss{\uS}}

\newcommand{\usssPl}{\sss{\uPl}}

\newcommand{\nS}{n_\usssS}

\newcommand{\uNL}{\mathrm{NL}}

\newcommand{\MeV}{\mathrm{MeV}}
\newcommand{\GeV}{\mathrm{GeV}}

\newcommand{\Mp}{M_\usssPl}

\newcommand{\fnl}{f_\uNL}

\newcommand{\efolds}{$e$-folds~}

\newcommand{\beq}{\begin{equation}}
\newcommand{\eeq}{\end{equation}}
\newcommand{\bea}{\begin{eqnarray}}
\newcommand{\eea}{\end{eqnarray}}

\newlength{\wsingfig}
\newlength{\wdblefig}
\newlength{\wquadfig}
\newlength{\wtriplefig}
\newcommand{\Eq}[1]{Eq.~(\ref{#1})}
\newcommand{\Eqs}[1]{Eqs.~(\ref{#1})}
\newcommand{\Fig}[1]{Fig.~{\ref{#1}}}

\newcommand{\Ref}[1]{Ref.~{\cite{#1}}}
\newcommand{\Refs}[1]{Refs.~{\cite{#1}}}
\newcommand{\Sec}[1]{Sec.~\ref{#1}}
\newcommand{\Secs}[1]{Secs.~\ref{#1}}
\newcommand{\App}[1]{Appendix~\ref{#1}}

\subheader{}

\title{Constraining Curvatonic Reheating}

\author[a]{Robert J. Hardwick,}
\author[a]{Vincent Vennin,}
\author[a]{Kazuya Koyama}
\author[a]{and David Wands}

\affiliation[a]{Institute of Cosmology \& Gravitation, University of Portsmouth, Dennis Sciama Building, Burnaby Road, Portsmouth, PO1 3FX, United Kingdom}

\emailAdd{robert.hardwick@port.ac.uk}
\emailAdd{vincent.vennin@port.ac.uk}
\emailAdd{kazuya.koyama@port.ac.uk}
\emailAdd{david.wands@port.ac.uk}

\date{today}

\begin{document}
\sloppy

\abstract{
We derive the first systematic observational constraints on reheating in models of inflation where an additional light scalar field contributes to primordial density perturbations and affects the expansion history during reheating. This encompasses the original curvaton model but also covers a larger class of scenarios. We find that, compared to the single-field case, lower values of the energy density at the end of inflation and of the reheating temperature are preferred when an additional scalar field is introduced. For instance, if inflation is driven by a quartic potential, which is one of the most favoured models when a light scalar field is added, the upper bound $T_\ureh<5\times 10^{4}\,\GeV$ on the reheating temperature $T_\ureh$ is derived, and the implications of this value on post-inflationary physics are discussed. The information gained about reheating is also quantified and it is found that it remains modest in plateau inflation (though still larger than in the single-field version of the model) but can become substantial in quartic inflation. The role played by the vev of the additional scalar field at the end of inflation is highlighted, and opens interesting possibilities for exploring stochastic inflation effects that could determine its distribution.
}



\maketitle

\section{Introduction}
\label{sec:intro}
Inflation~\cite{Starobinsky:1980te, Sato:1980yn, Guth:1980zm, Linde:1981mu, Albrecht:1982wi, Linde:1983gd} is the leading paradigm to describe the physical conditions that prevailed in the very early Universe. During this accelerated expansion epoch, cosmological perturbations are amplified from the vacuum quantum fluctuations of the gravitational and matter fields~\cite{Starobinsky:1979ty, Mukhanov:1981xt, Hawking:1982cz,  Starobinsky:1982ee, Guth:1982ec, Bardeen:1983qw}. This is why recent high-quality measurements~\cite{Ade:2013sjv, Adam:2015rua, Array:2015xqh} of these inhomogeneities in the Cosmic Microwave Background (CMB) have significantly improved our knowledge of inflation~\cite{Ade:2015lrj, Martin:2013tda, Martin:2013nzq, Vennin:2015eaa}. However, how inflation ends and is connected to the subsequent hot Big-Bang phase through the so-called ``reheating'' era is still poorly constrained. The main reason is that at linear order, in absence of entropic perturbations, curvature perturbations are preserved on large scales~\cite{Lukash:1980iv, Bardeen:1983qw}, hence their statistical properties at recombination time carry limited direct information about the microphysics at play during the reheating epoch. 

Nevertheless, the amount of expansion between the end of inflation and the onset of the radiation epoch determines the amount of expansion between the Hubble crossing time of the physical scales probed in the CMB and the end of inflation~\cite{Martin:2006rs, Martin:2010kz, Easther:2011yq, Dai:2014jja, Rehagen:2015zma, Domcke:2015iaa}. As a consequence, the kinematic properties of reheating set the time frame during which the fluctuations probed in cosmological experiments emerge, hence defining the location of the observational window along the inflationary potential. If inflation is realised with a single slowly-rolling field for instance, this effect can be used to extract constraints on a certain combination of the averaged equation-of-state parameter during reheating and the reheating temperature, the so-called ``reheating parameter'', yielding an information gain of about 1 bit on the reheating history~\cite{Martin:2014nya, Martin:2016oyk}. 

Since the reheating parameter is related to quantities such as the effective potential of the inflationary fields during reheating and the couplings between these fields and their decay products, this provides an indirect probe into the fundamental microphysical parameters of reheating~\cite{Drewes:2015coa}. Deriving such a relationship for concrete reheating models is therefore an important, although often laborious, task. Let us also notice that since the dependence of inflationary predictions on the reheating history is now of the same order as the accuracy of the data itself, different prescriptions for the reheating dynamics give rise to substantially different results regarding which inflationary models are preferred by the data~\cite{Martin:2014nya, Martin:2014rqa}. Therefore, improving our understanding of reheating has become crucial to derive meaningful constraints on inflation itself.
\begin{figure}[t]
\begin{center}
\includegraphics[width=14cm]{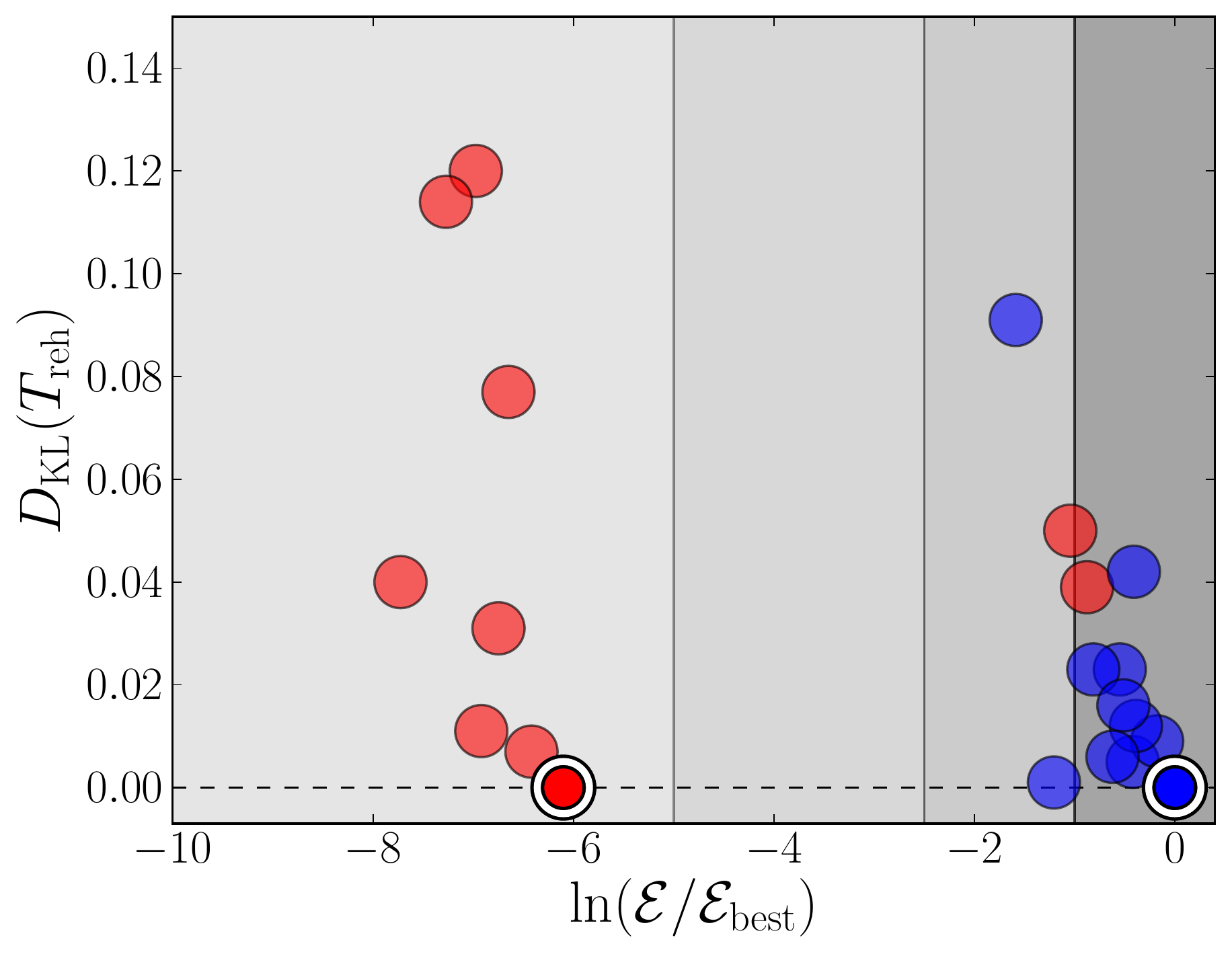}
\caption{Relative information gain between prior and posterior distributions over the reheating temperature $T_\ureh$, plotted against the Bayes factor normalised to the best model (single-field Higgs inflation). The white circled disks stand for purely single-field models (blue: Higgs inflation $\mathrm{HI}$, red: quartic inflation $\mathrm{LFI_4}$) and the other disks are the equivalent models with an additional light scalar field in the different reheating scenarios (blue: $\mathrm{MC}_i\mathrm{HI}$, red: $\mathrm{MC}_i\mathrm{LFI}_4$, for $i=1\cdots 10$). From left to right, the grey shading darkens, denoting the respective evidence ratios: strongly disfavoured, moderately disfavoured, weakly disfavoured and favoured. The two red disks that lie in the favoured region are reheating scenarios 5 and 8 for quartic inflation. A logarithmically flat prior has been used on the vev of the additional light scalar field at the end of inflation, see \Sec{sec:Bayesian}.}
\label{fig:mainresult}
\end{center}
\end{figure}

In this work, we follow this line of research and study the situation where inflation is driven by a single scalar inflaton field $\phi$, but an extra light (relative to the inflationary Hubble scale) scalar field $\sigma$ can also contribute to the total amount of curvature perturbations. This added field $\sigma$ is assumed to be subdominant during inflation but can store a substantial part of the energy budget of the Universe during reheating. In the limit where it is entirely responsible for the observed primordial curvature perturbations, the class of models this describes is essentially the curvaton scenario~\cite{Linde:1996gt, Enqvist:2001zp, Lyth:2001nq, Moroi:2001ct, Bartolo:2002vf}. Here however, we address the generic setup where both $\phi$ and $\sigma$ can a priori contribute to curvature perturbations~\cite{Dimopoulos:2003az, Langlois:2004nn, Lazarides:2004we, Moroi:2005np}. The reasons why we focus on these scenarios are threefold. First, from a theoretical perspective, most physical setups that have been proposed to embed inflation contain extra scalar fields that can play a role either during inflation or afterwards. This is notably the case in string theory models where extra light scalar degrees of freedom are usually considered~\cite{Turok:1987pg, Damour:1995pd, Kachru:2003sx, Krause:2007jk, Baumann:2014nda}. Second, from an observational point of view, these scenarios predict levels of non-Gaussianities that may lie within the reach of the next generation of cosmological surveys~\cite{Cooray:2006km, Pillepich:2006fj, Alvarez:2014vva, Munoz:2015eqa}. Their observational status is therefore likely to evolve in the coming years, which is why it is important to improve our understanding of these models. Third, at the practical level, these scenarios are interesting since the reheating parameter is an explicit function of the decay rates of both fields, the mass of the light field $\sigma$ and its vev at the end of inflation. This means that the same parameters determine the direct imprint of $\sigma$ on the statistics of curvature perturbations and the reheating kinematic effect on the location of the observational window along the inflaton potential. The associated increased sensitivity of the data to these parameters should allow us to better constrain them.

These scenarios have recently been brought into the full domain of Bayesian analysis in \Refs{Hardwick:2015tma, Vennin:2015vfa, Vennin:2015egh}. In this paper, we make use of the Bayesian inference techniques developed in these works to derive constraints on the inflationary energy scale and the reheating temperatures, and quantify the gain in information about these quantities from current observations. The paper is organised as follows. In \Sec{sec:Modeling}, we present in greater details the scenarios at hand and explain how information on reheating can be extracted using Bayesian inference. In \Sec{sec:results}, we provide our main results and analyse their implications for the physics of reheating and the amount of information that has been gained. In \Sec{sec:Discussion}, we extend the discussion by considering the role played by the inflationary energy scale in plateau potentials, the impact of gravitino overproduction bounds and the constraints on decay rates. We present our conclusions in \Sec{sec:Conclusions} and then end the paper with several appendices. In \App{Sec:DependencyTree}, we analyse the degeneracies and interdependences between the parameters of the problem. In \App{sec:DKL}, we present the Kullback-Leibler divergence as a tool to quantify information gain. In \App{Sec:IndividualScenarios}, we present our results for individual reheating scenarios. In \App{Sec:DKLDensity} finally, we discuss information gain densities.
\section{Method}
\label{sec:Modeling}
The method employed in this paper combines the analytical work of \Ref{Vennin:2015vfa} with the numerical tools developed in \Refs{Ringeval:2013lea, Martin:2013nzq, Vennin:2015egh}. In this section, we describe its main aspects and explain the use of Bayesian inference techniques and information gain quantification to analyse constraints on the parameters of reheating. 
\subsection{Curvaton and Reheating}
\label{sec:introscenarios}
\begin{figure}[t]
\begin{center}
\includegraphics[width=\textwidth]{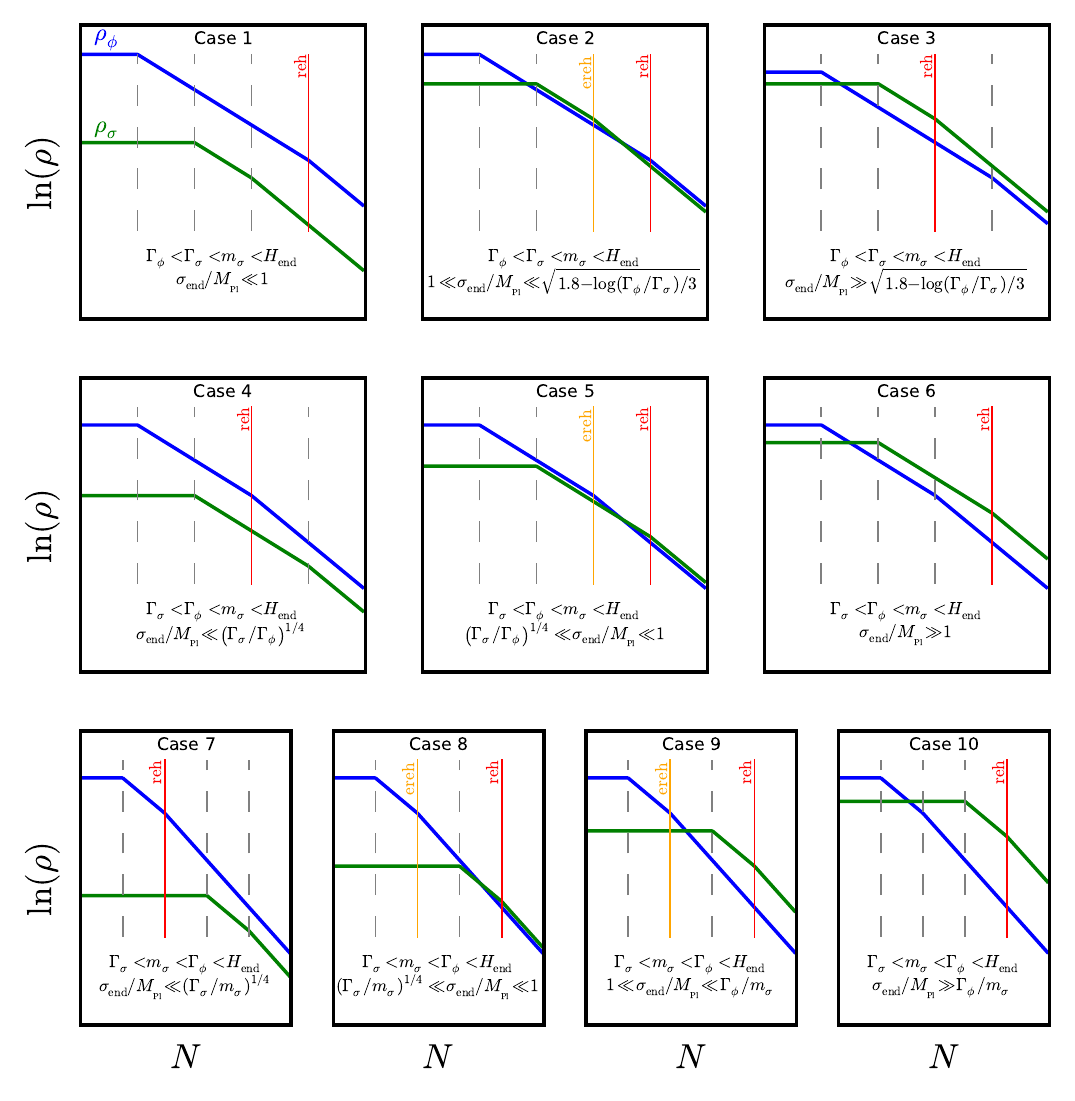}
\caption{Different possible reheating scenarios, depending on the values taken by $\Gamma_\sigma$, $m_\sigma$, $\Gamma_\phi$, $H_\uend$ and $\sigma_\uend$. Cases 1, 2 and 3 correspond to $\Gamma_\phi<\Gamma_\sigma<m_\sigma<H_\uend$; cases 4, 5 and 6 correspond to $\Gamma_\sigma<\Gamma_\phi<m_\sigma<H_\uend$; cases 7, 8, 9 and 10 correspond to $\Gamma_\sigma<m_\sigma<\Gamma_\phi<H_\uend$. Within each row, different cases are distinguished by $\sigma_\uend/\Mp$ which controls when $\sigma$ dominates the total energy density (the precise values for $\sigma_\uend$ at the limit between the different scenarios are given in \Ref{Vennin:2015vfa}). The blue curves stand for the energy density of $\phi$ while the green ones are for $\sigma$. The time at which the total energy density corresponds to the reheating temperature $T_{\rm reh}$ is marked out in red for each case, and early reheating temperature $T_{\rm ereh}$ are denoted in orange when two disconnected radiation phases exist.}
\label{fig:cases}
\end{center}
\end{figure}

As explained in \Sec{sec:intro}, we study the case where inflation is driven by a single field $\phi$ slowly rolling down its potential $U(\phi)$, and an extra light scalar field $\sigma$ (with mass $m_\sigma$ smaller than the inflationary Hubble scale) is present both during inflation and reheating. We therefore consider potentials of the type
\bea
V\left(\phi ,\sigma\right) = U\left(\phi\right) + \frac{1}{2}m_{\sigma}^2\sigma^2\,.
\eea
This extra field $\sigma$ is taken to be subdominant at the level of the background energy density during the whole inflationary epoch. Both fields are assumed to be slowly rolling during inflation, and eventually decay into radiation fluids with decay rates\footnote{
Here, $\Gamma_\phi$ (respectively $\Gamma_\sigma$) are effective values for which assuming instantaneous decay at $H=\Gamma_\phi$ (respectively $H=\Gamma_\sigma$) provides a good description of the full decay dynamics. 
} respectively denoted $\Gamma_\phi$ and $\Gamma_\sigma$, during reheating. While we require that $\phi$ becomes massive at the end of inflation, we do not make any assumption as to the ordering of the three events: $\sigma$ becomes massive, $\phi$ decays and $\sigma$ decays. Nor do we restrict the epochs during which $\sigma$ can dominate the energy content of the Universe. This leaves us with 10 possible cases (including situations where $\sigma$ drives a secondary phase of inflation~\cite{Langlois:2004nn, Moroi:2005kz, Ichikawa:2008iq, Dimopoulos:2011gb}), depending on the vev of $\sigma$ at the end of inflation $\sigma_\uend$. These ten ``reheating scenarios'' are listed and detailed in \Ref{Vennin:2015vfa} but are sketched in \Fig{fig:cases}. The usual curvaton scenario corresponds to case number 8 but one can see that a much wider class of models is covered by the present analysis.

In this work, we also assume that all particles are in full thermal equilibrium after $\phi$ and $\sigma$ decay. Therefore, there are no residual isocurvature modes~\cite{Lyth:2002my, Weinberg:2004kf}, that would otherwise give rise to additional constraints. Since such constraints depend on the specific processes of decay and thermalisation~\cite{Langlois:2004nn, Lemoine:2006sc, Langlois:2008vk, Lemoine:2008qj, Smith:2015bln}, we will consider this effect in a separate publication. Thermal equilibrium also allows us to relate energy densities $\rho_\urad$ contained in radiation fluids to temperatures through
\bea
T=\left(\frac{30\rho_\urad}{\pi^2 g_*}\right)^{1/4}\, ,
\label{eq:T:rho}
\eea
where $g_*$ is the effective number of degrees of freedom. When this expression is evaluated at the onset of the Big-Bang radiation epoch, it yields the ``reheating temperature'' $T_\ureh$. In reheating scenarios 1, 2, 4 and 7 (see \Fig{fig:cases}), this corresponds to the temperature of the thermalised decay products of $\phi$, while for scenarios 3, 5, 6, 8, 9 and 10, this corresponds to the decay products of $\sigma$. However, it can also happen that a transient radiation epoch takes place during reheating (as in reheating scenarios 2, 5, 8 and 9), in which case the energy density of the Universe at the beginning of this first radiation phase is called ``early reheating temperature'' and is noted $T_\uereh$. In reheating scenarios 5, 8 and 9, this corresponds to the decay products of $\phi$, while in scenario 2, this corresponds to the decay products of $\sigma$.

In \Ref{Vennin:2015vfa}, the $\delta N$ formalism~\cite{Starobinsky:1986fxa, Salopek:1990jq, Sasaki:1995aw, Sasaki:1998ug, Wands:2000dp, Lyth:2004gb, Lyth:2005fi} and the sudden decay approximation~\cite{Malik:2006pm,Sasaki:2006kq} were employed to relate observables of the models considered here to variations in the energy densities of both fields at the decay time of the last field. This allows one to calculate all relevant physical quantities by only keeping track of the background energy densities. Analytical expressions have been derived for all $10$ reheating scenarios, that have been implemented in the publicly available \texttt{ASPIC} library~\cite{aspic}. For a given inflaton potential, and from the values of $\Gamma_\phi$, $\Gamma_\sigma$, $m_\sigma$ and $\sigma_\uend$, this code returns the value of the first three slow-roll parameters (or equivalently at second order in the slow-roll approximation, of the scalar spectral index $\nS$ and its running, and of the tensor-to-scalar ratio $r$) and of the local-type non-Gaussianity parameter $f_{\mathrm{NL}}$. In \Ref{Vennin:2015egh}, this has been interfaced with the ``effective likelihood via slow-roll reparametrisation'' of \Ref{Ringeval:2013lea}, and Bayesian constraints were derived for the models that we consider here. The results presented in this paper are obtained from this numerical pipeline, where the Planck 2015 $TT$ data are combined with the high-$\ell$ $C_\ell^{TE}+C_\ell^{EE}$ likelihood and the low-$\ell$ temperature plus polarisation likelihood (PlanckTT,TE,EE+lowTEB in the notations of \Ref{Aghanim:2015xee}, see table~1 there), together with the BICEP2-Keck/Planck likelihood described in \Ref{Ade:2015tva}.

An important result of \Ref{Vennin:2015egh} is that the models favoured by the data are of two types: either the inflaton has a ``plateau potential'' (\ie is a monotonically increasing function of $\phi$ that asymptotes a constant positive value at infinity) and the reheating scenario can be any of the 10 cases listed in \Fig{fig:cases}, or the inflaton has a ``quartic potential'' (\ie is proportional to $\phi^4$) and reheating occurs in scenario 5 or 8. For this reason, we restrict the following analysis to these two kinds of potential. As an example of a plateau potential, we consider the one of Higgs inflation ($\mathrm{HI}$)
\bea
U\left(\phi\right)=M^4\left(1-e^{-\sqrt{\frac{2}{3}}\frac{\phi}{\Mp}}\right)^2\, ,
\label{eq:pot:hi}
\eea
which also matches the Starobinsky model~\cite{Starobinsky:1980te} (in \Sec{sec:KMIII}, another plateau potential is studied, ``K\"ahler moduli II inflation'', to investigate the role played by the inflationary energy scale in plateau models). The other potential we consider is the one of quartic inflation ($\mathrm{LFI}_4$)
\bea
\label{eq:pot:quartic}
U\left(\phi\right)=M^4\left(\frac{\phi}{\Mp}\right)^4\, .
\eea 
Here, ``$\mathrm{HI}$'' and ``$\mathrm{LFI}_4$'' refer to the terminology of \Ref{Martin:2013tda} and stand for the purely single-field versions of these models. When the prefix ``$\mathrm{MC}$'' is appended (for ``Massive Curvaton''), the index following the prefix refers to the reheating scenario number. For example, $\mathrm{MC}_5\mathrm{LFI}_4$ corresponds to the case where the inflaton potential is of the quartic type, and where the reheating scenario is of the fifth kind.

In \Fig{fig:mainresult}, some of the results of \Ref{Vennin:2015egh} have been summarised for the $\mathrm{HI}$ models (blue disks, where the white circled disk stands for the single-field version of the model and the other disks represent the $10$ reheating scenarios) and the $\mathrm{LFI}_4$ models (red disks). On the horizontal axis, the Bayesian evidence is displayed. One can see that for Higgs inflation, adding a light scalar field slightly decreases the Bayesian evidence of the model but at a level which is inconclusive for most reheating scenarios (and never more than weakly disfavoured). For quartic inflation, the single-field version of the model is strongly disfavoured and so are most of the reheating scenarios when a light scalar field is added. Two exceptions are to be noted however, namely cases 5 and 8, which lie in the favoured region. On the vertical axis, the information gained on $T_\ureh$ is displayed, as will be defined and analysed in \Sec{sec:InformationGain}.
\subsection{Inverse Problem for Reheating Parameters}
\label{sec:InversionProblem}
As mentioned in \Sec{sec:intro}, a specific feature of the models considered in this work is that the same parameters determine the expansion history during reheating as well as the contribution from the additional light scalar field to the total curvature perturbations. This is responsible for a high level of interdependency between these parameters, that plays an important role in shaping the constraints we obtain in \Sec{sec:results}. For this reason, it is important to first better understand their origin.

The number of \efolds $\Delta N_*$ elapsed between the Hubble exit time of the CMB pivot scale $k_{{}_\mathrm{P}}$ and the end of inflation is given by~\cite{Martin:2006rs, Martin:2010kz, Easther:2011yq}
\bea
\Delta N_* = \frac{1-3\bar{w}_\ureh}{12\left(1+\bar{w}_\ureh\right)}\ln\left(\frac{\rho_\ureh}{\rho_\uend}\right)
+\frac{1}{4}\ln\left(\frac{\rho_*}{9\Mp^4}\frac{\rho_*}{\rho_\uend}\right)
-\ln\left(\frac{k_{{}_\mathrm{P}}/a_\mathrm{now}}{\tilde{\rho}_{\gamma,\,\mathrm{now}}^{1/4}}\right)\, .
\label{eq:DeltaNstar}
\eea
In this expression, $\bar{w}_\ureh=\int_\ureh w(N)\dd N/N_\ureh$ is the averaged equation of state parameter during reheating, $\rho_\ureh$ is the energy density of the Universe at the end of reheating, $\rho_*$ is the energy density calculated $\Delta N_*$ \efolds before the end of inflation (all the quantities with a subscript ``*'' are evaluated at that time), $a_\mathrm{now}$ is the present value of the scale factor, and $\tilde{\rho}_{\gamma,\,\mathrm{now}}$ is the the energy density of radiation today rescaled by the number of relativistic degrees of freedom. Taking the pivot scale $k_{{}_\mathrm{P}}/a_\mathrm{now}$ to be $0.05\, \mathrm{Mpc}^{-1}$ and $\tilde{\rho}_{\gamma,\,\mathrm{now}}$ to its measured value, the last term is $N_0\equiv-\ln(k_{{}_\mathrm{P}}/a_\mathrm{now}/\tilde{\rho}_{\gamma,\,\mathrm{now}}^{1/4})\simeq 61.76$.

Let us first illustrate the use of \Eq{eq:DeltaNstar} to constrain reheating in the simple case of single-field quartic inflation, where the potential is given by \Eq{eq:pot:quartic} and there is no additional light scalar field $\sigma$. As mentioned above, we require that $\phi$ becomes massive at the end of inflation, so that in this case, one simply has $\bar{w}_\ureh=0$. Inflation ends by slow-roll violation at $\phi_\uend=2\sqrt{2}\Mp$, so that $\rho_\uend=3 U(\phi_\uend)/2 = 96 M^4$. On the other hand, the slow-roll trajectory is given by $\phi_*^2/\Mp^2 = 8(\Delta N_*+1)$, so that $\rho_*=U(\phi_*)= 64 M^4(\Delta N_*+1)^2$. For this reason, $\Delta N_*$ also appears in the right hand side of \Eq{eq:DeltaNstar} and this formula should be viewed as an implicit equation for $\Delta N_*$. In fact, this is all the more true since $M^4$ also implicitly depends on $\Delta N_*$. Indeed, this mass scale can be fixed by requiring that the correct scalar power spectrum amplitude $A_{{}_\mathrm{S}}=(M/\Mp)^4(\phi_*/\Mp)^6/(192\pi^2) $ is obtained (where $A_{{}_\mathrm{S}}$ has been evaluated at leading order in slow roll in quartic inflation). Making use of \Eq{eq:T:rho} to express $\rho_\ureh$ in terms of $T_\ureh$, one then obtains
\bea
\left.\Delta N_*\right\vert_{\mathrm{LFI}_4} = 
\frac{1}{12}\ln\left(\frac{512}{135}g_*\right)
+\frac{1}{2}\ln\left[\frac{64\pi}{3}\left(1+\Delta N_*\right)^3\right]+\frac{1}{3}\ln\left(\sqrt{A_{{}_\mathrm{S}}}\frac{T_\ureh}{\Mp}\right)
+N_0
\, .
\eea
This equation can be inverted using the $-1$ branch of the Lambert function $W$, and one finds
\bea
\left.\Delta N_*\right\vert_{\mathrm{LFI}_4} &=& -1 -\frac{3}{2}W_{-1}\left[-\frac{5^{1/18} \ee^{-\frac{2}{3}(1+N_0)}}{2^{3/2} 3^{1/2} \pi^{1/3} g_*^{1/18} A_{{}_\mathrm{S}}^{1/9}}\left(\frac{\Mp}{T_\ureh}\right)^{2/9}\right]
\\ 
& \simeq & -1-\frac{3}{2}W_{-1}\left[-4.11\times 10^{-14} \left(\frac{\mathrm{MeV}}{T_\ureh}\right)^{2/9}\right]
\simeq 
45.23+\frac{1}{3}\ln\left(\frac{T_\ureh}{\mathrm{MeV}}\right)\, , 
\eea
where in the second equality, we have used $A_{{}_\mathrm{S}}\simeq 2.2\times 10^{-9}$~\cite{Ade:2013sjv}, $g_*\simeq 106.75$ and the value given above for $N_0$, and the last expression corresponds to the limit $\Delta N_*\gg 1$. This makes explicit the dependence of $\Delta N_*$ on the reheating temperature $T_\ureh$. Since observable quantities such as the scalar spectral index $\nS$ or the tensor-to-scalar ratio $r$ depend on $\Delta N_*$ through $\phi_*$, this means that the reheating temperature is directly constrained by CMB measurements,
\bea
\label{eq:lfi4:nsr:Treh}
\left.\nS\right\vert_{\mathrm{LFI}_4} \simeq 1-\frac{3}{46.23+\frac{1}{3}\ln\left(\frac{T_\ureh}{\mathrm{MeV}}\right)}\, ,\quad
\left. r\right\vert_{\mathrm{LFI}_4} \simeq \frac{16}{46.23+\frac{1}{3}\ln\left(\frac{T_\ureh}{\mathrm{MeV}}\right)}\, .
\eea 
From these expressions, it is clear that observational constraints on $\nS$ and $r$ directly translate into constraints on the reheating temperature $T_\ureh$. As this simple calculation shows, this is the consequence of many interdependencies between the parameters of the problem, that are schematically summarised in \Fig{dependencytree} of \App{Sec:DependencyTree}, where the top panel displays the situation of purely single-field models that we just discussed. 

When a light scalar field is added, these dependencies are sketched in the bottom panel of \Fig{dependencytree} and one can see that the situation is substantially more complicated. For instance, the averaged equation of state parameter $\bar{w}_\ureh$ does not vanish anymore but is a non-trivial function of $\rho_\uend$, $\Gamma_\phi$, $\Gamma_\sigma$, $m_\sigma$ and $\sigma_\uend$, that is different for each of the 10 reheating scenarios of \Fig{fig:cases} (this function is given in Appendix~B of \Ref{Vennin:2015vfa}). Then, the mass scale of the potential $M^4$ is not simply related to the amplitude of the scalar power spectrum since $A_{{}_\mathrm{S}}$ also receives a contribution from the light scalar field $\sigma$, and this contribution depends on $\rho_\uend$, $\Gamma_\phi$, $\Gamma_\sigma$, $m_\sigma$ and $\sigma_\uend$. As a result, the dependency of observable quantities on these parameters is much more complicated than the one obtained for a purely single-field model, and the constraints one can infer on the reheating temperatures for instance are a priori much less trivial. The goal of this paper is precisely to derive these constraints.
\subsection{Bayesian Inference and Prior Choices}
\label{sec:Bayesian}
Starting from the data sets $\mathcal{D}$ mentioned in \Sec{sec:introscenarios}, our goal is to derive observational constraints on the energy scale of inflation $\rho_\uend$ and the reheating temperatures $T_\ureh$ and $T_\uereh$. This can be done using Bayesian inference techniques~\cite{Cox:1946,Jeffreys:1961,deFinetti:1974,Trotta:2005ar,Trotta:2008qt}. In this framework, assuming model $\mathcal{M}_i$, the posterior probability $p$ of its parameters $\theta_{ij}$ (labeled by $j$) is expressed as
\bea
\label{eq:posterior:def}
p\left(\theta_{ij}\vert\mathcal{D},\mathcal{M}_i\right)=\frac{\mathcal{L}
\left(\mathcal{D}\vert\theta_{ij},\mathcal{M}_i\right)
\pi\left(\theta_{ij}\vert \mathcal{M}_i\right)}{\mathcal{E}\left(\mathcal{D}\vert\mathcal{M}_i \right) } \, .
\eea
In this expression, $\mathcal{L}(\mathcal{D}\vert\theta_{ij},\mathcal{M}_i)$ is the likelihood and represents the probability of observing the data $\mathcal{D}$ assuming the model $\mathcal{M}_i$ is true and $\theta_{ij}$ are the actual values of its parameters, $\pi (\theta_{ij}\vert \mathcal{M}_i )$ is the prior distribution on the parameters $\theta_{ij}$, and $\mathcal{E}\left(\mathcal{D}\vert\mathcal{M}_i \right)$ is a normalisation constant called the Bayesian evidence and defined as
\bea
\label{eq:evidence:def}
\mathcal{E}\left(\mathcal{D}\vert\mathcal{M}_i \right) 
= \int\dd\theta_{ij}\mathcal{L}
\left(\mathcal{D}\vert\theta_{ij},\mathcal{M}_i\right)
\pi\left(\theta_{ij}\vert \mathcal{M}_i\right)\, .
\eea
The Bayesian evidence of the models considered in this work have been computed in \Ref{Vennin:2015vfa} and here, we are interested in the posterior distributions $p$ for the energy scale of inflation and the reheating temperatures. Notice that these quantities are not necessarily ``fundamental'' parameters that we start from but can be derived from them. For example, as stressed in \Sec{sec:InversionProblem}, $\rho_\uend$ is a complicated function of the parameters $\lbrace \theta_V \rbrace$ characterising the inflaton potential, $\Gamma_\phi$, $\Gamma_\sigma$, $m_\sigma$ and $\sigma_\uend$. In this case, for a derived parameter $\theta_d$ that can be expressed as $\theta_d=f_i(\theta_{ij})$, one marginalises the distribution obtained in \Eq{eq:posterior:def} according to
\bea
p\left(\theta_d\vert\mathcal{D},\mathcal{M}_i\right) = \int_{f(\theta_{ij})=\theta_d}p(\theta_{ij}\vert\mathcal{D},\mathcal{M}_i)\dd\theta_{ij}\, .
\eea

In this method, the priors are important quantities as they encode physical information one has ``a priori'' on the values of the parameters that describe the models. For the parameters of the potential $\lbrace \theta_V \rbrace$, we use the same priors as the ones proposed in \Ref{Martin:2013nzq}, based on \Ref{Martin:2013tda}. Because the extra field $\sigma$ is supposed to be still light at the end of inflation, its mass $m_\sigma$ must be smaller than the Hubble scale at the end of inflation, $H_\uend$. The same condition applies to the two decay rates, $\Gamma_\phi,\ \Gamma_\sigma<H_\uend$, since both fields decay after inflation. On the other hand, we want the Universe to have fully reheated before Big Bang Nucleosynthesis (BBN), which means that the two decay rates are also bounded from below by $H_{\mathrm{BBN}}\simeq (10\,\MeV)^2/\Mp$. The same lower bound applies to $m_\sigma$ since, assuming perturbative decay, $m_\sigma>\Gamma_\sigma$. Between these two values, the order of magnitude of $m_\sigma$ and of the two decay rates is a priori unknown, which is why a logarithmically flat prior (or ``Jeffreys prior'') is chosen:
\begin{align}
\ln H_{\mathrm{BBN}} < \ln \Gamma_\phi,\,\ln\Gamma_\sigma,\,\ln m_\sigma < \ln H_\uend\, .
\label{eq:prior:massscales}
\end{align}
The relative orderings identified in \Fig{fig:cases} then determine which of the 10 reheating scenarios is realised for a given set of parameters. For $\sigma_\uend$, two different priors are considered. The first one, denoted $\pilog$, is logarithmic and consists in assuming that the order of magnitude of $\sigma_\uend$ is unknown,
\begin{align}
\label{eq:sigmaend:LogPrior}
\ln\sigma_\uend^\umin < \ln\sigma_\uend < \ln \sigma_\uend^\umax \, .
\end{align}
Here, $\sigma_\uend^\umin$ and $\sigma_\uend^\umax$ are the boundary values given for each reheating case in \Fig{fig:cases}. For cases 1, 4 and 7, the lower bound is taken to be $\sigma_\uend^\umin=H_\uend/(2\pi)$, corresponding to the minimal quantum dispersion of the field, and for cases 3, 6 and 10, the upper bound $\sigma_\uend^\umax$ is set by the condition that the extra phase of inflation driven by $\sigma$ is sufficiently short so that the pivot scale $k_{{}_\mathrm{P}}$ exits the Hubble radius during the first phase of inflation, driven by $\phi$. The second prior relies on the equilibrium distribution of a light spectator field in a Sitter space-time with Hubble scale $H_\uend$~\cite{Starobinsky:1986fxa, Enqvist:2012},
\bea
\pisto \left(\sigma_\uend\right) \propto \exp\left(-\frac{4\pi^2 m_\sigma^2\sigma_\uend^2}{3H_\uend^4}\right)\, ,
\label{eq:sigmaend:GaussianPrior}
\eea
referred to as the ``stochastic'' prior on $\sigma_\uend$. A few words of caution regarding the use of this prior are in order here. In practice, the timescale of equilibration can be very large for small values of $m_\sigma$, and the initial conditions for spectator fields are not necessarily erased during inflation~\cite{Enqvist:2012}. Also note that in non-plateau models, the time variation of $H$, even in the slow-roll regime, is such that the distribution~(\ref{eq:sigmaend:GaussianPrior}) is not an equilibrium solution anymore, even approximatively. Moreover, since $H_\uend$ depends on $\sigma_\uend$ itself (see the bottom panel of \Fig{dependencytree} and the discussion of \Sec{sec:InversionProblem}), \Eq{eq:sigmaend:GaussianPrior} is not a simple Gaussian function of $\sigma_\uend$. This is why the use of \Eq{eq:sigmaend:GaussianPrior} should only be seen as a way to study the effects of picking a specific preferred scale for $\sigma_\uend$. In practice, we therefore implement this prior by simply rejecting realisations for which the argument of the exponential function in \Eq{eq:sigmaend:GaussianPrior} is smaller than $1/10$ or larger than $10$ (we have checked that when changing these arbitrary values to, say, $1/100$ and $100$, very similar results are obtained). In what follows, the inclusion of these two priors for $\sigma_\uend$ allow us to examine prior dependency of the reheating constraints.
\section{Results and Analysis}
\label{sec:results}
Let us now present our main results. In \Sec{sec:contraints}, we display and analyse the constraints obtained on the energy scale of inflation $\rho_\uend$ and the two reheating temperatures $T_\ureh$ and $T_\ureh$. In \Sec{sec:InformationGain}, we quantify how much information has been gained about these quantities.
\subsection{Constraints on Inflationary Energy and Reheating Temperatures}
\label{sec:contraints}
The posteriors on $\rho_\uend$, $T_\ureh$ and $T_\uereh$ for all 10 individual reheating scenarios (see \Fig{fig:cases}) are given in \App{Sec:IndividualScenarios}. In this section, for the sake of conciseness, as well as to allow direct comparison with purely single-field models, only the constraints averaged over the reheating scenarios are shown. Such distributions can be computed in the following manner. For the purpose of illustration, let us consider two toy models $\mathcal{M}_1$ and $\mathcal{M}_2$, that both depend on the same parameter $\theta$. In model $\mathcal{M}_1$, $\theta$ is assumed to lie within the range $[a,b]$ with a flat prior distribution, while in model $\mathcal{M}_2$, $\theta$ lies within the range $[b,c]$ with a flat prior distribution too. The model $\mathcal{M}_{1+2}$ is defined to be the ``union'' of $\mathcal{M}_1$ and $\mathcal{M}_2$, where $\theta$ lies in $[a,c]$ with a flat prior distribution, so that $\mathcal{M}_1$ and $\mathcal{M}_2$ are simply sub-models of $\mathcal{M}_{1+2}$ (in the same manner as all 10 reheating scenarios $\mathrm{MC}_i\mathrm{XXI}$, for $1\leq i \leq 10$ and some inflaton potential $\mathrm{XXI}$, are submodels of $\mathrm{MCXXI}$). From \Eq{eq:posterior:def}, one can see that
\bea
p\left(\theta\vert\mathcal{D},\mathcal{M}_{1+2}\right) = 
\frac{\pi\left(\theta\vert \mathcal{M}_{1+2}\right)}
{\mathcal{E}\left(\mathcal{D}\vert \mathcal{M}_{1+2}\right)}
\left[
\frac{\mathcal{E}\left(\mathcal{D}\vert \mathcal{M}_{1}\right)}{\pi\left(\theta\vert \mathcal{M}_{1}\right)}p\left(\theta\vert\mathcal{D},\mathcal{M}_1\right)
+
\frac{\mathcal{E}\left(\mathcal{D}\vert \mathcal{M}_{2}\right)}{\pi\left(\theta\vert \mathcal{M}_{2}\right)}p\left(\theta\vert\mathcal{D},\mathcal{M}_2\right)
\right]\, .
\label{eq:posterior:mixed}
\eea
In this expression, the Bayesian evidence of $\mathcal{M}_{1+2}$ can be evaluated with \Eq{eq:evidence:def}, which gives rise to
\bea
\mathcal{E}\left(\mathcal{D}\vert \mathcal{M}_{1+2}\right) = \frac{b-a}{c-a}\mathcal{E}\left(\mathcal{D}\vert \mathcal{M}_{1}\right) + \frac{c-b}{c-a}\mathcal{E}\left(\mathcal{D}\vert \mathcal{M}_{2}\right)\, .
\label{eq:evid:combined}
\eea
By combining \Eqs{eq:posterior:mixed} and~(\ref{eq:evid:combined}), the posterior distribution of the parameter $\theta$ within model $\mathcal{M}_{1+2}$ can be written as
\bea
p\left(\theta\vert\mathcal{D},\mathcal{M}_{1+2}\right) =
\frac{\mathcal{E}\left(\mathcal{D}\vert \mathcal{M}_{1}\right) (b-a) p\left(\theta\vert\mathcal{D},\mathcal{M}_{1}\right)
+
\mathcal{E}\left(\mathcal{D}\vert \mathcal{M}_{2}\right) (c-b) p\left(\theta\vert\mathcal{D},\mathcal{M}_{2}\right)}
{\mathcal{E}\left(\mathcal{D}\vert \mathcal{M}_{1}\right) (b-a)
+
\mathcal{E}\left(\mathcal{D}\vert \mathcal{M}_{2}\right) (c-b)}\, .
\label{eq:posterior:averaged}
\eea
In other words, it is given by the averaged sum of the posterior distributions within each sub-model, weighted by the product of the Bayesian evidence and the fractional prior volume of the sub-models. These fractional prior volumes can be viewed as priors for the sub-models themselves. In particular, one can check that \Eq{eq:posterior:averaged} is correctly normalised. 

The above formula can easily be generalised for arbitrary priors and arbitrary number of sub-models. In practice, the Bayesian evidence and fractional prior volumes of all 10 reheating scenarios are given in \Ref{Vennin:2015egh} for the inflaton potentials considered here, and we compute posterior distributions averaged over reheating scenarios adopting this approach. They correspond to the constraints one would obtain starting from the priors~(\ref{eq:prior:massscales}), without the ordering conditions of \Fig{fig:cases}, and simply computing observables according to the reheating scenario in which each sampled point falls.
\subsubsection{Energy Density at the End of Inflation}
\label{sec:result:rhoend}
\begin{figure}[t]
\figpilogsto
\begin{center}
\includegraphics[width=0.45\textwidth]{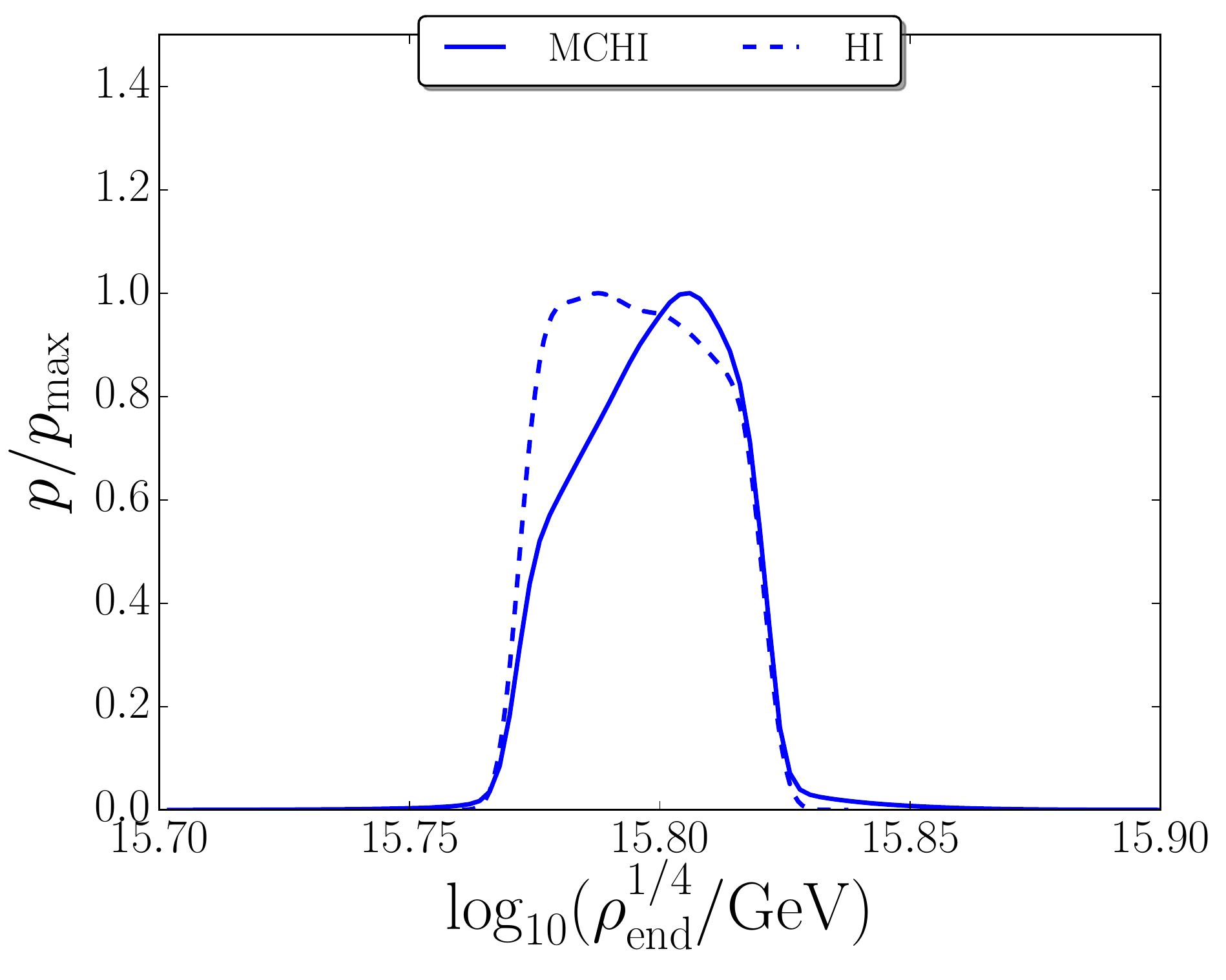}
\includegraphics[width=0.45\textwidth]{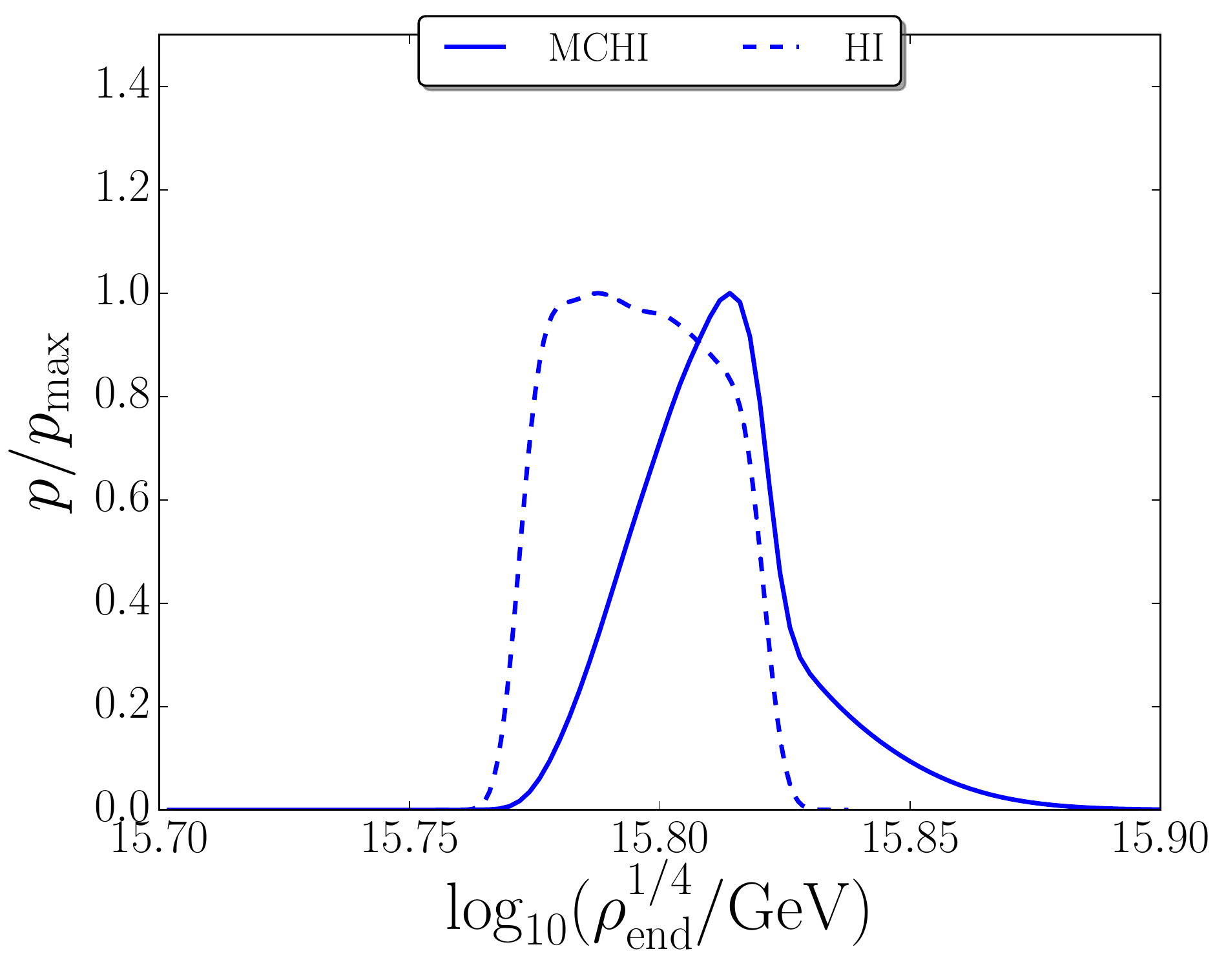}

\includegraphics[width=0.45\textwidth]{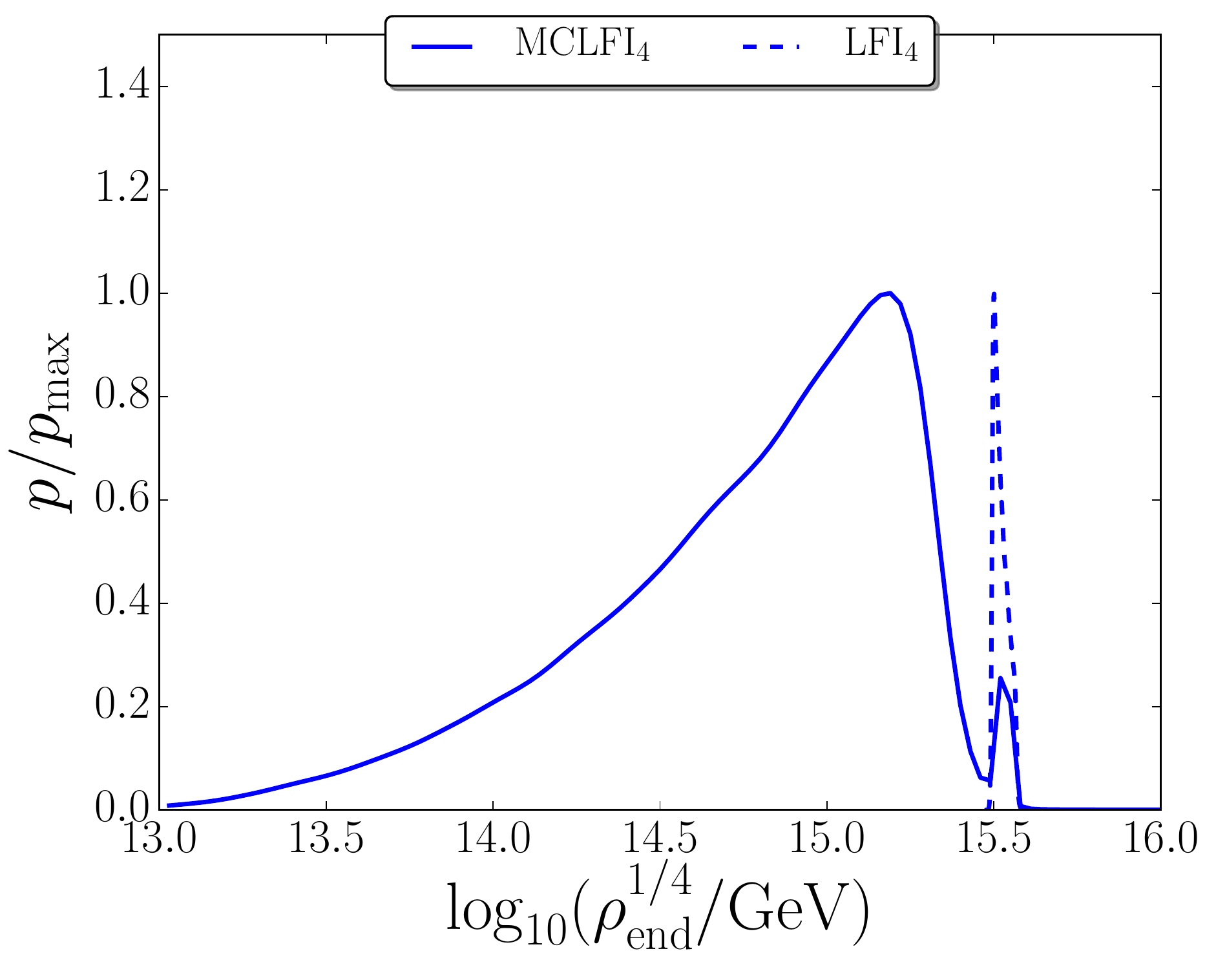}
\includegraphics[width=0.45\textwidth]{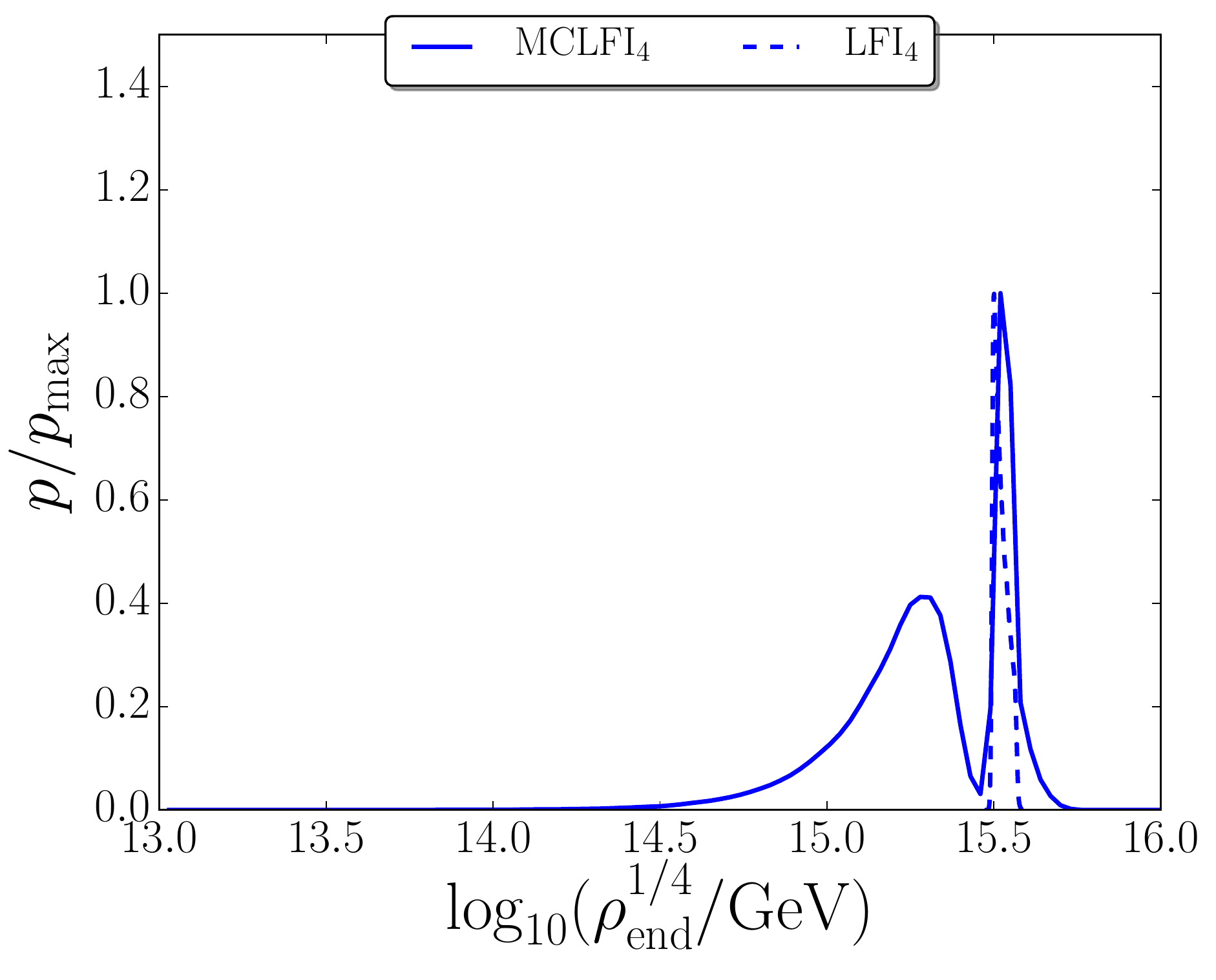}
\caption{Posterior distributions on the energy density at the end of inflation with the plateau potential~(\ref{eq:pot:hi}) of Higgs inflation (top panels) and the quartic potential~(\ref{eq:pot:quartic}) (bottom panels). The left panels correspond to the logarithmically flat prior~(\ref{eq:sigmaend:LogPrior}) $\pilog$ on $\sigma_\uend$, and the right panels stand for the stochastic prior~(\ref{eq:sigmaend:GaussianPrior}) $\pisto$ derived from the equilibrium distribution of a light scalar field in a de Sitter space-time with Hubble scale $H_\uend$. The dashed lines correspond to the single-field versions of the models, while the solid lines stand for the averaged posterior distributions over all $10$ reheating scenarios.}
\label{fig:post:rhoend:averaged}
\end{center}
\end{figure}
In \Fig{fig:post:rhoend:averaged}, the posterior distributions on $\rho_\uend$, the energy density at the end of inflation, is displayed. If the inflaton potential is of the plateau type (Higgs inflation, top panels), the difference between the purely single-field result and the one with an extra light scalar field, averaged over all $10$ reheating scenarios, is very small. One can check that this is also the case at the level of the individual posterior distributions for the different reheating scenarios in \Fig{fig:post:rhoend:individual} of \App{sec:app:rhoend:individual}. This is consistent with the generic robustness of plateau models under the introduction of extra light scalar fields noticed in \Ref{Vennin:2015egh}. In particular, the range of values allowed for $\rho_\uend$ is remarkably narrow. The stochastic prior tends to favour slightly larger values of the energy density. This is because this prior samples larger values of $\sigma_\uend$, hence larger contributions of $\sigma$ to the total curvature power spectrum~\cite{Vennin:2015egh}, hence bluer values of $\nS$. This effect can be compensated for by increasing $\Delta N_*$, hence $\rho_\uend$ [see \Eq{eq:DeltaNstar}], which decreases $\nS$ back into the data's sweet spot~\cite{Vennin:2015vfa}.

The situation is quite different for the quartic potential (bottom panels). In this case, the single-field version of the model provides a very poor fit to the data due to values of the tensor-to-scalar ratio $r$ that are too large. When a light scalar field is introduced, $r$ is typically decreased, and so is $\rho_\uend$. In scenarios where the amount of non-Gaussianities remains small, \ie scenarios 5 and 8, this explains why lower values of $\rho_\uend$ are favoured, see \Fig{fig:post:rhoend:individual}. In other cases, $\fnl$ increases when $r$ decreases, and the trade-off between both effects leads to bimodal posterior distributions. Since scenarios 5 and 8 are favoured however (see \Fig{fig:mainresult}), the clear preference is for lower values of $\rho_\uend$. If a stochastic prior on $\sigma_\uend$ is used, the maximum of the distribution is switched back to the single-field prediction, but all reheating scenarios are moderately or strongly disfavoured in this case anyway~\cite{Vennin:2015egh}.
\subsubsection{Reheating Temperature}
\label{sec:result:Treh}
\begin{figure}[t]
\figpilogsto
\begin{center}
\includegraphics[width=0.45\textwidth]{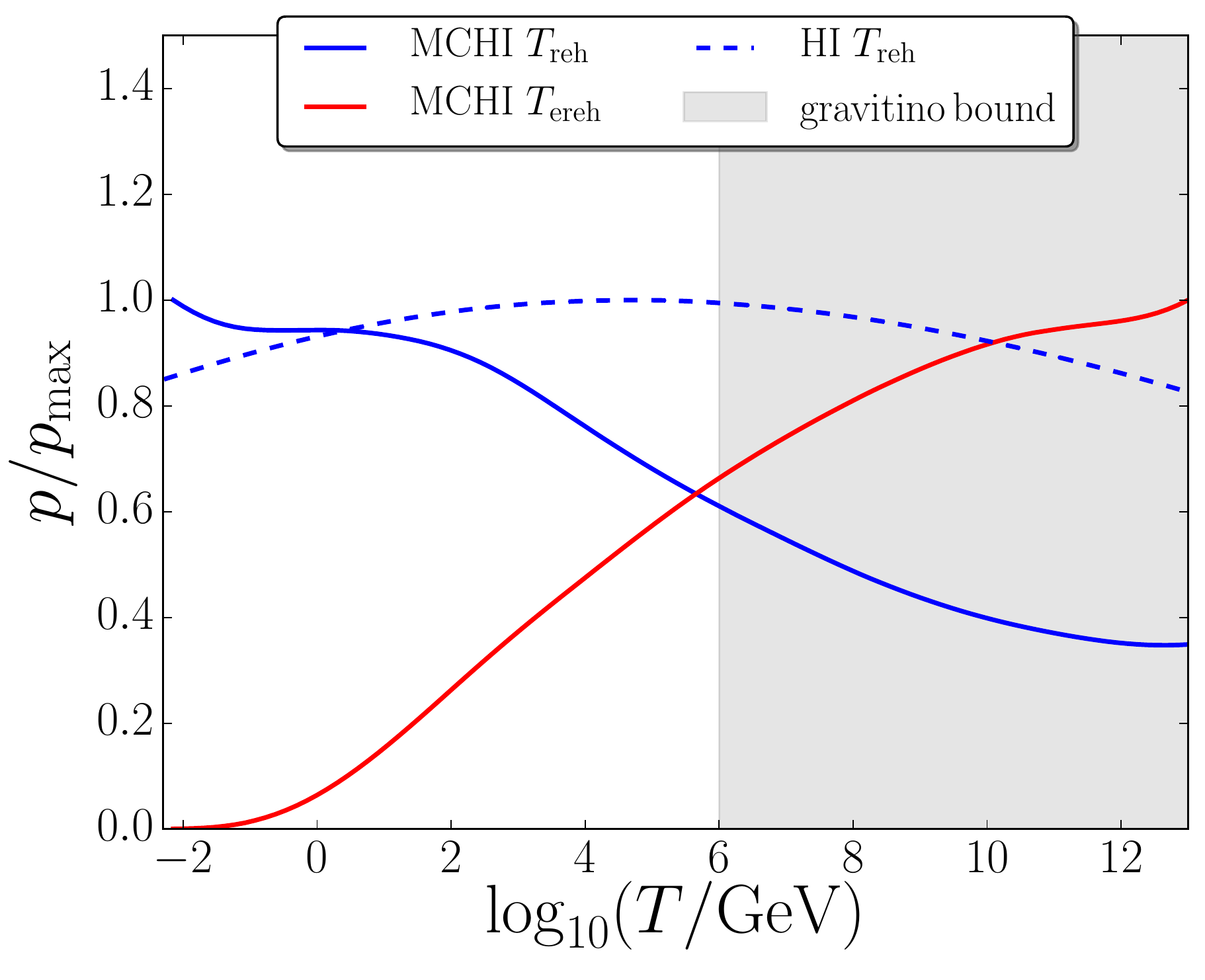}
\includegraphics[width=0.45\textwidth]{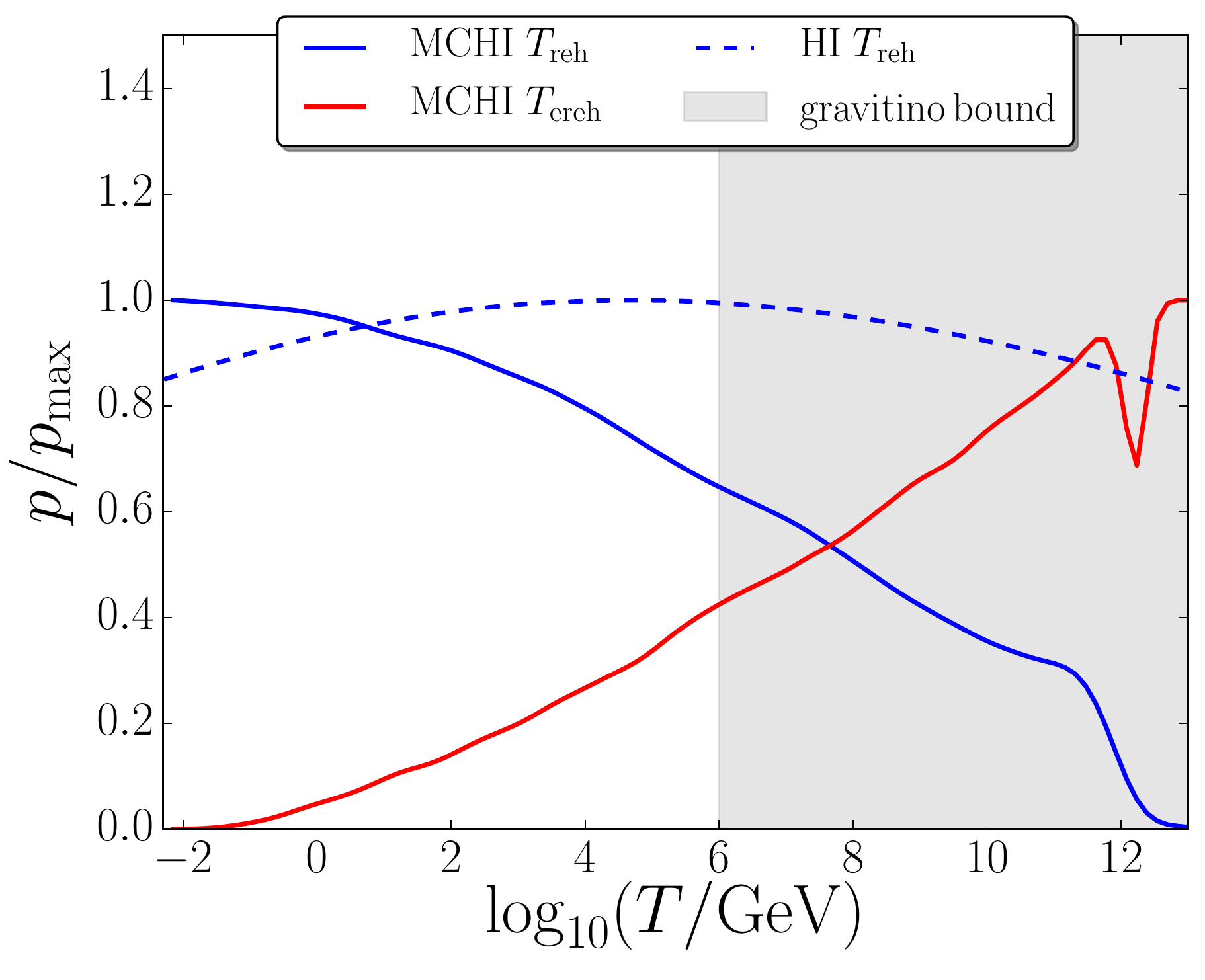}
\includegraphics[width=0.45\textwidth]{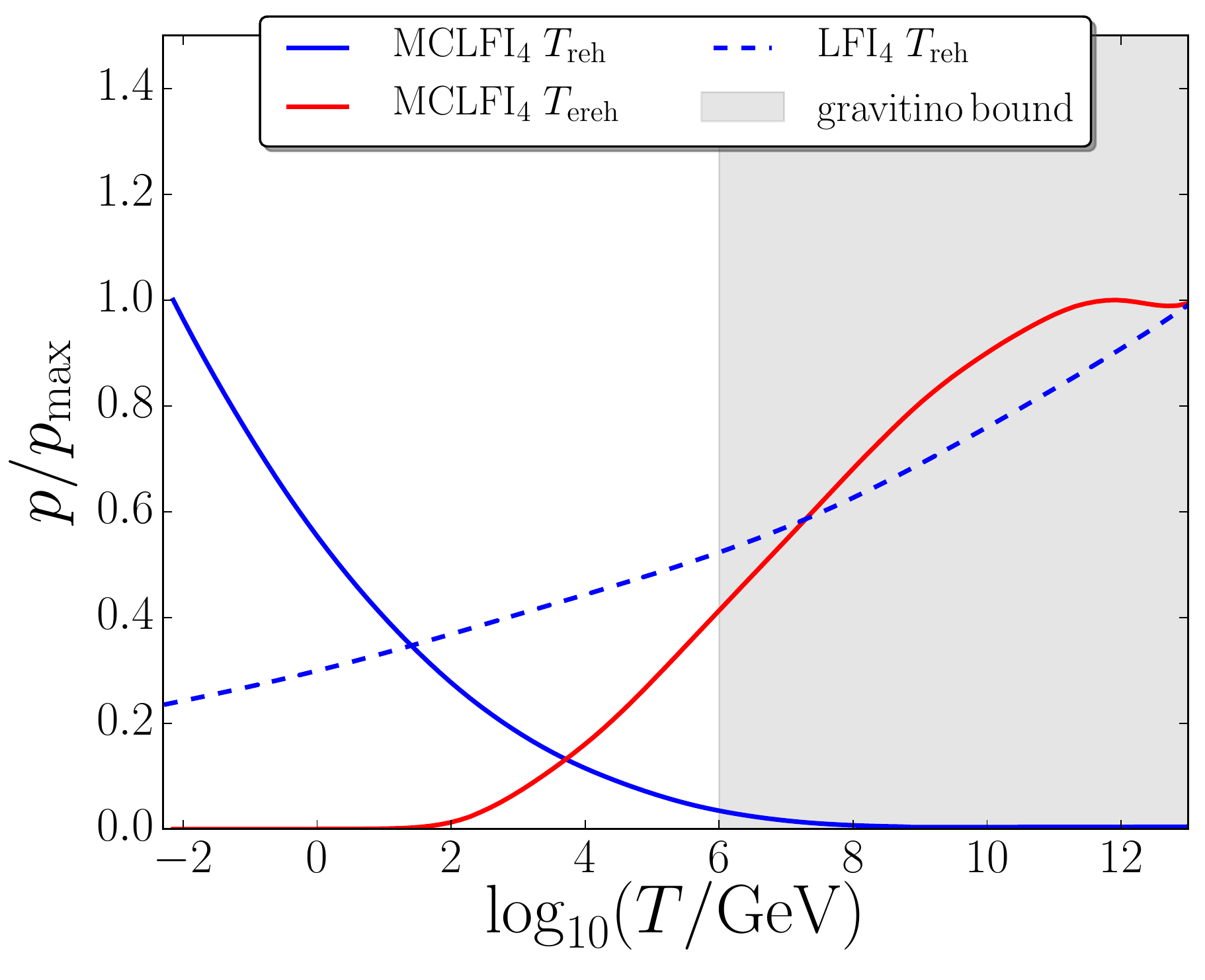}
\includegraphics[width=0.45\textwidth]{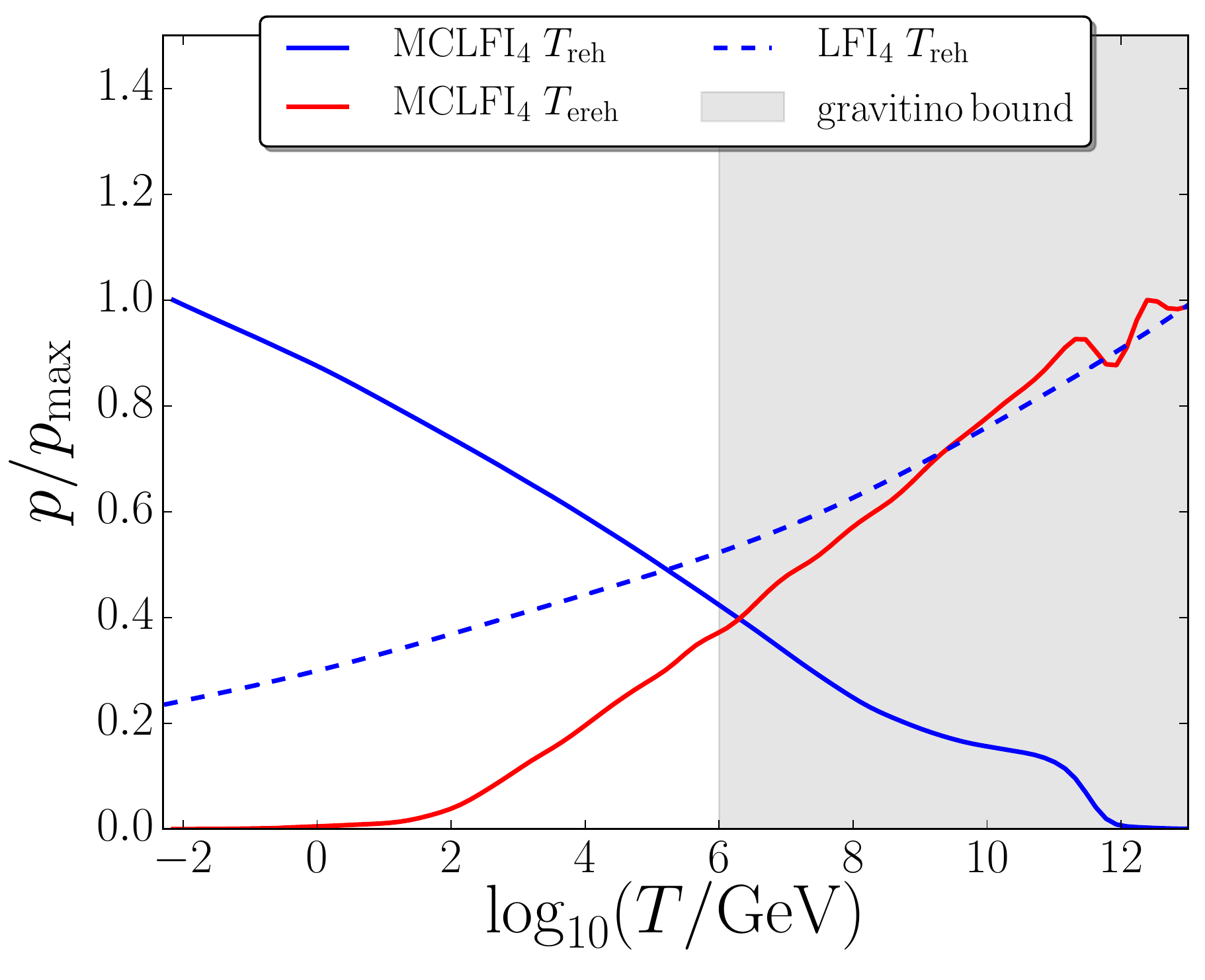}
\caption{Posterior distributions on the reheating temperature $T_\ureh$ and early reheating temperature $T_\uereh$ with the plateau potential~(\ref{eq:pot:hi}) of Higgs inflation (top panels) and the quartic potential~(\ref{eq:pot:quartic}) (bottom panels). The left panels correspond to the logarithmically flat prior~(\ref{eq:sigmaend:LogPrior}) $\pilog$ on $\sigma_\uend$, and the right panels stand for the stochastic prior~(\ref{eq:sigmaend:GaussianPrior}) $\pisto$ derived from the equilibrium distribution of a light scalar field in a de Sitter space-time with Hubble scale $H_\uend$. The dashed blue lines correspond to the single-field constraints on $T_\ureh$, while the solid lines stand for the averaged posterior distributions on $T_\ureh$ (blue) and $T_\uereh$ (red) when an extra light scalar field is added. The grey shaded region corresponds to reheating temperatures that would be excluded by gravitino production, see \Sec{sec:gravitino}.}
\label{fig:post:Trehs:averaged}
\end{center}
\end{figure}
In \Fig{fig:post:Trehs:averaged}, the posterior distributions on the reheating temperature $T_\ureh$ are displayed. In the single-field version of the plateau model of Higgs inflation, the reheating temperature is rather unconstrained. This is because all reheating temperatures can accommodate the data equally well for this model (at least when $\bar{w}_\ureh=0$, see \Ref{Martin:2016oyk} otherwise). When a light scalar field is introduced however, a slight preference is found for lower reheating temperatures. Looking at \Fig{fig:post:Treh:individual} of \App{sec:app:Treh:individual}, one can see that in the case of the logarithmic prior on $\sigma_\uend$, this trend is mostly due to reheating scenarios 1, 2, 5, 6, 8 and 9, for which $T_\ureh$ is bounded from above. For scenarios 3, 4, and 10 however, the distributions have a maximum around the scale $T_\ureh\sim 10^4\, \mathrm{GeV}$, and for scenario 7, larger values of $T_\ureh$ are even preferred. A similar dichotomy is observed with the stochastic prior on $\sigma_\uend$ where scenarios 1, 2, 5 and 6 prefer smaller values of $T_\ureh$, scenarios 3, 7, 9 and 10 prefer larger values of $T_\ureh$, and scenarios 4 and 8 leave $T_\ureh$ unconstrained. When averaging over the 10 reheating scenarios, the resulting distributions show preference for lower values of $T_\ureh$, but because of these opposite individual behaviours, the constraint is not very strong.

For the single-field version of quartic inflation, larger values of the reheating temperature are preferred since they lead to smaller values for the tensor-to-scalar ratio $r$ as well as larger values of $\nS$ that are in better agreement with the data, as shown explicitly in \Eq{eq:lfi4:nsr:Treh}. When a light scalar field is introduced, one can note in \Fig{fig:post:Treh:individual} that the same variety of individual behaviours of the 10 reheating scenarios is obtained as with Higgs inflation. However, since scenarios 5 and 8 strongly dominate the averaged posterior distribution due to their large Bayesian evidence, and since they both show preference for lower values of $T_\ureh$, better constraints are obtained from the averaged posterior distribution than with a plateau potential. In practice, an upper bound on the reheating temperature can be derived,
\bea
\left.T_\ureh\right\vert_{\mathrm{MCLFI}_4} < 5\times 10^4\,\GeV \ \mathrm{at}\ 95\%\,\mathrm{C.L.}\, .
\label{eq:LFI4:constraint:Treh}
\eea
This value has been obtained with the logarithmic prior $\pilog$ on $\sigma_\uend$. With the stochastic prior, the constraint would be much weaker, but one should remember that this prior is not well motivated in that case and that $\mathrm{MCLFI}_4$ is strongly disfavoured~\cite{Vennin:2015egh} when $\pisto$ is used anyway. 
\subsubsection{Early Reheating Temperature}
\label{sec:result:Tereh}
The weighted posterior distributions on the early reheating temperature $T_{\rm ereh}$ are displayed as the solid red lines in \Fig{fig:post:Trehs:averaged}. Obviously, these distributions are averaged over the scenarios for which $T_\ureh$ is defined only, that is to say cases 2, 5, 8 and 9, and the individual posteriors are given in \Fig{fig:post:Tereh:individual} in \App{sec:app:Tereh:individual} for these scenarios. Contrary to the reheating temperature discussed in \Sec{sec:result:Treh}, one can see that larger values are preferred and that lower bounds on $T_\uereh$ can be obtained,
\bea
\left.T_\uereh\right\vert_{\mathrm{MCHI}} > 251\,\GeV \ \mathrm{at}\ 95\%\,\mathrm{C.L.}\, ,
\quad\quad
\left.T_\uereh\right\vert_{\mathrm{MCLFI}_4} > 10^5\,\GeV \ \mathrm{at}\ 95\%\,\mathrm{C.L.}\, ,
\eea
with a logarithmic flat prior on $\sigma_\uend$. In this case, from \Fig{fig:post:Tereh:individual}, one can see that the constraint mostly comes from scenarios 8 and 9, while the posterior distribution for scenarios 2 and 5 has a maximum around $10^7\, \mathrm{GeV}$ for Higgs inflation and $10^9\, \mathrm{GeV}$ for quartic inflation. If one uses the stochastic prior instead, one obtains
\bea
\left.T_\uereh\right\vert_{\mathrm{MCHI}} > 501\,\GeV \ \mathrm{at}\ 95\%\,\mathrm{C.L.}\, ,
\quad\quad
\left.T_\uereh\right\vert_{\mathrm{MCLFI}_4} > 2.5\times 10^4\,\GeV  \ \mathrm{at}\ 95\%\,\mathrm{C.L.}\, .
\eea
In this case, one can check in \Fig{fig:post:Tereh:individual} that all reheating scenarios favour large values for $T_\uereh$.
\subsection{Information Gain}
\label{sec:InformationGain}
\begin{figure}[t]
\begin{center}
\includegraphics[width=9cm]{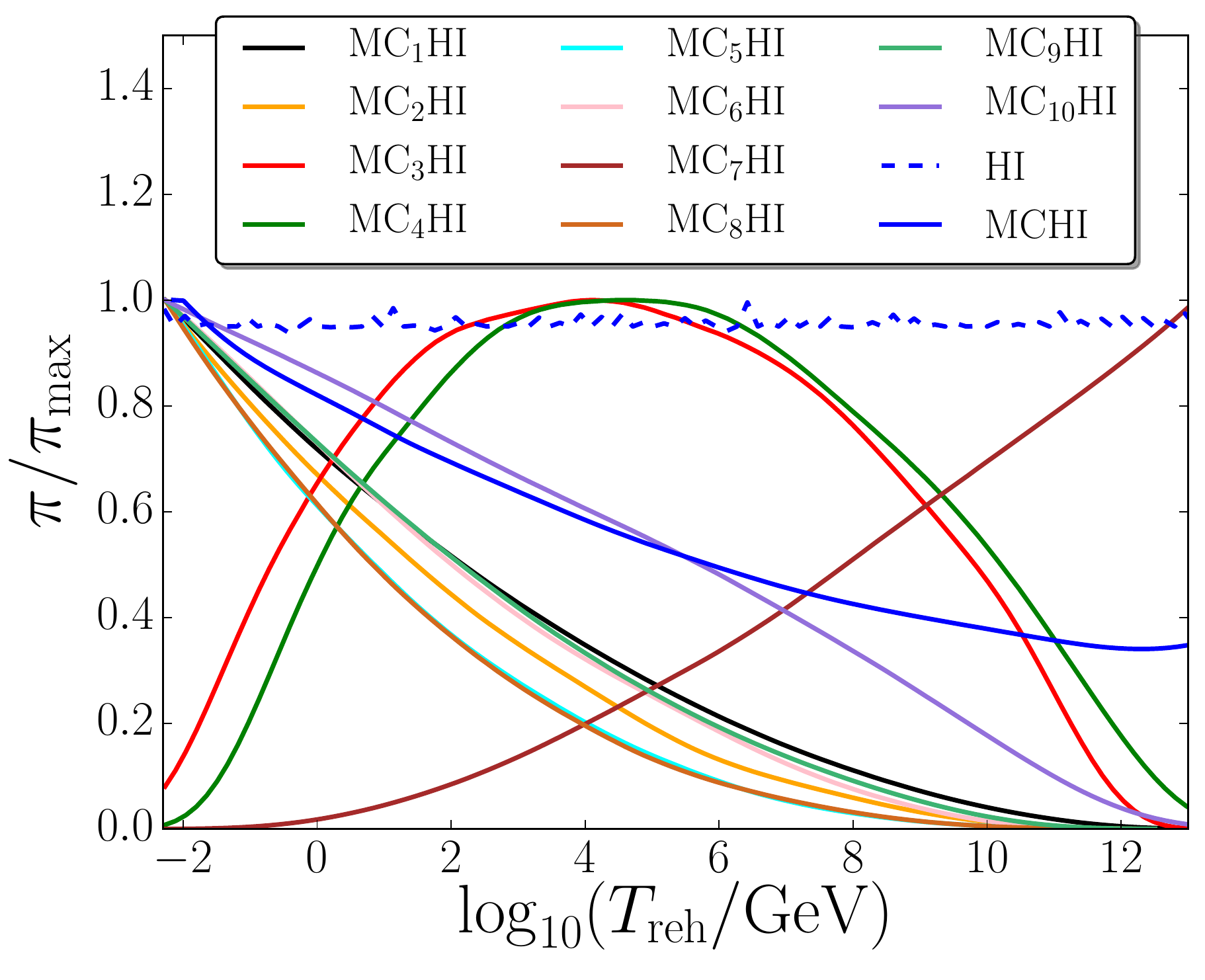}
\caption{Induced prior on $\log_{10}(T_{\rm reh}/\mathrm{GeV})$ for the Higgs inflation potential and using the logarithmic prior~(\ref{eq:sigmaend:LogPrior}) on $\sigma_{\rm end}$. The dashed blue line stands for the single-field version of the model for which the prior is flat, the solid coloured lines correspond to the 10 reheating scenarios when a light scalar field is added and the solid blue line is the averaged prior distribution over all reheating scenarios.}
\label{fig:prior:Treh}
\end{center}
\end{figure}
In \Sec{sec:contraints}, the posterior distributions on $\rho_\uend$, $T_\ureh$ and $T_\uereh$ have been displayed and it was shown that, compared to single-field models, different constraints are obtained when a light scalar field is included. In \Sec{sec:InversionProblem}, we explained that both situations are indeed qualitatively different, since in the later case the same parameters define both the contribution from $\sigma$ to the curvature power spectrum and the kinematic properties of reheating that determine the location of the observational window along the inflaton potential. This leads to an increased interdependence between these parameters and observations, which yields more information about these quantities. This is why in this section, we quantify the information gain on reheating parameters to quantitatively describe this effect.

A first remark is that since the induced priors on $\rho_\uend$, $T_\ureh$ and $T_\uereh$ are not logarithmically flat, information gain cannot be simply assessed by measuring how the distributions of \Sec{sec:contraints} are peaked, or more generally deviate from a flat profile. For example, in \Fig{fig:prior:Treh}, the induced priors\footnote{
In practice, induced priors are reconstructed using a fiducial, constant likelihood in our Bayesian inference code (so that the posteriors we extract correspond to the actual induced priors), where only the value of $A_{{}_\mathrm{S}}$ is used to normalise the mass scale $M^4$ appearing in the inflaton potentials. This is because $A_{{}_\mathrm{S}}$ is so accurately measured that it effectively reduces the support of the posterior to a hypersurface in parameter space, and distributions are considered along this hypersurface only.} on $\log T_\ureh$ are  displayed in the case of Higgs inflation, for the single-field model (dashed blue line), for the 10 reheating scenarios (coloured solid lines), and when averaged over all reheating scenarios (solid blue line). The prior is exactly flat in the single-field case since $T_\ureh$ is directly related to $\Gamma_\phi$ in this case, over which the logarithmically flat prior $ \ln H_{\mathrm{BBN}} < \ln \Gamma_\phi < \ln H_\uend$ is chosen. When a light scalar field is added however, $T_\ureh$ is either related to $\Gamma_\phi$ (in cases 1, 2, 4 and 7) or to $\Gamma_\sigma$ (in cases 3, 5, 6, 8, 9 and 10). Since the ordering conditions of \Fig{fig:cases} are further imposed on top of the logarithmically flat priors for these quantities, the non-flat induced priors of \Fig{fig:prior:Treh} are obtained. 

This is why the posterior distributions are not sufficient to estimate the information gain, but one needs to compute the relative information between the prior and posterior distributions. This can be done using the Kullback-Leibler divergence~\cite{kullback1951} $\DKL$ between the prior $\pi (\theta )$ and the posterior $p(\theta )$ of some parameter $\theta$ [here, for display convenience, the notations of \Sec{sec:Bayesian} are simplified, $p(\theta)\equiv p(\theta\vert\mathcal{D},\mathcal{M}_i)$, etc.],
\bea
\DKL\left(p \vert\vert \pi\right) \equiv \int^{\infty}_{-\infty} {p}\left(\theta \right) \log_2 \left[\frac{{p}\left(\theta \right)}{\pi \left(\theta \right)} \right] \dd \theta \equiv \int^{\infty}_{-\infty} \delta \DKL\left(\theta\right)\dd \theta \, .
\label{eq:DKL}
\eea
This is a measure of the amount of information provided by the data about the parameter $\theta$ (here, $\theta$ will be either $\rho_\uend$, $T_\ureh$ or $T_\uereh$). In \App{sec:DKL}, it is shown that the Kullback-Leibler divergence is invariant under any reparametrisation $\theta^\prime = f(\theta)$ (other properties of $\DKL$ are also discussed in this appendix). Since it uses a logarithmic score function as in the Shannon's entropy, it is therefore a well-behaved measure of information~\cite{bernardo:2008}.

The second equality in \Eq{eq:DKL} also defines the information density $\delta \DKL$, that can be viewed as the information gained in each bin $\dd\theta$ of the parameter $\theta$ under consideration. Contrary to $\DKL$, it is parameterisation dependent, but it indicates where information is mostly gained and lost. This quantity is displayed in \App{Sec:DKLDensity} for $\rho_\uend$, $T_\ureh$ and $T_\uereh$, for the 10 different reheating scenarios of \Fig{fig:cases}, for the three potentials considered in this work (Higgs inflation, quartic inflation and K\"ahler moduli II inflation - see \Sec{sec:KMIII}) and when the logarithmically flat prior $\pilog$ or the stochastic prior~(\ref{eq:sigmaend:GaussianPrior}) $\pisto$ on $\sigma_\uend$ are used. 

In this section, only the integrated Kullback-Leibler divergences are discussed. The numbers obtained for all models previously discussed are given in table~\ref{table:DKL} in \App{Sec:DKLDensity}. In table~\ref{table:DKL:summary}, the results are summarised and the divergence obtained in the single-field versions of the models are compared with the ones obtained from the averaged distributions over all 10 reheating scenarios. The averaged posterior distribution has been defined in \Sec{sec:contraints}, and the averaged prior distribution is simply the averaged sum of all prior distributions weighted by the fractional prior volume of the sub-models. Let us note that this divergence cannot be obtained by a simple weighted summation over each individual value. For instance, in table~\ref{table:DKL}, one can check that the divergence between averaged distribution can be larger than all individual divergences, as further discussed in \Sec{sec:InformationGain:Treh}.
\begin{table}[t]
\centering
\begin{tabular}{|l||*{3}{c|}}\hline
\backslashbox{Model}{$\DKL$}
&\makebox[3em]{$\rho_\uend$}&\makebox[3em]{$T_\ureh$}&\makebox[3em]{$T_\uereh$}
\\\hline\hline
$\mathrm{HI}$ & 1.370 & 0.004 & -\\\hline
$\mathrm{MCHI}(\pilog)$  & 0.114 & 0.005 & 0.018 \\\hline
$\mathrm{MCHI}(\pisto)$ & 0.224 & 0.006 & 0.014 \\\hline
$\mathrm{LFI}_4$ & 1.171 & 0.108 & - \\\hline
$\mathrm{MCLFI}_4(\pilog)$  & 3.104 & 0.656 & 0.181 \\\hline
$\mathrm{MCLFI}_4(\pisto)$ & 4.780 & 0.111 & 0.281 \\\hline
\end{tabular}
\caption{Kullback-Leibler divergences $\DKL$ on $\rho_\uend$, $T_\ureh$ and $T_\uereh$ for Higgs inflation and quartic large field inflation. The result is given for the single-field versions of the models and for the averaged priors and posteriors over the 10 reheating scenarios, when a logarithmically flat prior $\pilog$ on $\sigma_\uend$ is used, and with the stochastic prior $\pisto$ of \Eq{eq:sigmaend:GaussianPrior} as well. Note that the early reheating temperature $T_\ureh$ is not defined for single-field models, which is why no value is displayed.} 
\label{table:DKL:summary}
\end{table}
\subsubsection{Energy Density at the End of Inflation}
\label{sec:InformationGain:rhoend}
In table~\ref{table:DKL:summary}, one can see that more than one bit of information is gained on $\rho_\uend$ for the two single-field models considered here, $\mathrm{HI}$ and $\mathrm{LFI}_4$. The main reason is that, since these single-field potentials have no free parameters (apart from the overall mass scale $M^4$), as shown in \Sec{sec:InversionProblem}, $\rho_\uend$ is entirely fixed by $A_{{}_\mathrm{S}}$, up to a small dependence on $T_\ureh$. In this case, the support of both the priors and the posteriors on $\rho_\uend$ are very narrow, and even a small difference between their preferred values is enough to yield a large Kullback-Leibler divergence, see the discussion around \Eq{eq:DKL:gaussian} in \App{sec:DKL}. However, as soon as another free parameter is introduced in the inflaton potential for instance, this effect disappears as will be explicitly checked in \Sec{sec:KMIII}. Therefore, these large values of $\DKL$ for $\mathrm{HI}$ and $\mathrm{LFI}_4$ are mostly a consequence of the very sharp measurement on $A_{{}_\mathrm{S}}$.

When a light scalar field is added, a few tenths of bits of information on $\rho_\uend$ are typically gained with the plateau potential of Higgs inflation. This number can be larger for individual reheating scenarios, see for instance $\mathrm{MC}_{3}\mathrm{HI}$ and $\mathrm{MC}_{10}\mathrm{HI}$ in table~\ref{table:DKL} where, depending on the prior chosen for $\sigma_\uend$, one gains between one and two bits of information. The situation is particularly interesting for quartic inflation, where the by far favoured reheating scenarios are 5 and 8 (see \Fig{fig:mainresult}). For these models, one typically obtains one bit of information with the logarithmic prior on $\sigma_\uend$ and $3.5$ bits with the stochastic prior, see table~\ref{table:DKL}. This is because, as explained in \Sec{sec:result:rhoend}, the data favours regions of parameter space where $\sigma$ provides the main contribution to curvature perturbations and $\rho_\uend$ is smaller than its single-field counterpart, yielding non-trivial information about the energy density at the end of inflation. The divergence between the averaged distributions displayed in table~\ref{table:DKL:summary} is even larger, the additional information coming from the update in the relative degrees of belief between the different reheating scenarios, namely the fact that the data strongly favours scenarios 5 and 8. 

\subsubsection{Reheating Temperature}
\label{sec:InformationGain:Treh}
For the reheating temperature, very little information is gained with the single-field versions of the models. One may wonder whether this is consistent with \Ref{Martin:2016oyk}, where it is found that almost one bit of information is obtained on the reheating parameter of single-field models, on average. This is in fact the case since, in \Ref{Martin:2016oyk}, $\bar{w}_{\ureh}$ is allowed to vary between $-1/3$ and $1$. In \Eq{eq:DeltaNstar}, one can see that the dependence of $\Delta N_*$ on $T_\ureh$ is maximal when $\bar{w}_{\ureh}=-1/3$ [that is to say, the multiplying factor $(1-3\bar{w}_{\ureh})/(1+\bar{w}_{\ureh})$ between $\rho_\ureh$ and $\Delta N_*$ is maximal when $\bar{w}_{\ureh}=-1/3$], which explains why most of the information measured in \Ref{Martin:2016oyk} is gained close to $\bar{w}_{\ureh}=-1/3$. In the present work however, one imposes $\bar{w}_{\ureh}=0$ in the single-field models, to allow fair comparison with the situation where an extra light scalar field is introduced where it is assumed that the inflaton is massive between the end of inflation and its decay.

For the plateau potential of Higgs inflation, although more information on $T_\ureh$ is gained once an extra light scalar field is introduced, the Kullback-Leibler divergences remain small. With a quartic potential however, $0.66$ bits of information are obtained with the logarithmic prior on $\sigma_\uend$, which is a sizeable value. Looking at table~\ref{table:DKL}, one can see that it is in fact much more than any individual reheating scenario for the quartic potential. This means that these $0.66$ bits of information mostly correspond to the selection of scenarios 5 and 8 amongst all 10 possible reheating scenarios, similarly to what was discussed in \Sec{sec:InformationGain:rhoend} for $\rho_\uend$.

The values of the individual Kullback-Leibler divergences on $T_\ureh$ are also shown in \Fig{fig:mainresult}, together with the Bayesian evidence of the models they correspond to.
\subsubsection{Early Reheating Temperature}
\label{sec:InformationGain:Tereh}
The early reheating temperature is defined only for scenarios 2, 5, 8 and 9. One obtains small information gains with plateau potentials, and depending on the prior one uses on $\sigma_\uend$, $0.2$ or $0.3$ bits with the quartic potential. 

In summary, one finds that more information about reheating can be extracted from the data in models where an extra light scalar field is added than in purely single-field setups. In particular, the Kullback-Leibler divergences on the reheating temperatures can be substantial if the inflaton potential is quartic, and are more modest for a plateau potential.
\section{Discussion}
\label{sec:Discussion}
In \Sec{sec:results}, constraints were derived on the energy scale of inflation, the reheating temperature and the early reheating temperature. In this section, we extend the discussion in a few directions to investigate the physical implications of the constraints we obtained.
\subsection{Inflationary Energy Scale in Plateau Models}
\label{sec:KMIII}
\begin{figure}[t]
\figpilogsto
\begin{center}
\includegraphics[width=0.45\textwidth]{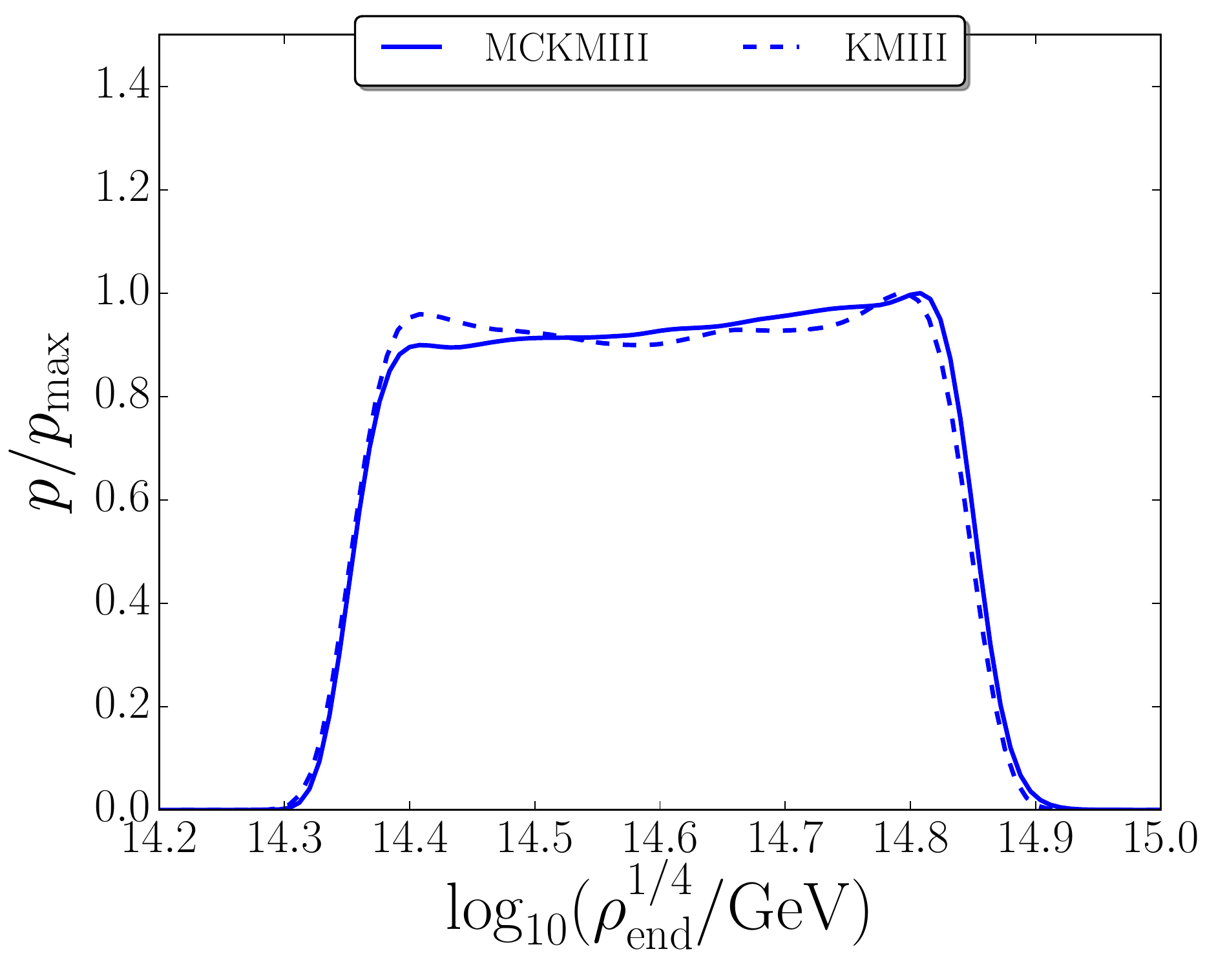}
\includegraphics[width=0.45\textwidth]{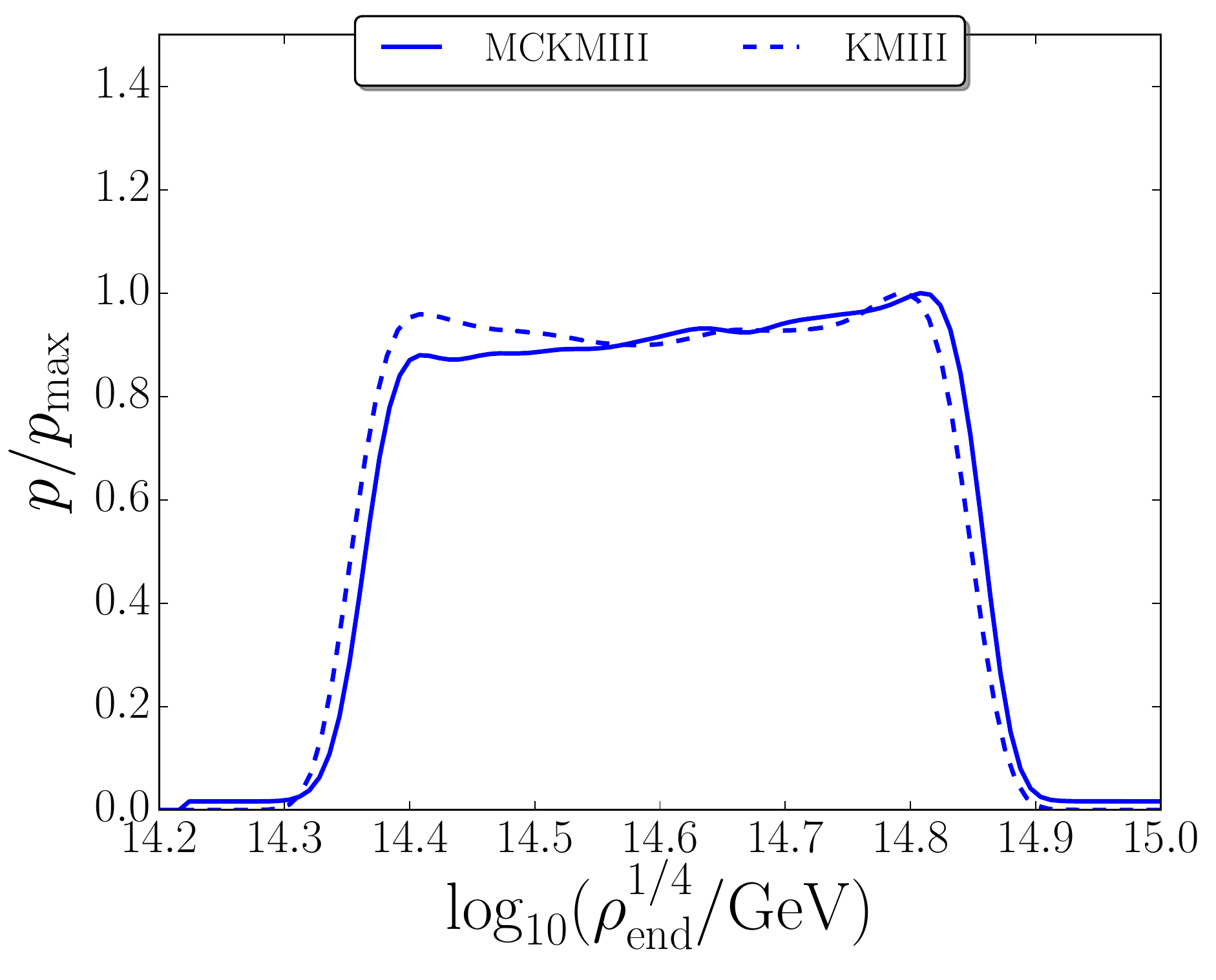}
\includegraphics[width=0.45\textwidth]{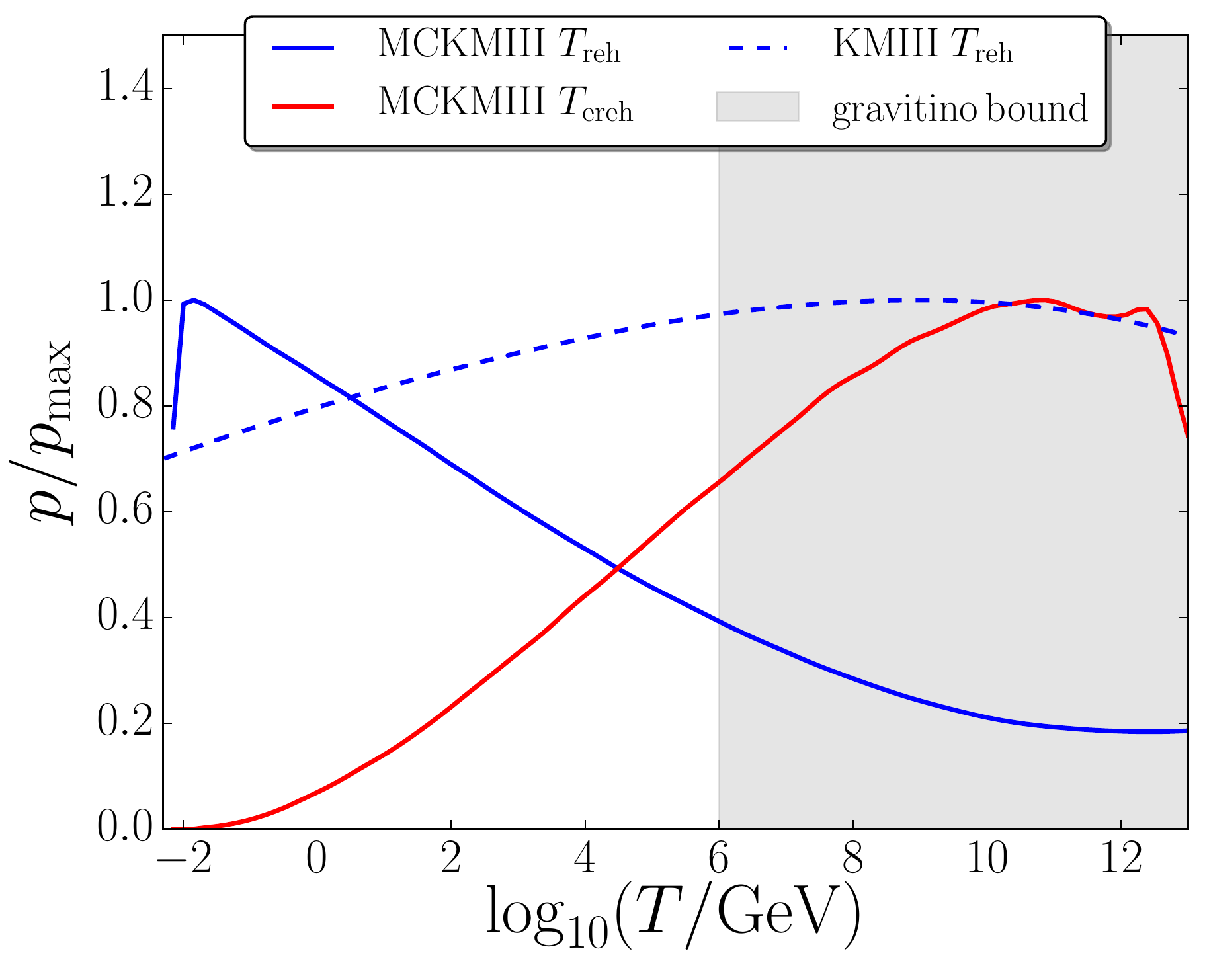}
\includegraphics[width=0.45\textwidth]{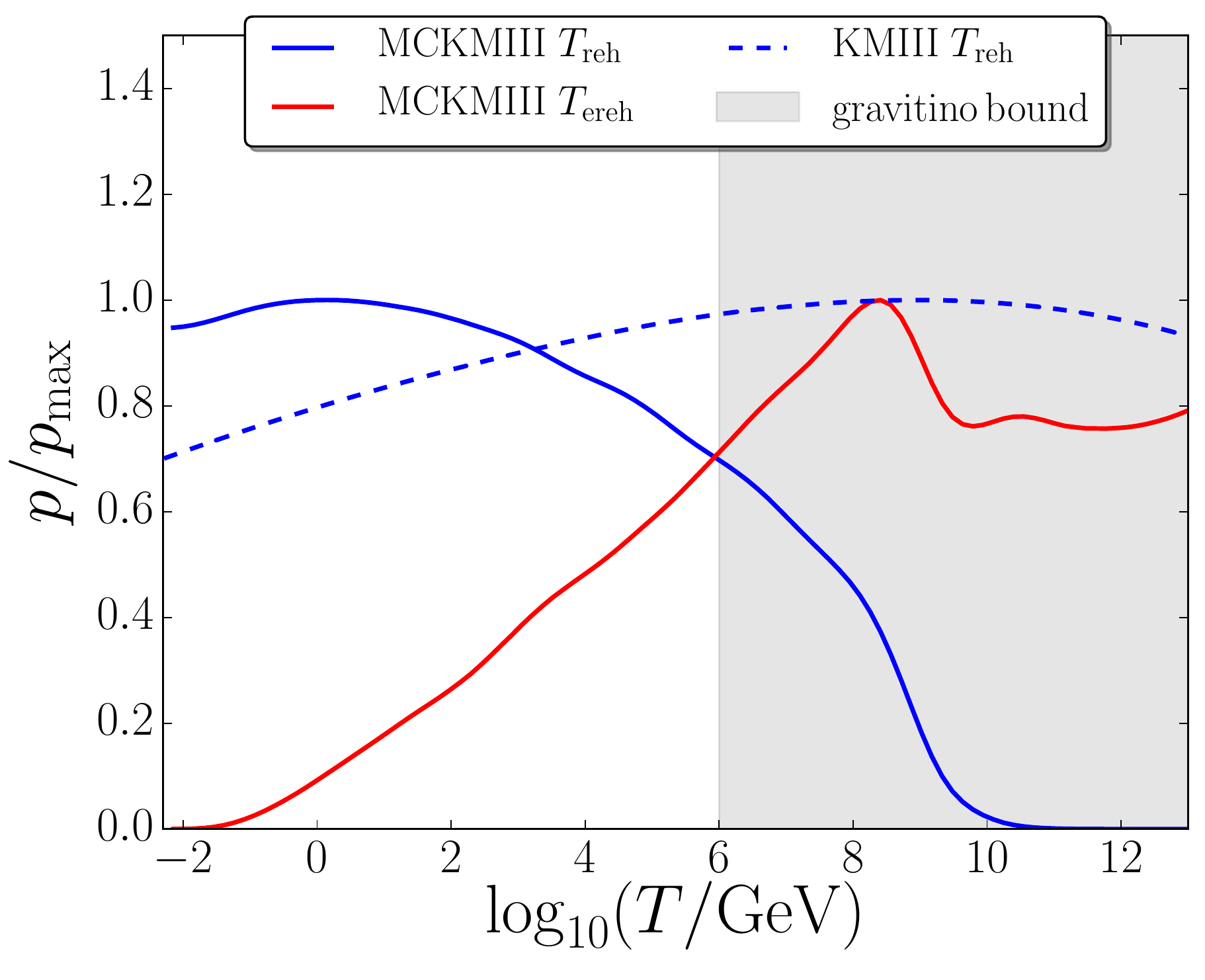}
\caption{Posterior distribution on the energy density at the end of inflation (top panels) and the reheating temperatures (bottom panels) in the K\"ahler moduli II potential~(\ref{eq:pot:KMIII})  of inflation. The left panels correspond to the logarithmically flat prior~(\ref{eq:sigmaend:LogPrior}) $\pilog$ on $\sigma_\uend$, and the right panels stand for the stochastic prior~(\ref{eq:sigmaend:GaussianPrior}) $\pisto$ derived from the equilibrium distribution of a light scalar field in a de Sitter space-time with Hubble scale $H_\uend$. The dashed lines correspond to the single-field version of the model, while the solid lines stand for the averaged posterior distributions over all $10$ reheating scenarios.}
\label{fig:KMIII}
\end{center}
\end{figure}
As explained in \Sec{sec:intro}, the Bayesian model comparison program applied to the scenarios discussed in the present paper show that~\cite{Vennin:2015egh} the models favoured by the data are of two types: either plateau potentials, in any of the 10 reheating scenarios, or quartic potentials in scenarios 5 and 8. Quartic potentials are rather uniquely defined but several versions of plateau inflation have been proposed in the literature. So far, the potential of Higgs inflation (or equivalently the Starobinsky model) has been used to study these models. As noticed in \Fig{fig:post:rhoend:averaged} and further commented on in \Sec{sec:InformationGain:rhoend}, this leads to very sharp constraints on $\rho_\uend$, which, in the absence of any other free parameter in the potential, is mostly fixed by $A_{{}_\mathrm{S}}$. However, plateau potentials exist where inflation can be realised at different energies. To study how the conclusions drawn above are dependent on the specific shape (and energy scale) of the plateau potential considered, in this section, we include another plateau potential in our analysis, K\"ahler moduli II inflation (KMIII in the terminology of \Ref{Martin:2013tda}),
\bea
U(\phi)&= M^4\left[1-\alpha\left(\frac{\phi}{\Mp}\right)^{4/3}\ee^{-\beta\left(\frac{\phi}{\Mp}\right)^{4/3}}\right]\, .
\label{eq:pot:KMIII}
\eea
The posterior constraints on $\rho_\uend$, $T_\ureh$ and $T_\uereh$ are shown in \Fig{fig:KMIII}, and the individual reheating scenarios are displayed in \App{Sec:IndividualScenarios}. 

Compared to \Fig{fig:post:rhoend:averaged}, one can see that inflation proceeds at lower energy, with a wider range of allowed energy scales due to the presence of the free parameters $\alpha$ and $\beta$ in \Eq{eq:pot:KMIII}. This leads to a much smaller Kullback-Leibler divergence on $\rho_\uend$ than in the case of single-field Higgs or quartic inflation, see table~\ref{table:DKL}. However, one still notices that the $\rho_\uend$ posteriors when an extra light scalar field is added are very close to the single-field constraints. For the reheating temperatures, the same remarks apply as in \Secs{sec:result:Treh} and~\ref{sec:result:Tereh} for Higgs inflation. In particular, small reheating temperatures and large early reheating temperatures are preferred. Therefore, apart from the large value of $\DKL$ for $\rho_\uend$, the results obtained above for Higgs inflation seem to characterise plateau potentials in general.
\subsection{Gravitino Overproduction Bounds}
\label{sec:gravitino}
Reheating affects cosmology in different ways. First, as explained in \Sec{sec:InversionProblem}, it contributes to the expansion history through its averaged equation-of-state parameter and its energy density at completion. This is the effect we used to constrain reheating in single-field models. Second, it may produce additional features (such as gravitational waves, magnetic fields, topological defects, baryon asymmetries or dark matter, etc.), and enhance the contribution from light scalar fields (that are otherwise spectator fields during inflation) to curvature perturbations. This is the case of the scenarios considered in the present work and this additional effect is the one we have used to constrain reheating in these setups. Third, it affects the subsequent thermal history of the Universe, since it determines the temperature at the onset of the radiation dominated epoch. 

To illustrate how this last effect can be important to constrain reheating, in this section, we consider gravitinos, the gauge fermion supersymmetric partners of the graviton of supergravity theories. Gravitinos are produced from scatterings in the hot plasma during reheating, and their abundance is directly related to the magnitude of the reheating temperature~\cite{Terada:2014uia}. Their lifetime depends on their mass $m_{3/2}$, and if they survive long enough, their decay products can produce spectral distortions of the CMB. Combining current constraints on CMB spectral distortions and BBN, upper bounds can be derived on $T_\ureh$. In \Ref{Dimastrogiovanni:2015wvk}, it is found that, with $m_{3/2} \sim {\cal O}(100\,\GeV)$, one typically obtains the most stringent constraint $T_\ureh<10^{6}\,\GeV$. 

This value is shown in \Fig{fig:post:Trehs:averaged} and the bottom panels of \Fig{fig:KMIII} where the posterior distributions on $T_\ureh$ and $T_\uereh$ are displayed. One can see that it excludes a large set of possible temperatures. However, scenarios where an extra light scalar field is added seem to more easily evade the gravitino overproduction bound than their single-field counterpart. For instance, in quartic inflation with an additional light field, the reheating temperature is typically smaller than $10^6 \, \GeV$ [see the bottom left panel of \Fig{fig:post:Trehs:averaged} and \Eq{eq:LFI4:constraint:Treh}], which is not the case of the single-field versions of Higgs inflation, quartic inflation or even K\"ahler moduli III inflation in \Fig{fig:KMIII}. On the other hand, since large early reheating temperatures are preferred in general, the gravitino problem might be worsened if gravitinos are generated from the decay products of the first decaying field in scenarios 2, 5, 8 and 9. 

Interestingly, this also shows that if gravitinos exist, they provide a powerful indirect way to further constrain the models discussed in this work. In particular, gravitino production bounds seem to yield less additional constraints for quartic models than for plateau models (with an extra light scalar field in both cases). If they were explicitly included in the set of observations, they would therefore probably lead to a slight preference of the former against the later. 
\subsection{Decay Mediation Scale}
\label{sec:DecayMediationScale}
\begin{figure}[t]
\figpilogsto
\begin{center}
\includegraphics[width=0.45\textwidth]{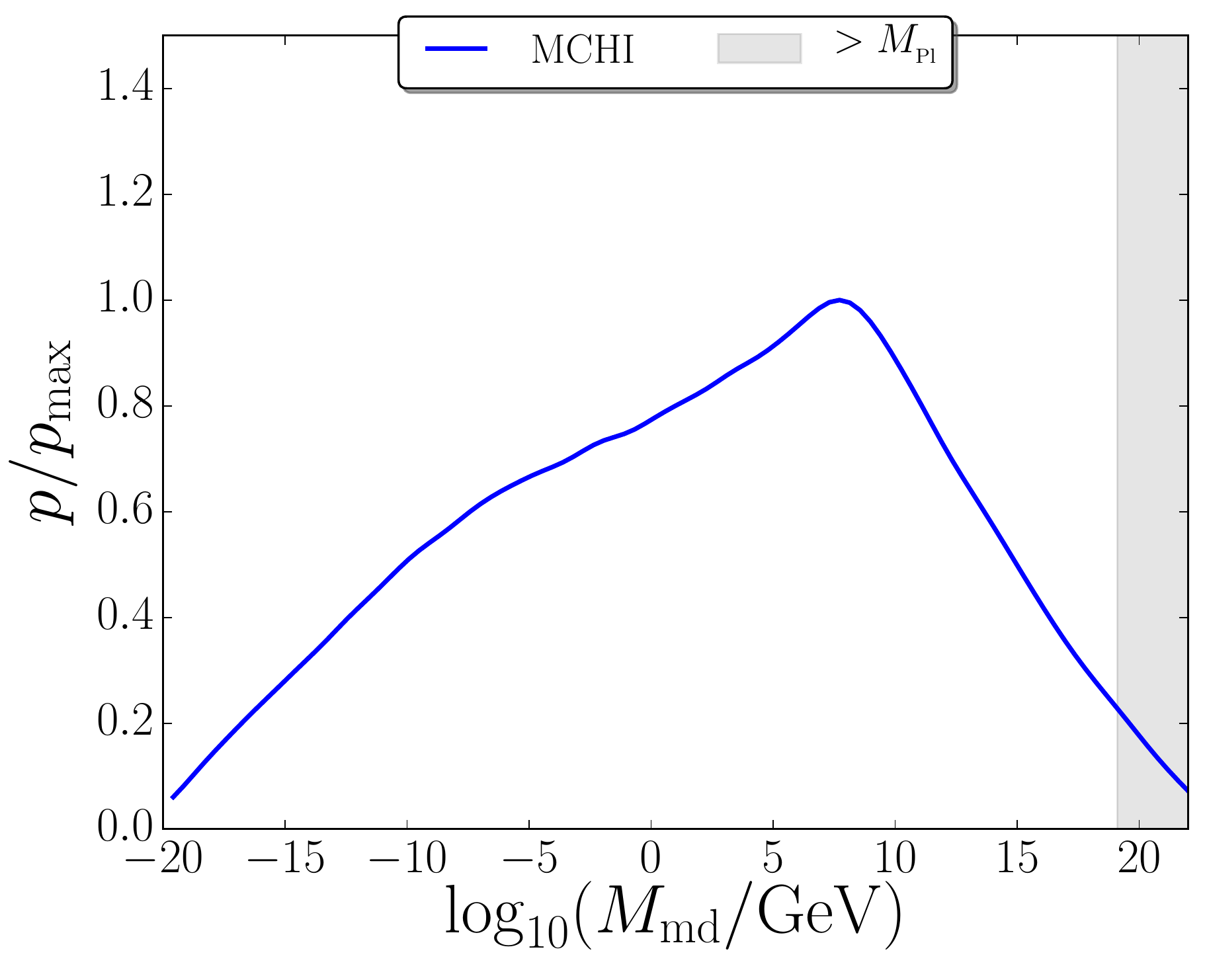}
\includegraphics[width=0.45\textwidth]{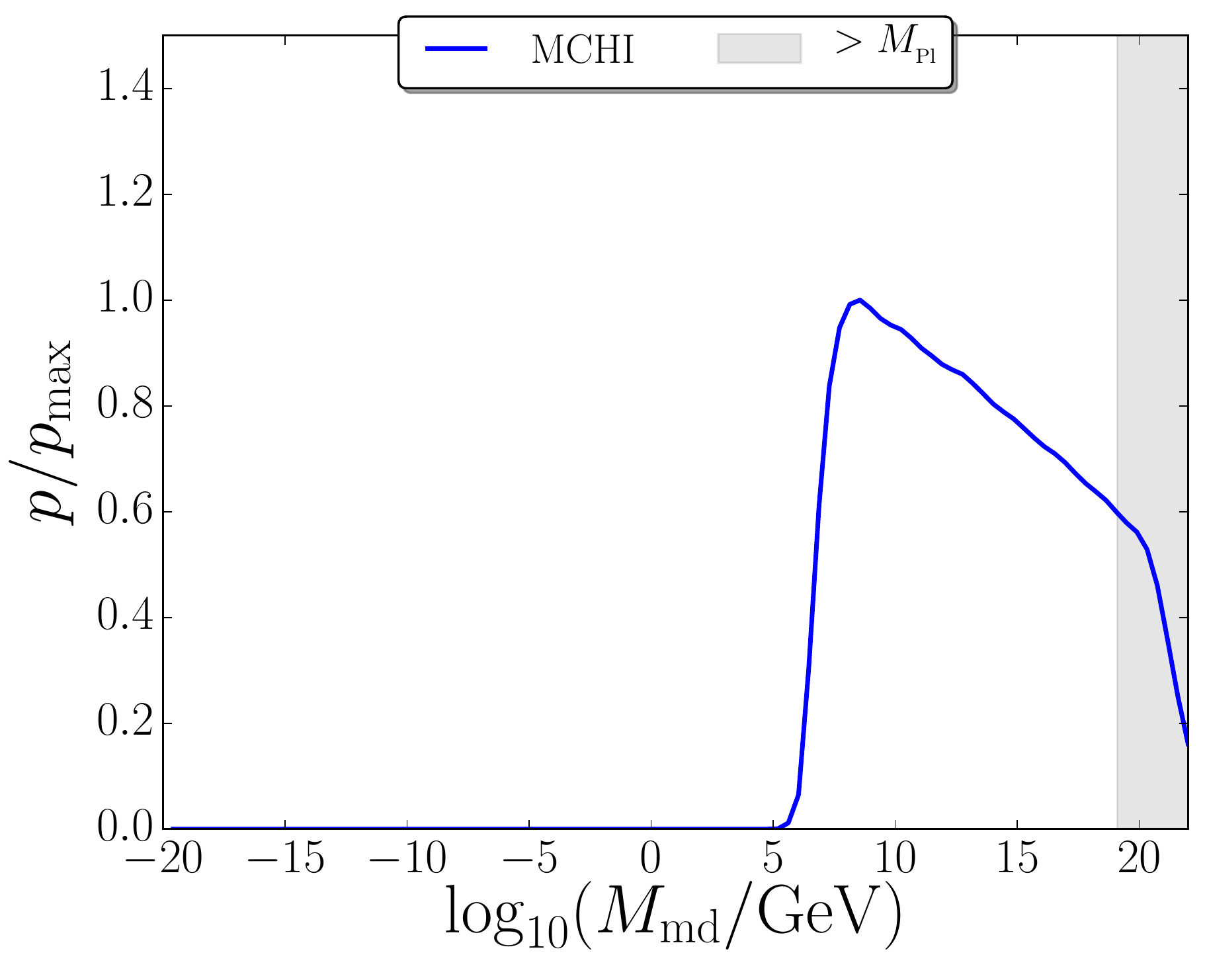}
\includegraphics[width=0.45\textwidth]{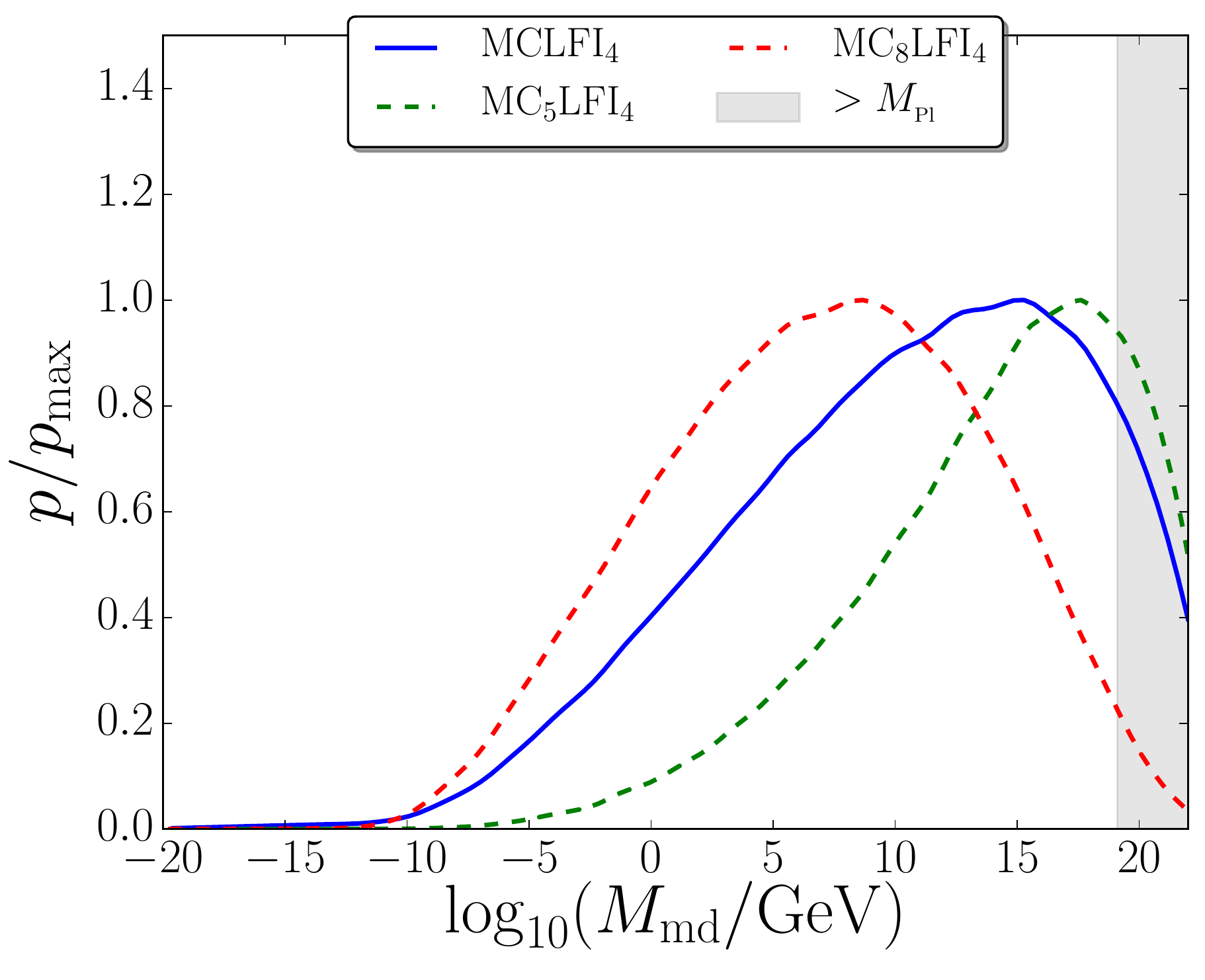}
\includegraphics[width=0.45\textwidth]{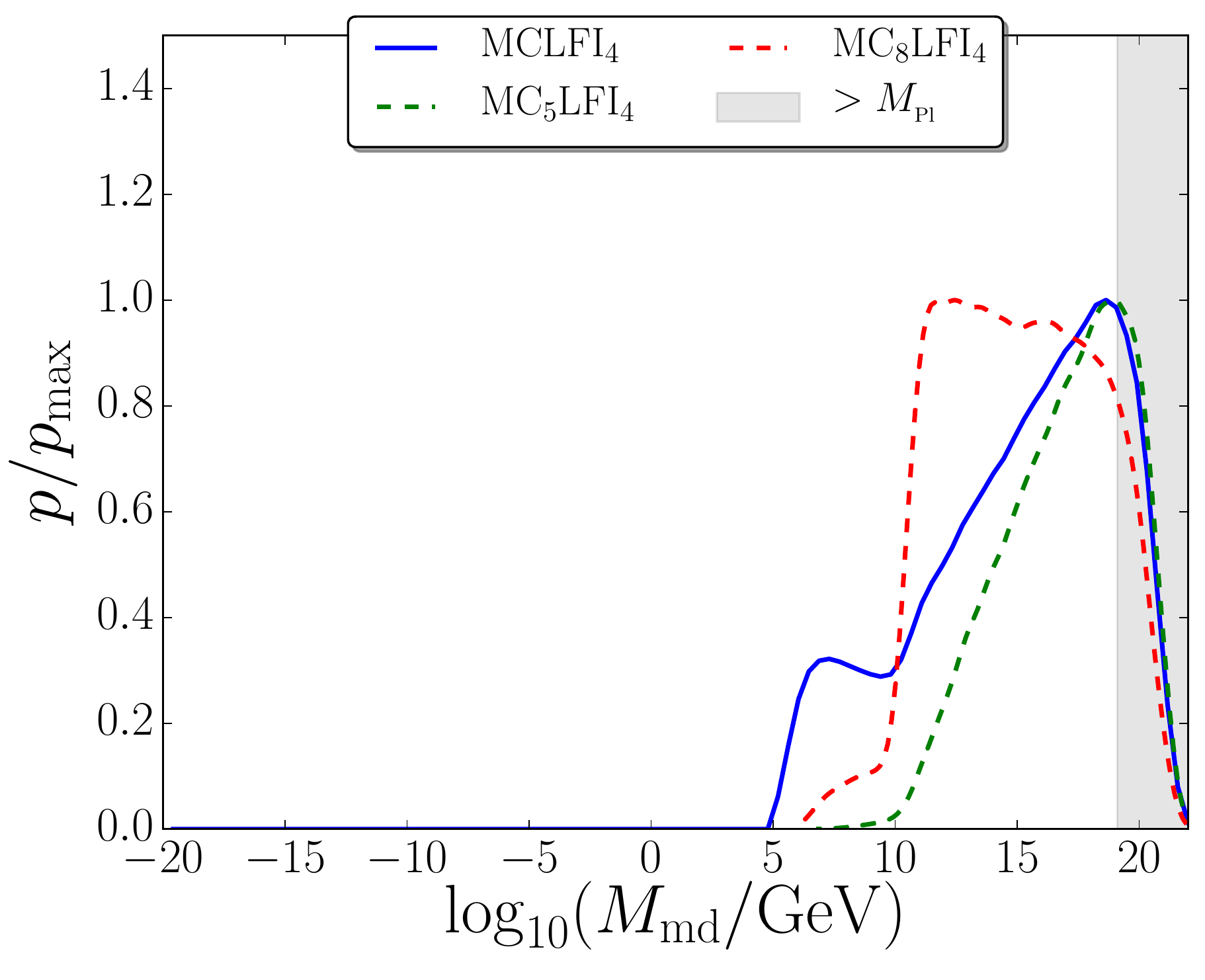}
\caption{Posterior distributions of the mass mediation scale $\Mmd$ defined for $\sigma$ in \Eq{eq:Mstar}, for Higgs inflation (top panels) and quartic inflation (bottom panels), when the logarithmically flat prior (left panels) and the stochastic prior (right prior) are used on $\sigma_\uend$. The distributions are averaged over all 10 reheating scenarios, and for quartic inflation, the individual posteriors for scenarios 5 and 8 are shown since these are the only non-ruled out scenarios. The region corresponding to decay rates that are smaller than gravitational mediation, $\Mmd>\Mp$, is shaded in grey.}
\label{fig:MediationScale}
\end{center}
\end{figure}
So far, the decay rate of the additional scalar field $\sigma$, $\Gamma_\sigma$, and its mass $m_\sigma$, have been assumed to be independent (up to the ordering conditions of \Fig{fig:cases}). However, these scales may be related by the physics of the decay of $\sigma$, and in this section we study the implications of the results we obtained on such processes. More specifically, we consider the case where spontaneous decay of $\sigma$ by dimension $5$ operators is mediated by some scale $\Mmd$. The decay rate and the mass are then related through~\cite{Harigaya:2013vwa}
\bea
\label{eq:Mstar}
\Gamma_{\sigma} \simeq \frac{m_{\sigma}^3}{\Mmd^2}\,.
\eea
Let us study which values of $\Mmd$ are typically predicted by the scenarios considered in this work. In \Fig{fig:MediationScale}, the averaged (over reheating scenarios) posterior distributions for $\Mmd$ are displayed. For electroweak suppressed decay for instance, one should have $\Mmd\sim 100\, \mathrm{GeV}$. Although such values are well within the distributions when a logarithmically flat prior on $\sigma_\uend$ is used,  higher mediation scales are typically preferred, which is in agreement with the standard curvaton picture where  gravitationally mediated decay~\cite{Enqvist:2013gwf} is assumed. 

One can also see that the large-field quartic models favour slightly higher mediation scales than the plateau potential of Higgs inflation (that has a very similar posterior on $\Mmd$ as K\"ahler moduli II inflation introduced in \Sec{sec:KMIII}, which is why this other plateau potential is not displayed here). This is due to the fact that the most likely scenarios for the quartic potential, cases 5 and 8, yield large values compared to the other scenarios, while for plateau potentials all reheating cases contribute to the distributions plotted in \Fig{fig:MediationScale}. For this reason, individual cases 5 and 8 are also displayed (green and red dashed lines respectively) for the quartic potential in \Fig{fig:MediationScale}. Interestingly, while these two scenarios are indistinguishable with respect to all criteria discussed so far, they give slightly different preferred values for $\Mmd$, which suggests that requiring specific mediated decay scales may be a way to distinguish between these cases.

When the stochastic prior on $\sigma_\uend$ is used, one notices that mediated decay cannot happen for $\Mmd$ below $10^5\,\GeV$. This is in sharp contrast with the result obtained with a logarithmically flat prior on $\sigma_\uend$ and can be understood as follows. When $\sigma_\uend$ is super-Planckian, $\sigma$ drives a second phase of inflation (cases 6, 9 and 10 in \Fig{fig:cases}), the duration of which is roughly given by $\sigma_\uend^2/(4\Mp^2)$ in numbers of $e$-folds. Therefore, $\sigma_\uend$ cannot be much larger than, say, $20\Mp$. Therefore the stochastic prior on $\sigma_\uend$, which implies that $m_\sigma\sim H_\uend^2/\sigma_\uend$ [see \Eq{eq:sigmaend:GaussianPrior}], yields a lower bound on $m_\sigma$, that does not exist when a logarithmically flat prior on $\sigma_\uend$ is used. This explains why higher values of $m_\sigma$, hence of $\Mmd$, are obtained with the stochastic prior.
\section{Conclusion}
\label{sec:Conclusions}
In this work, we have presented the first systematic observational constraints on reheating in scenarios where inflation is driven by a single scalar inflaton field $\phi$, but an extra light scalar field $\sigma$ can also contribute to the total amount of curvature perturbations. Following the results of \Ref{Vennin:2015egh}, the analysis was performed in the two classes of models that are favoured by the data, where the inflationary potential is either of the plateau or the quartic type.  

Bayesian inference techniques were employed to derive posterior constraints on the energy density at the end of inflation $\rho_\uend$, and the temperature of the Universe $T_\ureh$  (and $T_\uereh$) at the onset of the radiation dominated epoch(s). If inflation is realised with a plateau potential, it was found that the constraints on $\rho_\uend$ are scarcely altered by the introduction of a light scalar field (compared to the purely single-field case), in agreement with the strong robustness of these models under the introduction of an additional scalar field noted in \Ref{Vennin:2015egh}. For a quartic inflationary potential however, it was found that lower values of $\rho_\uend$ are favoured with an extra light scalar field. Indeed, quartic inflation predicts a value of the tensor-to-scalar ratio that is not too large only when the extra field provides the dominant contribution to curvature perturbations, in which case $\rho_\uend$ is smaller than in the single-field scenario. 

For the reheating temperature, plateau potentials yield constraints on $T_\ureh$ that depend on the reheating scenario (these scenarios are listed in \Fig{fig:cases} and the constraints are given in \Fig{fig:post:Treh:individual}). For quartic inflation, the only favoured reheating scenarios are 5 and 8, and both show a preference for lower reheating temperatures than with the single-field counterpart of the model. When a logarithmically flat prior on the vev of the extra light field at the end of inflation is used, one obtains the averaged $95\,\%\ \mathrm{CL}$ upper bound $T_\ureh<5\times 10^{4}\,\GeV$. In reheating scenarios 2, 5, 8 and 9, the Universe undergoes a transient early radiation dominated epoch during a two-stage reheating process and the constraints on the temperature at its onset, $T_\uereh$, were also derived. Contrary to $T_\ureh$, lower bounds can be derived on $T_\uereh$, typically larger than $\sim 10^2\,\GeV$ for a plateau potential and larger than $\sim 10^5\,\GeV$ for the quartic potential.

In general, it was observed that tighter constraints on reheating are derived with an additional light scalar field than without, in agreement with the results of \Ref{Vennin:2015egh} where Bayesian complexity was used to quantify the number of unconstrained parameters. Indeed, when the extra field is present, the same parameters define both its contribution to curvature perturbations and to the expansion history of reheating that determines the location of the observational window along the inflationary potential. More information about reheating can therefore be gained in scenarios with an additional scalar field, compared to the single-field case where only the later effect allows one to constrain reheating from observations. This information gain was quantified by computing the Kullback-Leibler divergence between the prior and posterior distributions of $\rho_\uend$, $T_\ureh$ and $T_\uereh$. Even if the information gain remains modest when the inflationary potential is of the plateau type, it becomes substantial in quartic inflation (where, for instance, more than 3 bits of information are gained on the energy density at the end of inflation).

Since the process of reheating determines the temperature of the Universe at the onset of the radiation dominated epoch, it affects its subsequent thermal history. The constraints we derived thus have implications for post-inflationary physics. For instance, we have considered gravitino overproduction bounds and shown that since models with an additional scalar field predict lower reheating temperatures, they evade those bounds more easily than their single-field counterpart. This is particularly true if the inflationary potential is of the quartic type, so that if gravitino bounds were explicitly included in the set of observations used to constrain the models, they would probably lead to a slight preference of quartic inflation with an extra light scalar field (in reheating scenarios 5 and 8) over all other models, including the single-field plateau ones. 

The sensitivity to the microphysics of reheating has also been demonstrated with the mass mediation scale of the extra scalar field decay, on which constraints have been derived. Notably, it was found that reheating scenarios 5 and 8 in quartic inflation, otherwise indistinguishable with respect to all other criteria discussed in this paper, give slightly different preferred values for this mass scale.

In this analysis, the crucial role played by the prior on the vev of the extra light scalar field at the end of inflation, $\sigma_\uend$, has also been highlighted. Even though the main conclusions quoted above are robust under changes of priors on $\sigma_\uend$, the detailed constraints on reheating and the relative parameter space volume associated to the 10 reheating scenarios depend on the assumptions one makes about its value. In particular, for quartic inflation, which, in reheating scenarios 5 and 8, is one of the most favoured models, if $\sigma_\uend$ is set by the quantum diffusion effects during inflation, one finds that the Gaussian distribution~(\ref{eq:sigmaend:GaussianPrior}) is not an equilibrium solution of the stochastic dynamics of $\sigma$. In fact, there is no equilibrium solution in this case, and the typical value acquired by the additional scalar field at the end of inflation both depends on its value at the onset of inflation and on the total duration of inflation. This may be relevant to the question~\cite{Starobinsky:1986fxa, Linde:2005yw, Enqvist:2012, Burgess:2015ajz} whether observations can give access to scales beyond the classical horizon, and we plan to study this question further in a future work.
\acknowledgments
This work is supported by STFC grants ST/K00090X/1 and ST/N000668/1. In addition, RH is supported by STFC grant ST/K502248/1 and KK is supported by the European Research Council grant through 646702 (CosTesGrav). The authors would like to thank Chris Byrnes and Jes\'{u}s Torrado for interesting and enjoyable discussions.

\begin{appendix}
\newpage

\section{Parameter Dependency Trees}
\label{Sec:DependencyTree}
\begin{figure}[!h]
\begin{center}
\includegraphics[width=13cm]{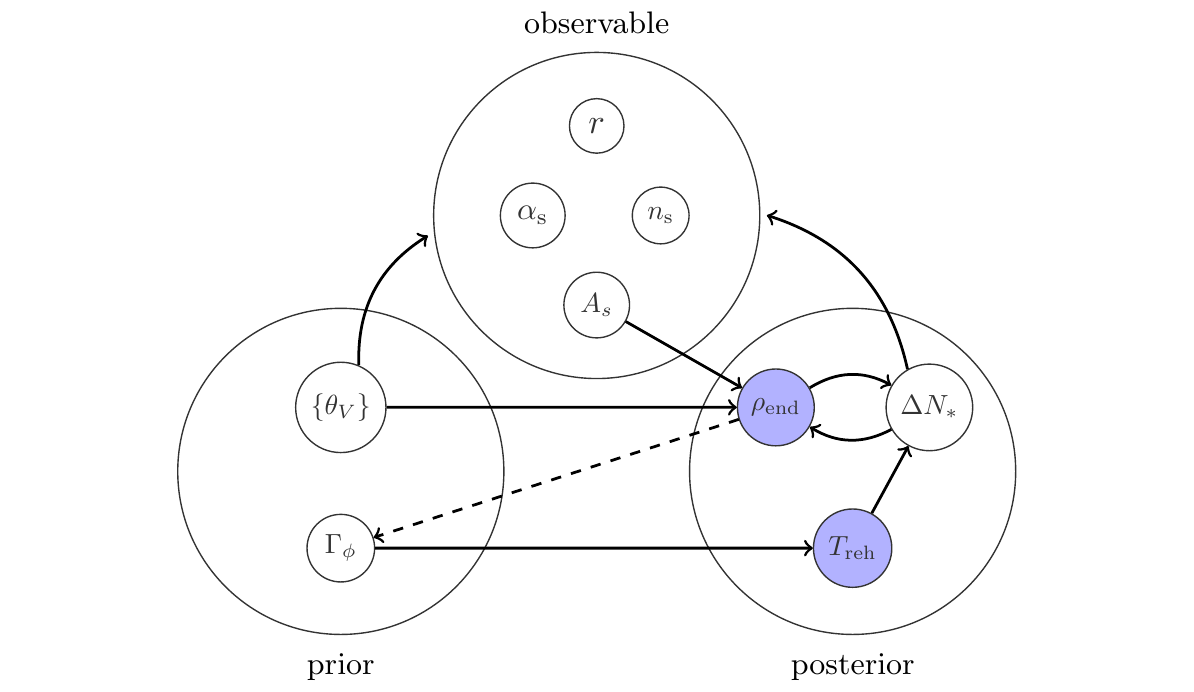} 
\quad\\ \quad\\
\includegraphics[width=13cm]{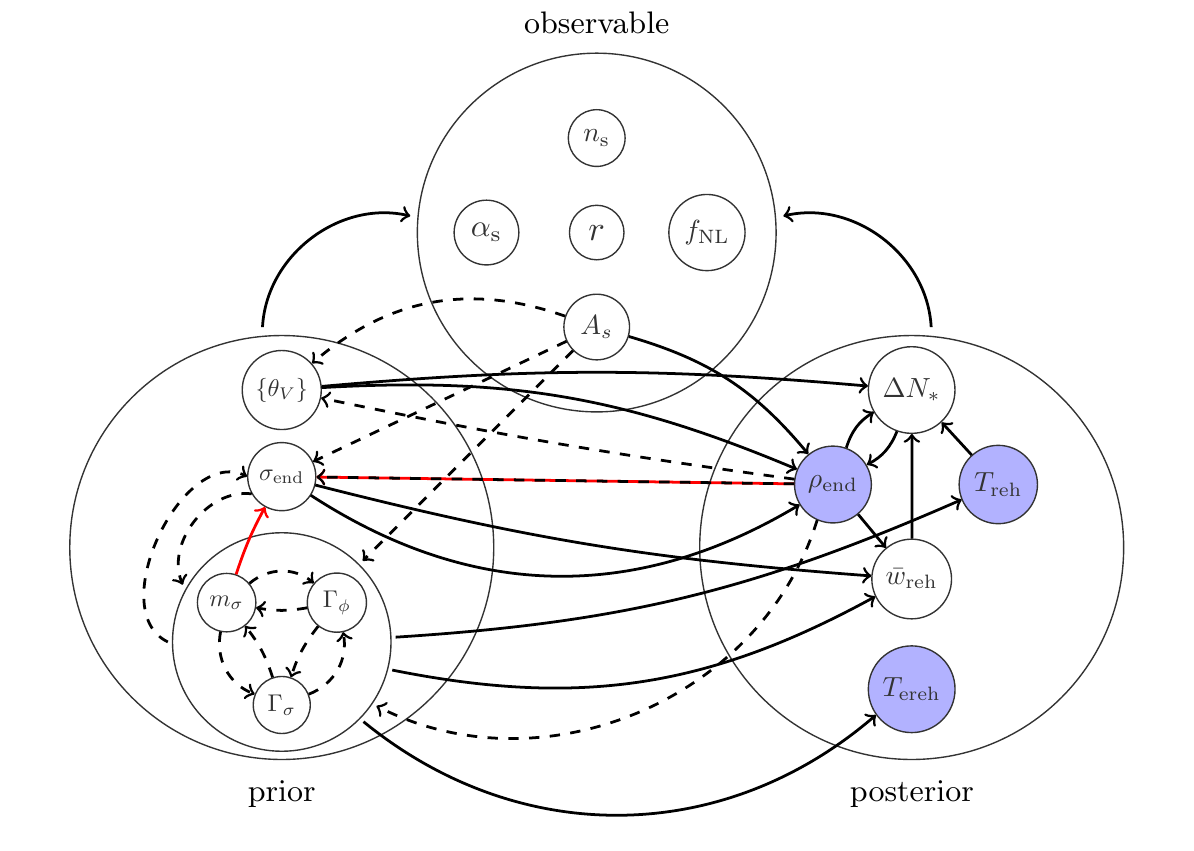}
\caption{Parameter dependency trees featuring the observable, prior and posterior parameters. These trees illustrate the degree of complexity and inter-connectivity of parameters within the purely single-field models (top panel) and the scenarios where an extra light scalar field is added (bottom panel). Solid black arrows correspond to direct dependencies, dashed arrows are for weak hard prior cutoff dependencies and solid red arrows stand for dependencies that only exist in the case of a stochastic prior on $\sigma_{\rm end}$ (see \Sec{sec:Bayesian}). The derived parameters for which posteriors are displayed in this paper are filled in light blue.}
\label{dependencytree}
\end{center}
\end{figure}
\section{Kullback-Leibler Divergence}
\label{sec:DKL}
The relative (or conditional) entropy between the prior $\pi(\theta)$ and posterior $p(\theta)$ distributions on some parameter $\theta$ is called the Kullback-Leibler divergence, and defined as  
\bea
\label{eq:app:dkl}
\DKL\left(p \vert\vert \pi\right) \equiv \int^{\infty}_{-\infty} {p}\left(\theta \right) \log_2 \left[\frac{{p}\left(\theta \right)}{\pi \left(\theta \right)} \right] \dd \theta\, .
\eea
This quantity is used in \Sec{sec:InformationGain} to assess the gain of information in the energy scale of inflation and the reheating temperatures. In this appendix, we derive some of its important properties.
 
A first property of the Kullback-Leibler divergence is that it is invariant under a generic reparameterisation $\theta\rightarrow \theta^\prime$. Indeed, the prior and posterior on $\theta^\prime$ can be calculated according to
\bea
\pi (\theta ){\rm d}\theta = \bar{\pi}(\theta'){\rm d}\theta'\, ,\quad\quad
p (\theta ){\rm d}\theta = \bar{p}(\theta'){\rm d}\theta' \, ,
\eea
which leads to
\bea
\bar{{p}}(\theta') \log_2 \left[ \frac{\bar{p}(\theta')}{\bar{\pi} (\theta')} \right] {\rm d}\theta' = {p}(\theta ) \log_2 \left[ \frac{{p}(\theta )}{\pi (\theta )} \right] {\rm d}\theta \, .
\eea
 
Another important property of the Kullback-Leibler divergence is that it is always positive, due to Gibbs' inequality which states that for two continuous normalised distributions $\pi(\theta)$ and $p(\theta)$, one has
\bea
\int p(\theta)\log_2 p(\theta) \dd\theta \geq \int p(\theta)\log_2 \pi(\theta) \dd\theta\, .
\eea

Finally, in order to gain some insight on the way that the Kullback-Leibler divergence is affected by the shape of the prior and posterior distributions, let us compute its value in the case where both distributions are Gaussian with mean values $\mu_\pi$ and $\mu_p$ respectively, and standard deviation $\sigma_\pi$ and $\sigma_p$ respectively. Denoting $\delta\mu=\mu_p-\mu_\pi$, one obtains
\bea
\label{eq:DKL:gaussian}
\DKL = \frac{1}{2\ln 2}\left[ \frac{(\delta \mu )^2}{\sigma_{\pi}^2} + \frac{\sigma_{p}^2}{\sigma_{\pi}^2}+ 2\ln \left( \frac{\sigma_{\pi}}{\sigma_{p}}\right) -1\right]\,.
\eea
In this expression, the first term accounts for the update in the preferred value and can be understood as follows: if the change in the preferred value is large compared to the uncertainty level of the prior, then non-trivial information is gained and the value of $\DKL$ is large. The other terms depend on the ratio $\sigma_p/\sigma_\pi$ only, and yield a contribution to $\DKL$ that increases when $\sigma_p/\sigma_\pi$ decreases (while being smaller than one), corresponding to improved measurements of the parameter under consideration.

In table~\ref{table:DKL}, the Kullback-Leibler divergences on the energy scale of inflation and the reheating temperatures are given for the three potentials considered in this work (Higgs inflation, quartic inflation and K\"ahler moduli II inflation), for the single-field  versions of the model as well as for all 10 reheating scenarios, where the divergence between the averaged priors and posteriors are also given. The left tables were obtained with a logarithmically flat prior on $\sigma_\uend$, and the right priors with the stochastic prior~(\ref{eq:sigmaend:GaussianPrior}).
%
%
\begin{table}[!h]
  \small
  \centering
  \resizebox{6cm}{!}{ 
\begin{tabular}{ | l | l | l | l |}
\hline
  $\pi_{\log} (\sigma_{\rm end})$   & \multicolumn{2}{l}{\qquad \qquad $\DKL$} & \\ \hline
    Model & $\rho_{\rm end}$ & $T_{\rm reh}$ & $T_{\rm ereh}$   \\ \hline \hline
    HI & 1.370 & 0.004 & - \\ \hline
    MCHI & 0.114 & 0.005 & 0.018 \\ \hline \hline
    ${\rm MC}_1$HI & 0.107 & 0.005 & -\\ \hline
    ${\rm MC}_2$HI & 0.009 & 0.009 & 0.001 \\ \hline
    ${\rm MC}_3$HI & 1.059 & 0.001 & -\\ \hline
    ${\rm MC}_4$HI & 0.061 & 0.042 & -\\ \hline
    ${\rm MC}_5$HI & 0.504 & 0.023 & 0.039 \\ \hline
    ${\rm MC}_6$HI & 0.687 & 0.023 & -\\ \hline
    ${\rm MC}_7$HI & 0.280 & 0.012 & - \\ \hline
    ${\rm MC}_8$HI & 0.587 & 0.016 & 0.015 \\ \hline
    ${\rm MC}_9$HI & 0.548 & 0.006 & 0.001 \\ \hline
    ${\rm MC}_{10}$HI & 1.539 & 0.091 & -\\ \hline \hline
        ${\rm LFI}_4$ & 1.171 & 0.108 & - \\ \hline
    MC${\rm LFI}_4$ & 3.104 & 0.656 & 0.181 \\ \hline \hline
    ${\rm MC}_1{\rm LFI}_4$ & 0.000 & 0.120 & -\\ \hline
    ${\rm MC}_2{\rm LFI}_4$ & 0.080 & 0.077 & 0.019 \\ \hline
    ${\rm MC}_3{\rm LFI}_4$ & 0.971 & 0.039 & -\\ \hline
    ${\rm MC}_4{\rm LFI}_4$  & 0.190 & 0.011 & -\\ \hline 
    ${\rm MC}_5{\rm LFI}_4$  & 0.911 & 0.039 & 0.125 \\ \hline
    ${\rm MC}_6{\rm LFI}_4$ & 0.425 & 0.114 & -\\ \hline
    ${\rm MC}_7{\rm LFI}_4$ & 0.317 & 0.007 & - \\ \hline
    ${\rm MC}_8{\rm LFI}_4$  & 1.093 & 0.050 & 0.044 \\ \hline
    ${\rm MC}_9{\rm LFI}_4$  & 0.719 & 0.031 & 0.044 \\ \hline
    ${\rm MC}_{10}{\rm LFI}_4$ & 1.195 & 0.223 & -\\ \hline \hline
       KMIII & 0.083 & 0.008 & - \\ \hline
    MCKMIII & 0.121 & 0.015 & 0.010 \\ \hline \hline
    ${\rm MC}_1$KMIII & 0.092 & 0.021 & -\\ \hline
    ${\rm MC}_2$KMIII & 0.000 & 0.102 & 0.006 \\ \hline
    ${\rm MC}_3$KMIII & 0.072 & 0.022 & -\\ \hline
    ${\rm MC}_4$KMIII & 0.089 & 0.002 & -\\ \hline
    ${\rm MC}_5$KMIII & 0.095 & 0.003 & 0.002 \\ \hline
    ${\rm MC}_6$KMIII & 2.584 & 0.125 & -\\ \hline
    ${\rm MC}_7$KMIII & 0.095 & 0.000 & - \\ \hline
    ${\rm MC}_8$KMIII & 0.095 & 0.002 & 0.000 \\ \hline
    ${\rm MC}_9$KMIII & 0.000 & 0.012 & 0.011 \\ \hline
    ${\rm MC}_{10}$KMIII & n.c. & n.c. & -\\ \hline
    \end{tabular}
    }
      \resizebox{6cm}{!}{
\begin{tabular}{ | l | l | l | l |}
\hline
   $\pisto (\sigma_{\rm end})$  & \multicolumn{2}{l}{\qquad \qquad $\DKL$} & \\ \hline
    Model & $\rho_{\rm end}$ & $T_{\rm reh}$ & $T_{\rm ereh}$   \\ \hline \hline
    HI & 1.370 & 0.004 & - \\ \hline
    MCHI & 0.224 & 0.006 & 0.014 \\ \hline \hline
    ${\rm MC}_1$HI & 0.060 & 0.004 & -\\ \hline
    ${\rm MC}_2$HI & 0.058 & 0.006 & 0.000 \\ \hline
    ${\rm MC}_3$HI & 1.077 & 0.007 & -\\ \hline
    ${\rm MC}_4$HI & 0.087 & 0.001 & -\\ \hline
    ${\rm MC}_5$HI & 0.015 & 0.002 & 0.000 \\ \hline
    ${\rm MC}_6$HI & 0.800 & 0.032 & -\\ \hline
    ${\rm MC}_7$HI & - & - & - \\ \hline
    ${\rm MC}_8$HI & 0.046 & 0.002 & 0.000 \\ \hline
    ${\rm MC}_9$HI & 0.606 & 0.028 & 0.015 \\ \hline
    ${\rm MC}_{10}$HI & 2.069 & 0.130 & -\\ \hline \hline
     ${\rm LFI}_4$ & 1.171 & 0.108 & - \\ \hline
    MC${\rm LFI}_4$ & 4.780 & 0.111 & 0.281 \\ \hline \hline
    ${\rm MC}_1{\rm LFI}_4$ & 0.176 & 0.110 & -\\ \hline
    ${\rm MC}_2{\rm LFI}_4$ & 0.167& 0.088 & 0.010 \\ \hline
    ${\rm MC}_3{\rm LFI}_4$ & 1.337 & 0.049 & -\\ \hline
    ${\rm MC}_4{\rm LFI}_4$  & 0.141 & 0.105 & -\\ \hline
    ${\rm MC}_5{\rm LFI}_4$  & 3.499 & 0.051 & 0.097 \\ \hline
    ${\rm MC}_6{\rm LFI}_4$ & 1.197 & 0.117 & -\\ \hline
    ${\rm MC}_7{\rm LFI}_4$ & - & - & - \\ \hline
    ${\rm MC}_8{\rm LFI}_4$  & 3.695 & 0.157 & 0.028 \\ \hline
    ${\rm MC}_9{\rm LFI}_4$  & 1.035 & 0.174 & 0.016 \\ \hline
    ${\rm MC}_{10}{\rm LFI}_4$ & 1.528 & 0.175 & -\\ \hline \hline
        KMIII & 0.083 & 0.008 & - \\ \hline
    MCKMIII & 0.162 & 0.011 & 0.010 \\ \hline \hline
    ${\rm MC}_1$KMIII & 0.098 & 0.016 & -\\ \hline
    ${\rm MC}_2$KMIII & n.c. & n.c. & n.c. \\ \hline
    ${\rm MC}_3$KMIII & n.c. & 0.021 & -\\ \hline
    ${\rm MC}_4$KMIII & 0.099 & 0.006 & -\\ \hline
    ${\rm MC}_5$KMIII & 0.095 & 0.011 & 0.001 \\ \hline
    ${\rm MC}_6$KMIII & n.c. & n.c. & -\\ \hline
    ${\rm MC}_7$KMIII & - & - & - \\ \hline
    ${\rm MC}_8$KMIII & 0.079 & 0.004 & 0.001 \\ \hline
    ${\rm MC}_9$KMIII & n.c. & n.c. & n.c. \\ \hline
    ${\rm MC}_{10}$KMIII & n.c. & 0.133 & -\\ \hline
    \end{tabular}
    } 
\caption{Kullback-Leibler divergences $\DKL$ on $\rho_\uend$, $T_\ureh$ and $T_\uereh$ for Higgs (top row), quartic large field (middle row) and K\"{a}hler moduli II (bottom row) inflation. The result is given for the single-field versions of the model and for the 10 reheating scenarios of \Fig{fig:cases} as well. The divergence between the averaged (over reheating scenarios) priors and posteriors is also displayed. The left tables were obtained with a logarithmically flat prior on $\sigma_\uend$, and the right tables with the stochastic prior~(\ref{eq:sigmaend:GaussianPrior}). Note that the early reheating temperature $T_\ureh$ is defined only for scenarios 2, 5, 8 and 9, and that scenario 7 cannot be sampled when a stochastic prior is used. For some of the K\"{a}hler moduli II cases, denoted n.c. (for ``not converged''), numerically robust results could not be obtained.} 
\label{table:DKL}
\end{table}
%
%
\section{Individual Reheating Scenarios Constraints}
\label{Sec:IndividualScenarios}
In this appendix, we display the posterior constraints on $\rho_\uend$, $T_\ureh$ and $T_\uereh$, for the individual 10 reheating scenarios of \Fig{fig:cases}, for the three potentials considered in this work (Higgs inflation, quartic inflation and K\"ahler moduli II inflation) and when the logarithmically flat prior or the stochastic prior~(\ref{eq:sigmaend:GaussianPrior}) on $\sigma_\uend$ are used. For the K\"{a}hler moduli II cases denoted ``n.c.'' in table~\ref{table:DKL}, well-converged distributions could not be inferred due to the numerical difficulty in sampling these scenarios.
\newpage
\subsection{Energy Density at the End of Inflation}
\label{sec:app:rhoend:individual}
\begin{figure}[!h]
\figpilogsto
\begin{center}
\includegraphics[width=0.45\textwidth]{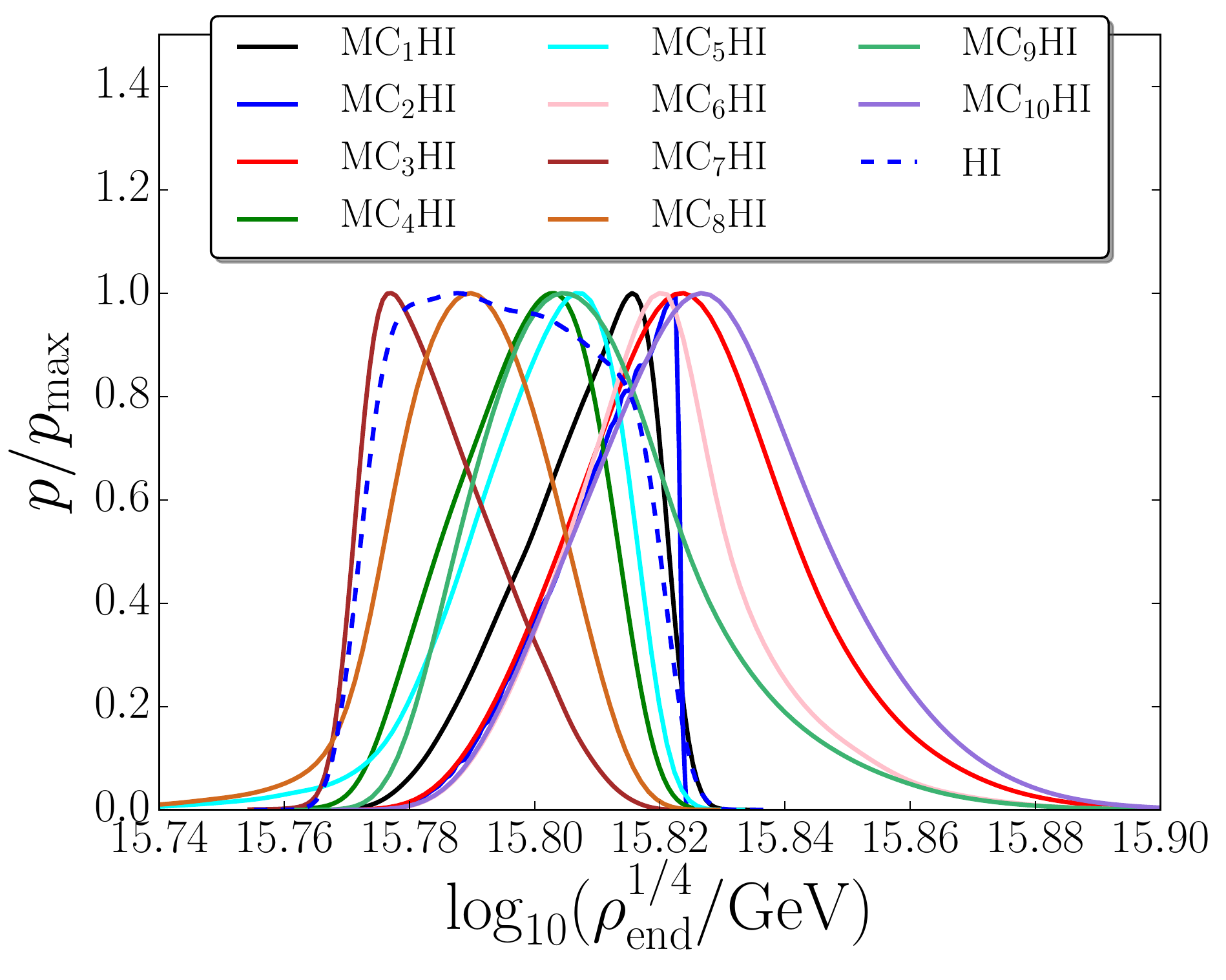}
\includegraphics[width=0.45\textwidth]{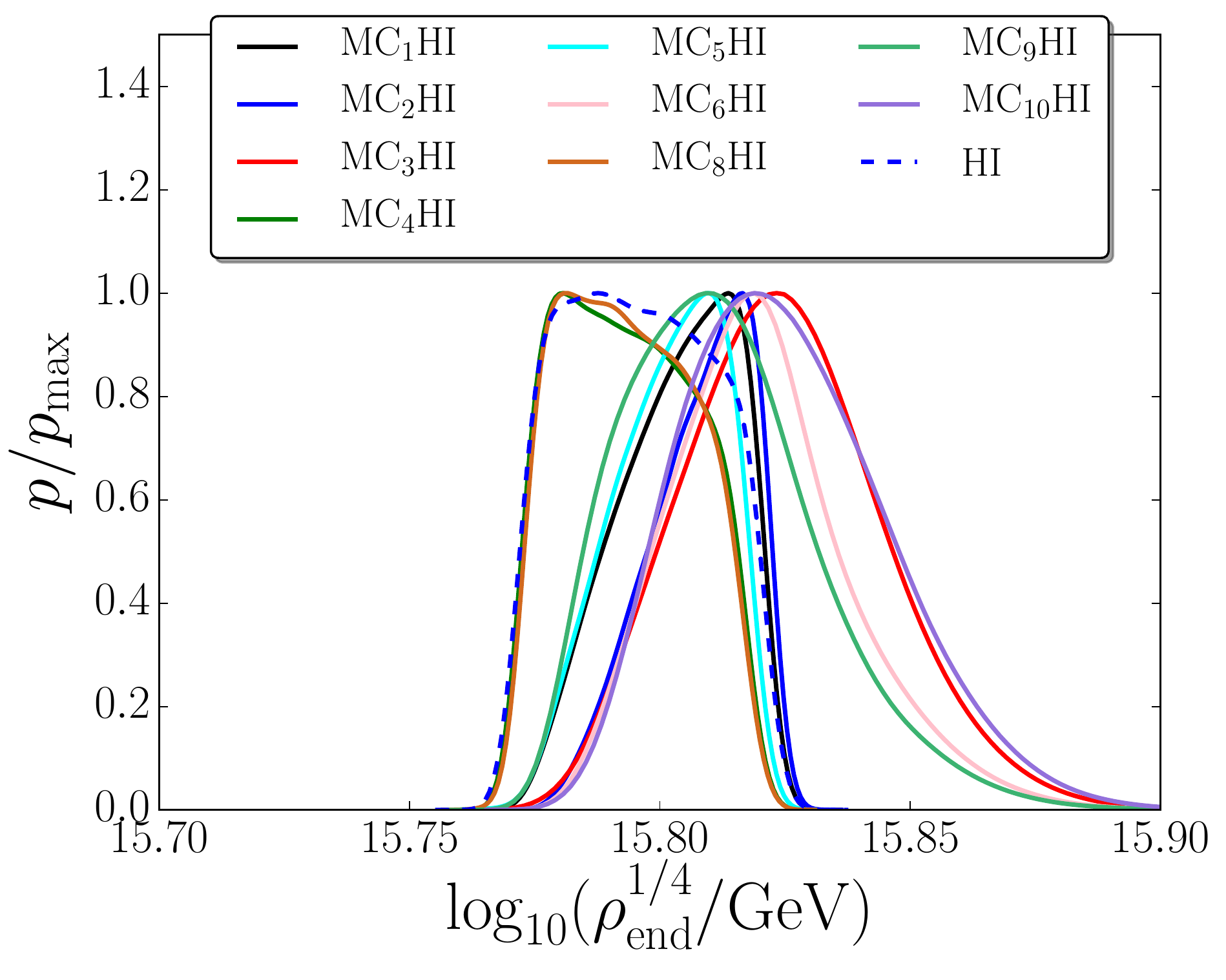}
\includegraphics[width=0.45\textwidth]{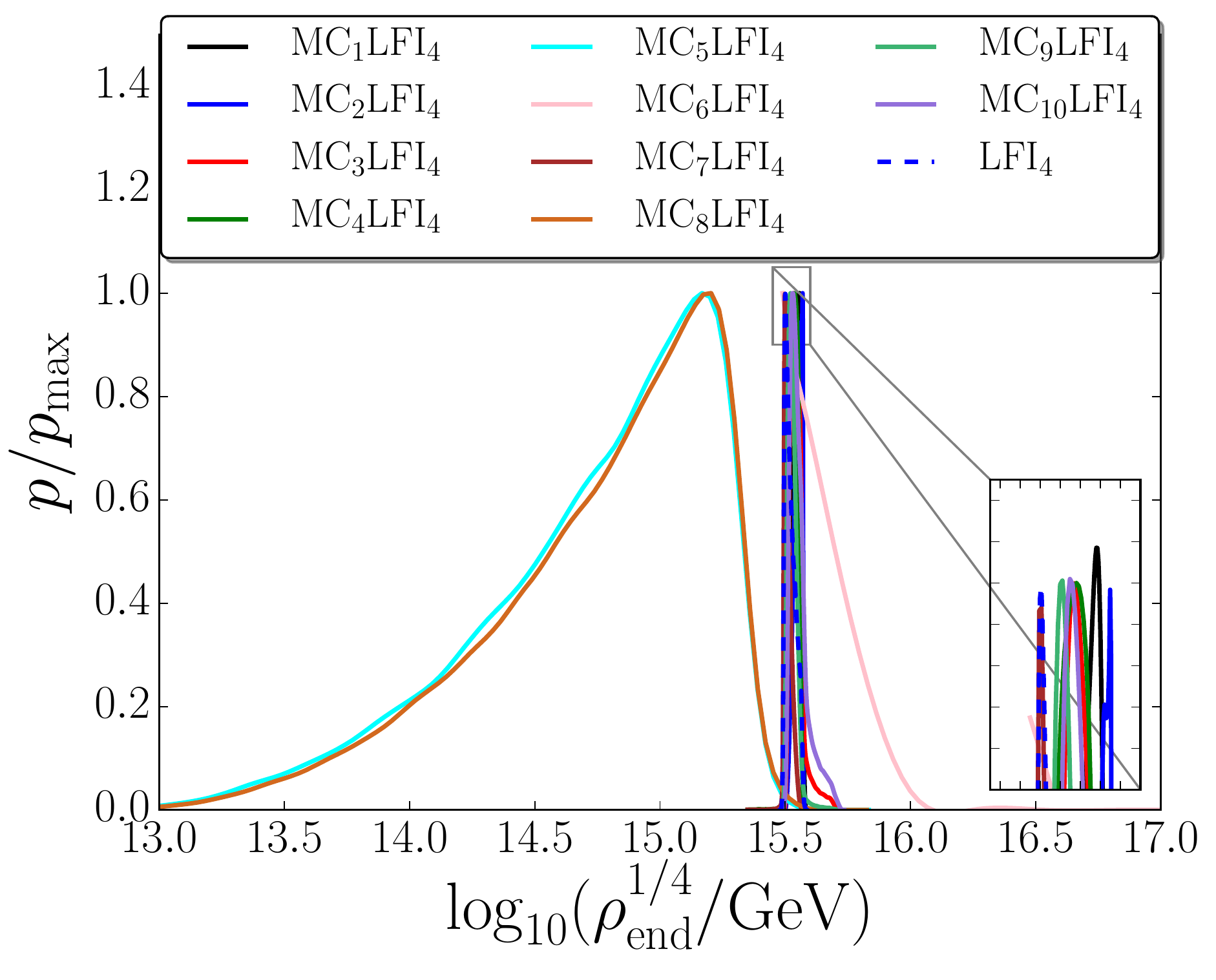}
\includegraphics[width=0.45\textwidth]{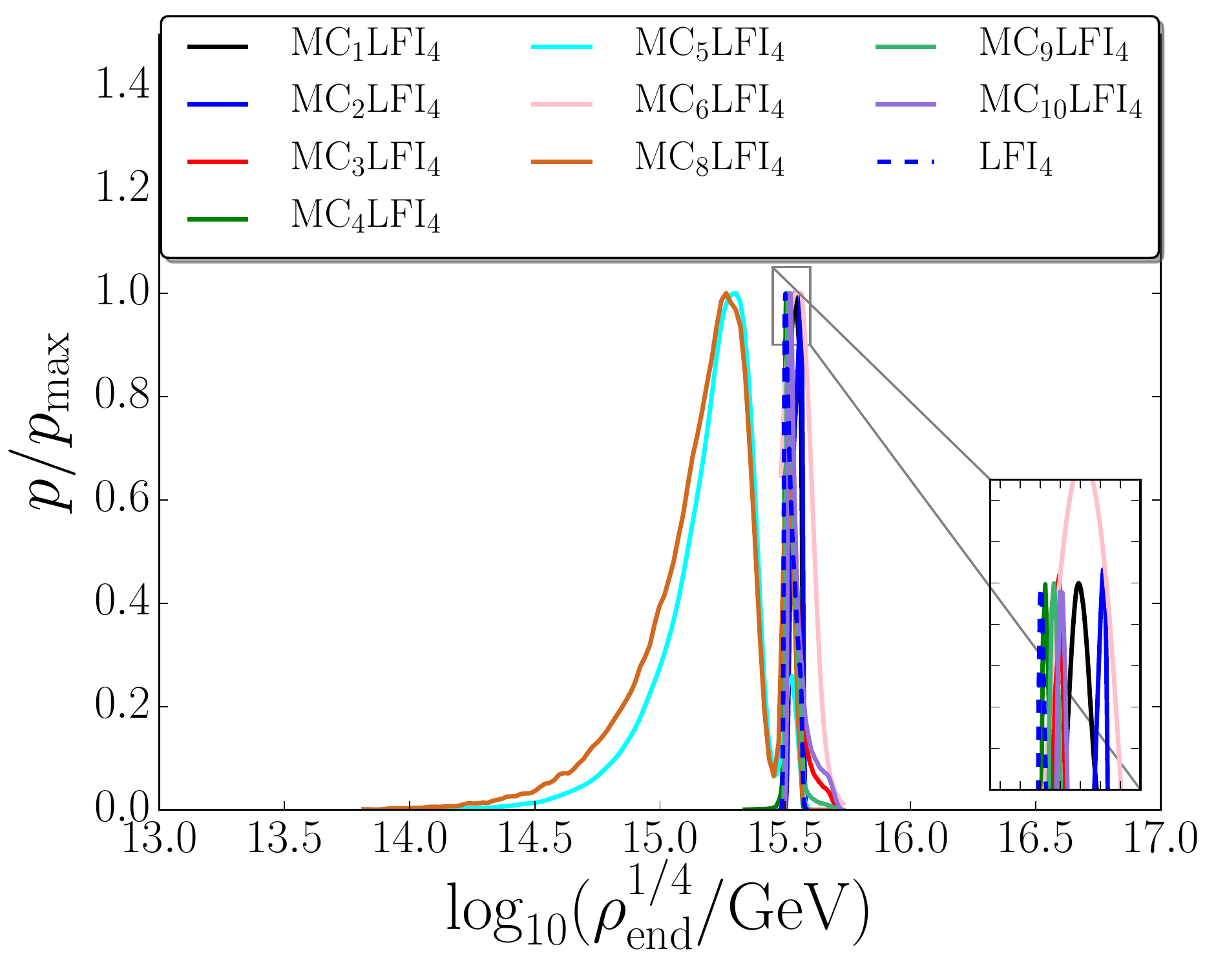}
\includegraphics[width=0.45\textwidth]{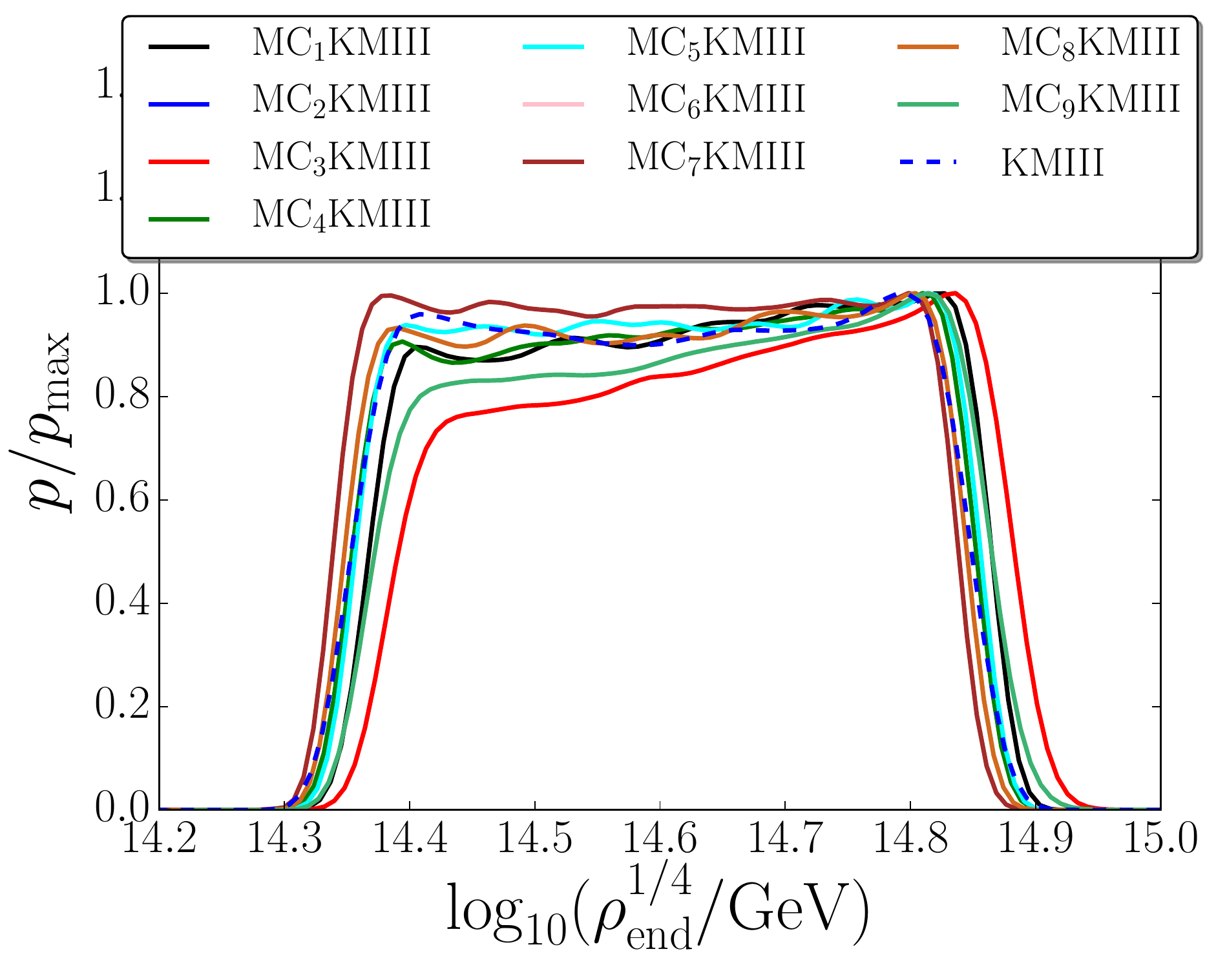}
\includegraphics[width=0.45\textwidth]{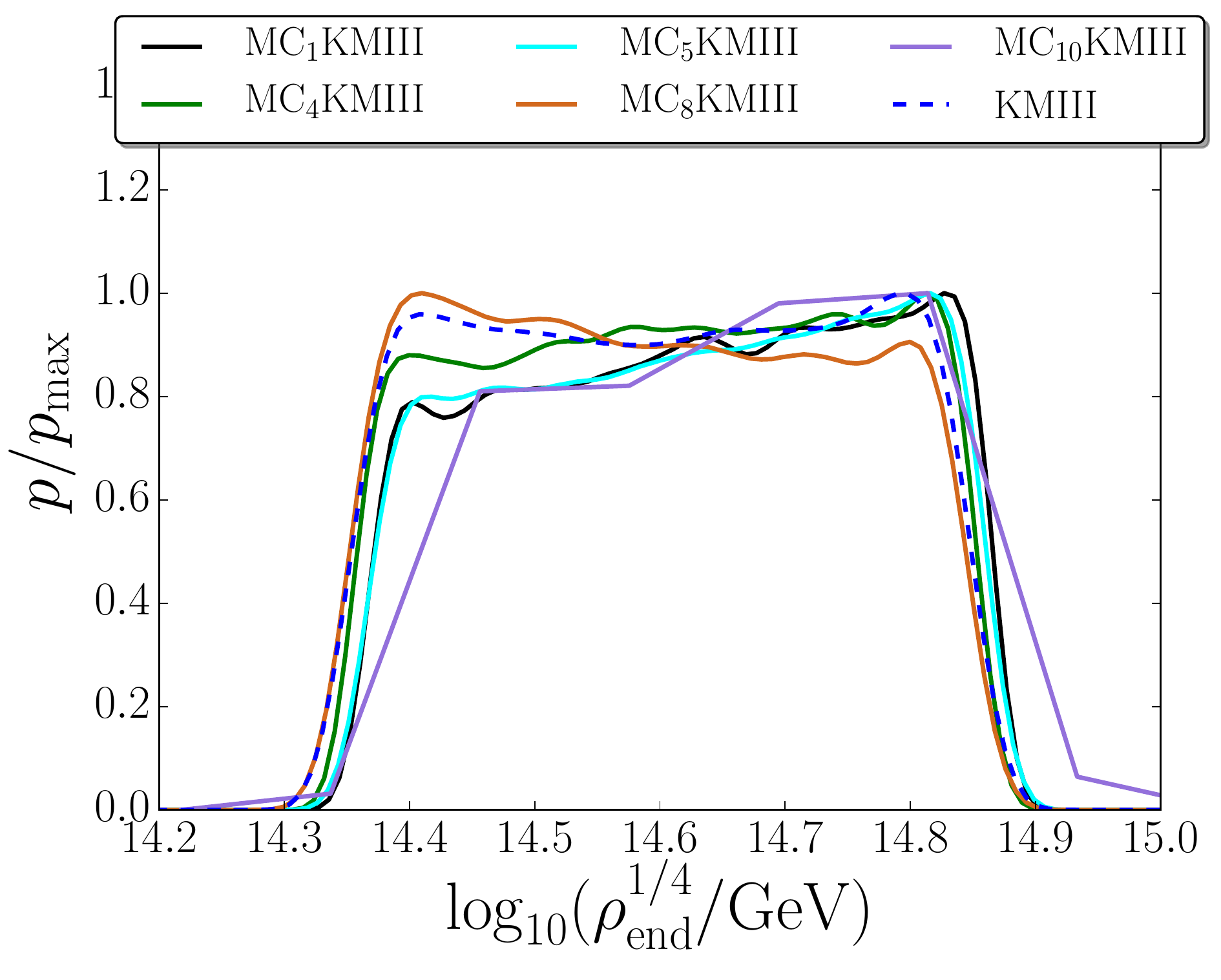}
\caption{Posterior distributions on the energy density at the end of inflation with the plateau potential~(\ref{eq:pot:hi}) of Higgs inflation (top panels), the quartic potential~(\ref{eq:pot:quartic}) (middle panels), and the plateau potential~(\ref{eq:pot:KMIII}) of K\"ahler moduli inflation II (bottom panels). The left panels correspond to the logarithmically flat prior~(\ref{eq:sigmaend:LogPrior}) on $\sigma_\uend$, and the right panels stand for the stochastic prior~(\ref{eq:sigmaend:GaussianPrior}) derived from the equilibrium distribution of a light scalar field in a de Sitter space-time with Hubble scale $H_\uend$. The dashed blue lines correspond to the single-field versions of the models, while the solid coloured lines stand for the $10$ reheating scenarios of \Fig{fig:cases} when an extra light scalar field is present.}
\label{fig:post:rhoend:individual}
\end{center}
\end{figure}
\newpage
\subsection{Reheating Temperature}
\label{sec:app:Treh:individual}
\begin{figure}[!h]
\figpilogsto
\begin{center}
\includegraphics[width=0.45\textwidth]{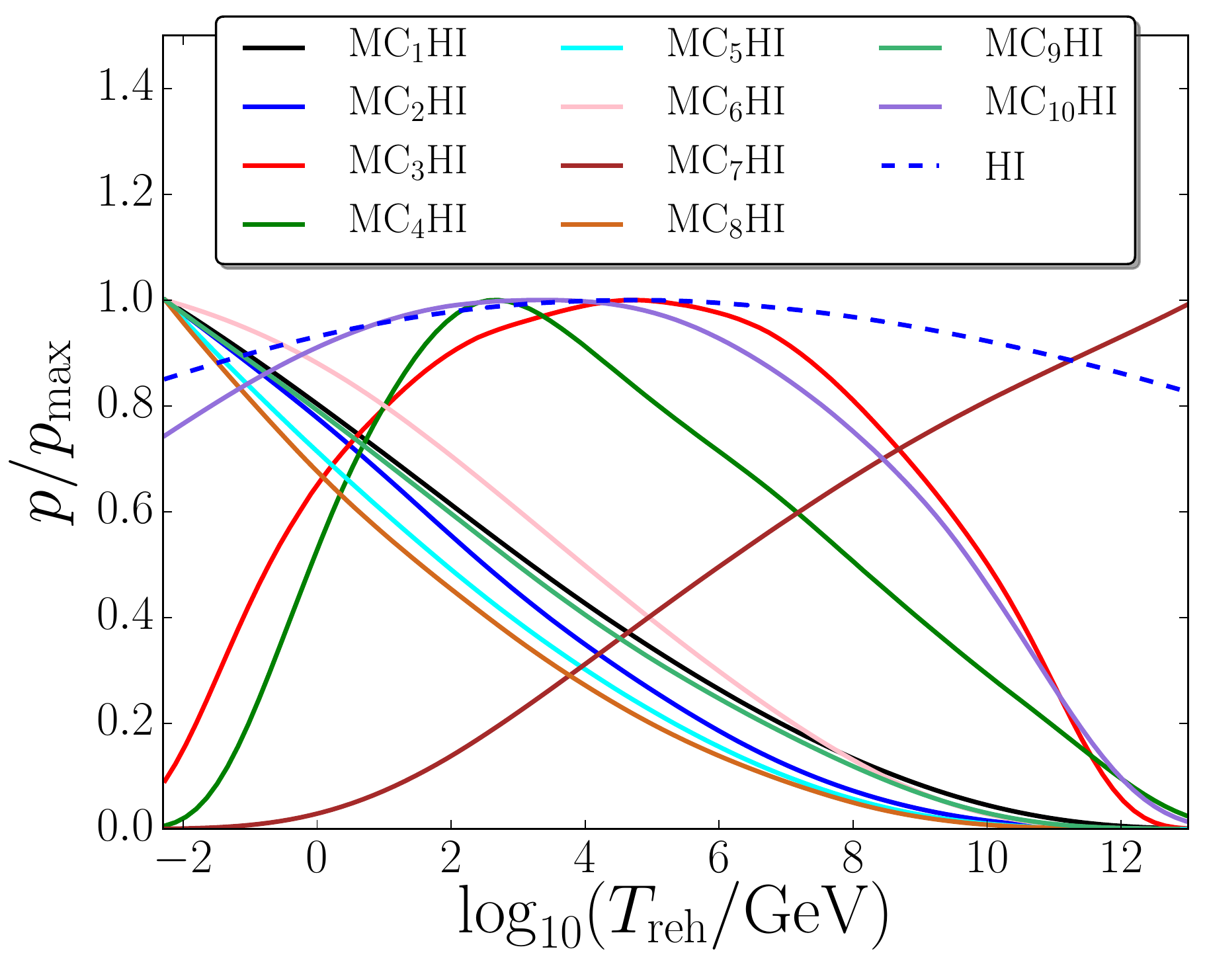}
\includegraphics[width=0.45\textwidth]{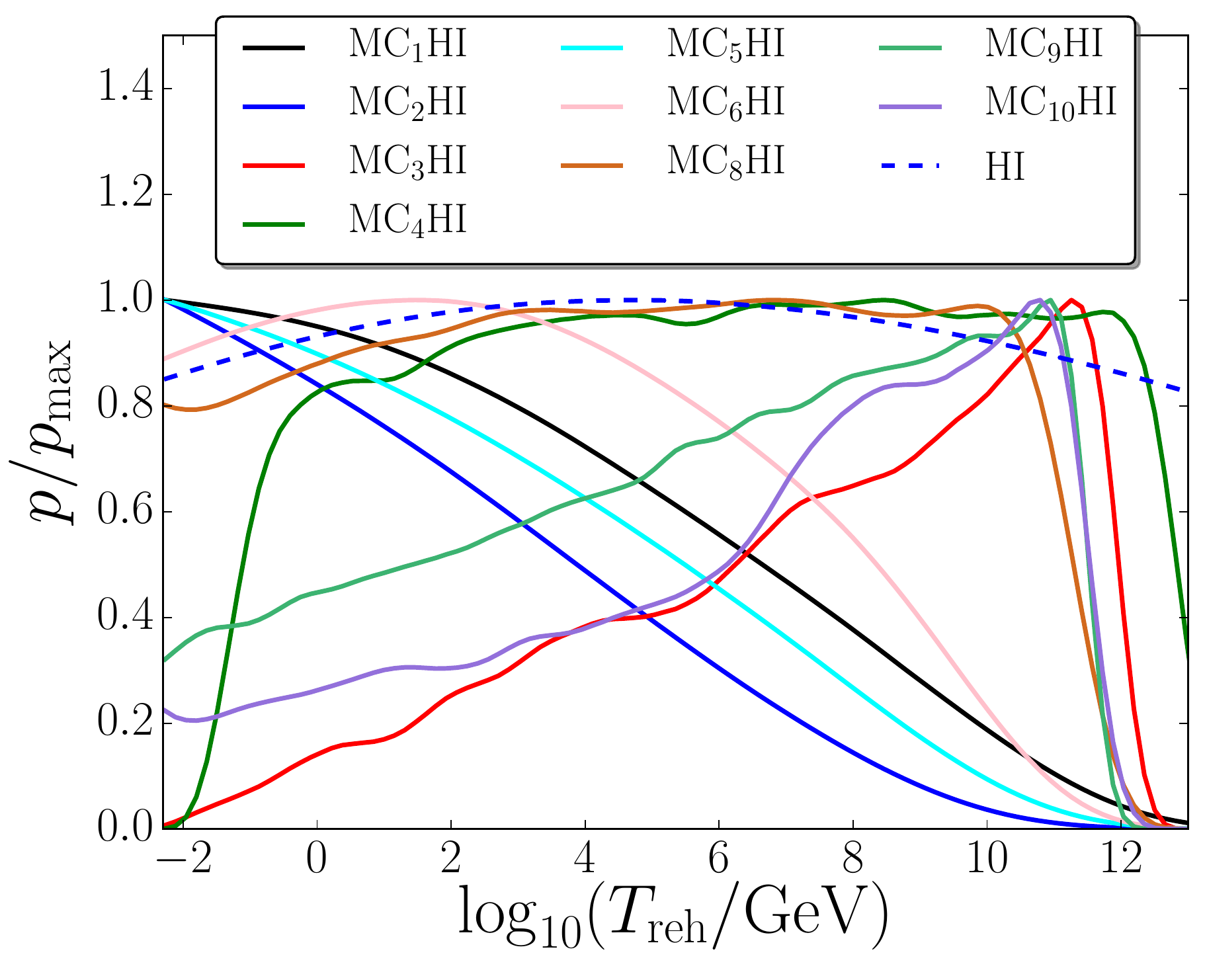}
\includegraphics[width=0.45\textwidth]{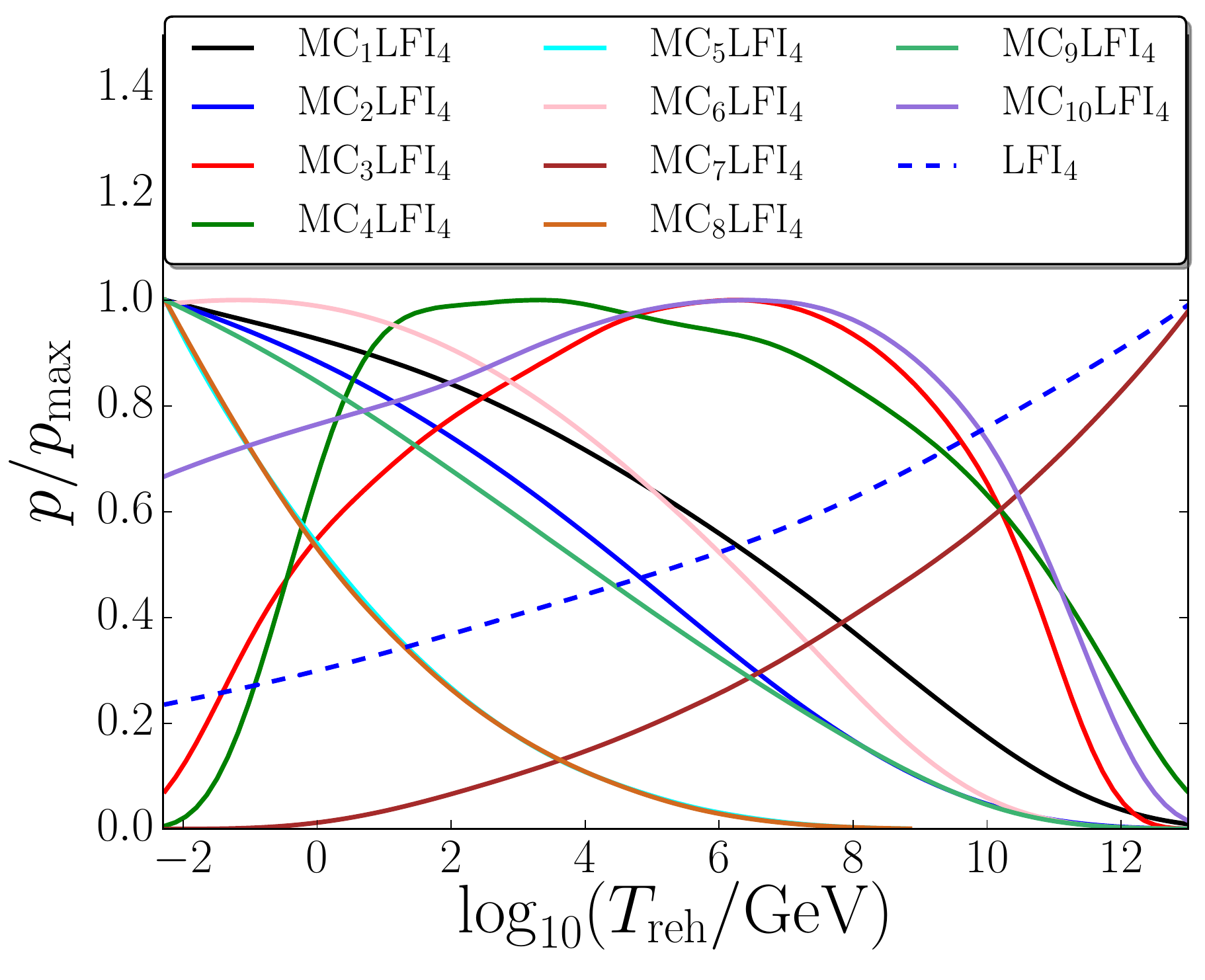}
\includegraphics[width=0.45\textwidth]{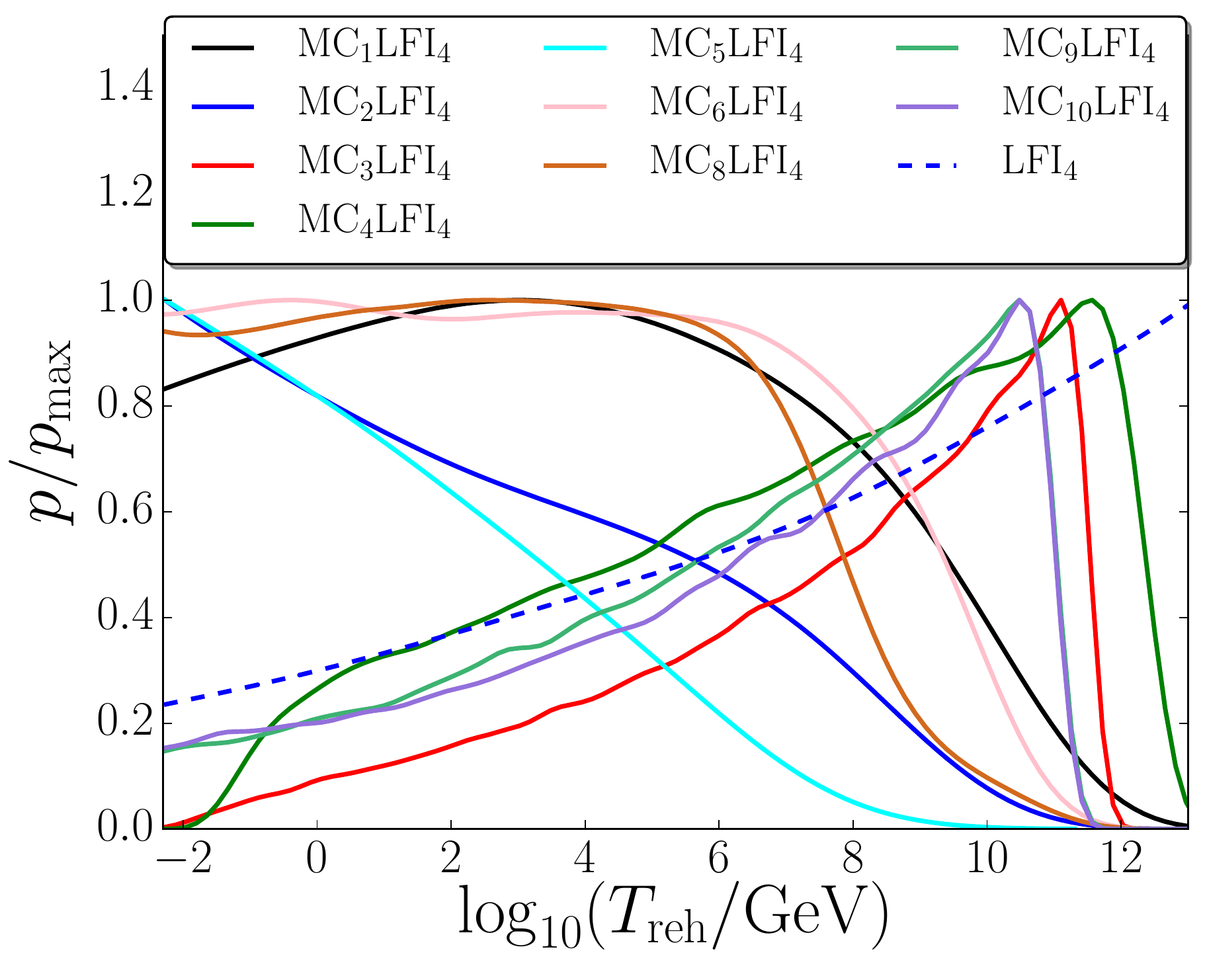}
\includegraphics[width=0.45\textwidth]{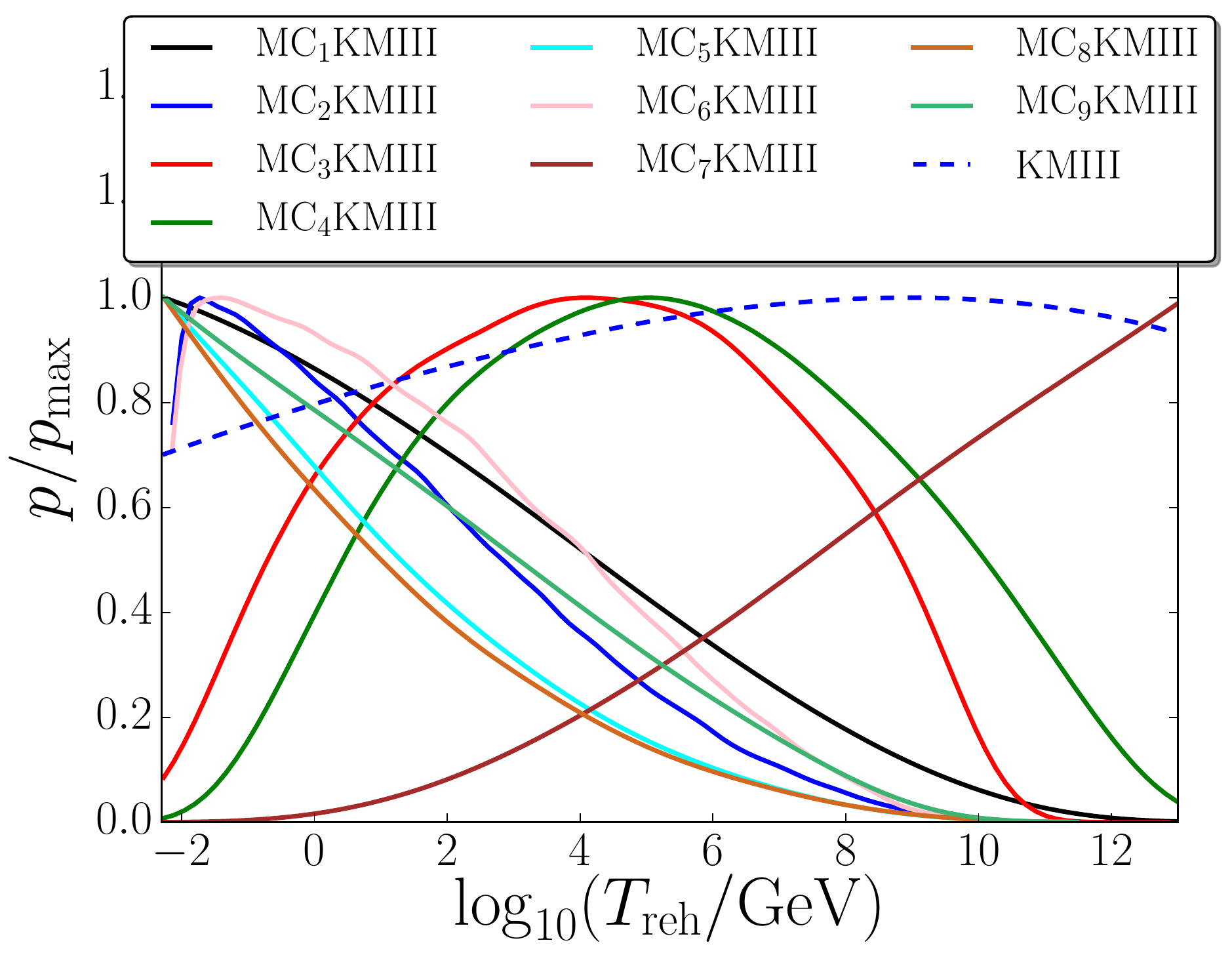}
\includegraphics[width=0.45\textwidth]{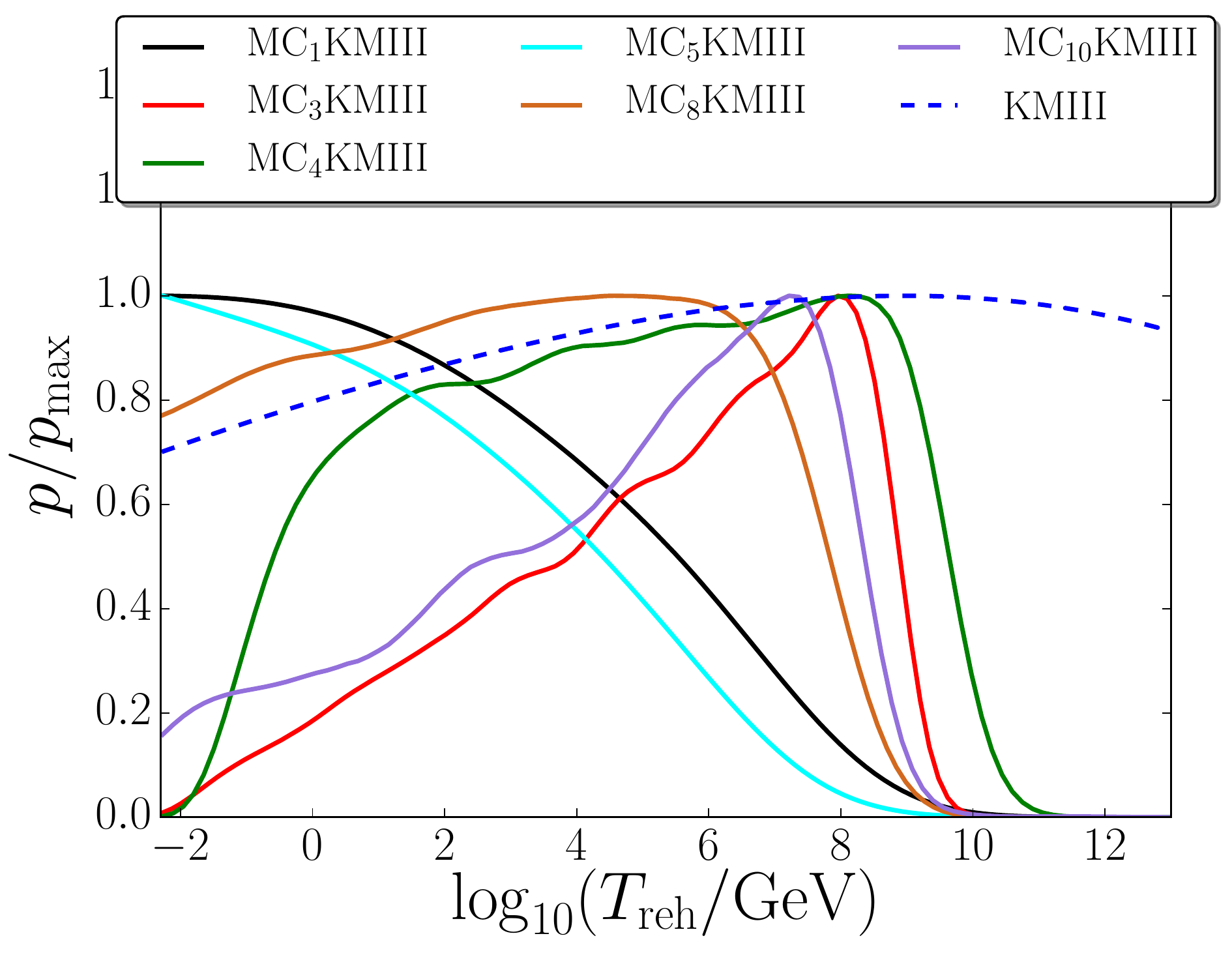}
\caption{Posterior distributions on the reheating temperature $T_\ureh$ with the plateau potential~(\ref{eq:pot:hi}) of Higgs inflation (top panels), the quartic potential~(\ref{eq:pot:quartic}) (middle panels), and the plateau potential~(\ref{eq:pot:KMIII}) of K\"ahler moduli inflation II (bottom panels). The left panels correspond to the logarithmically flat prior~(\ref{eq:sigmaend:LogPrior}) on $\sigma_\uend$, and the right panels stand for the stochastic prior~(\ref{eq:sigmaend:GaussianPrior}) derived from the equilibrium distribution of a light scalar field in a de Sitter space-time with Hubble scale $H_\uend$. The dashed blue lines correspond to the single-field versions of the models, while the solid coloured lines stand for the $10$ reheating scenarios of \Fig{fig:cases} when an extra light scalar field is present.}
\label{fig:post:Treh:individual}
\end{center}
\end{figure}
\newpage
\subsection{Early Reheating Temperature}
\label{sec:app:Tereh:individual}
\begin{figure}[!h]
\figpilogsto
\begin{center}
\includegraphics[width=0.45\textwidth]{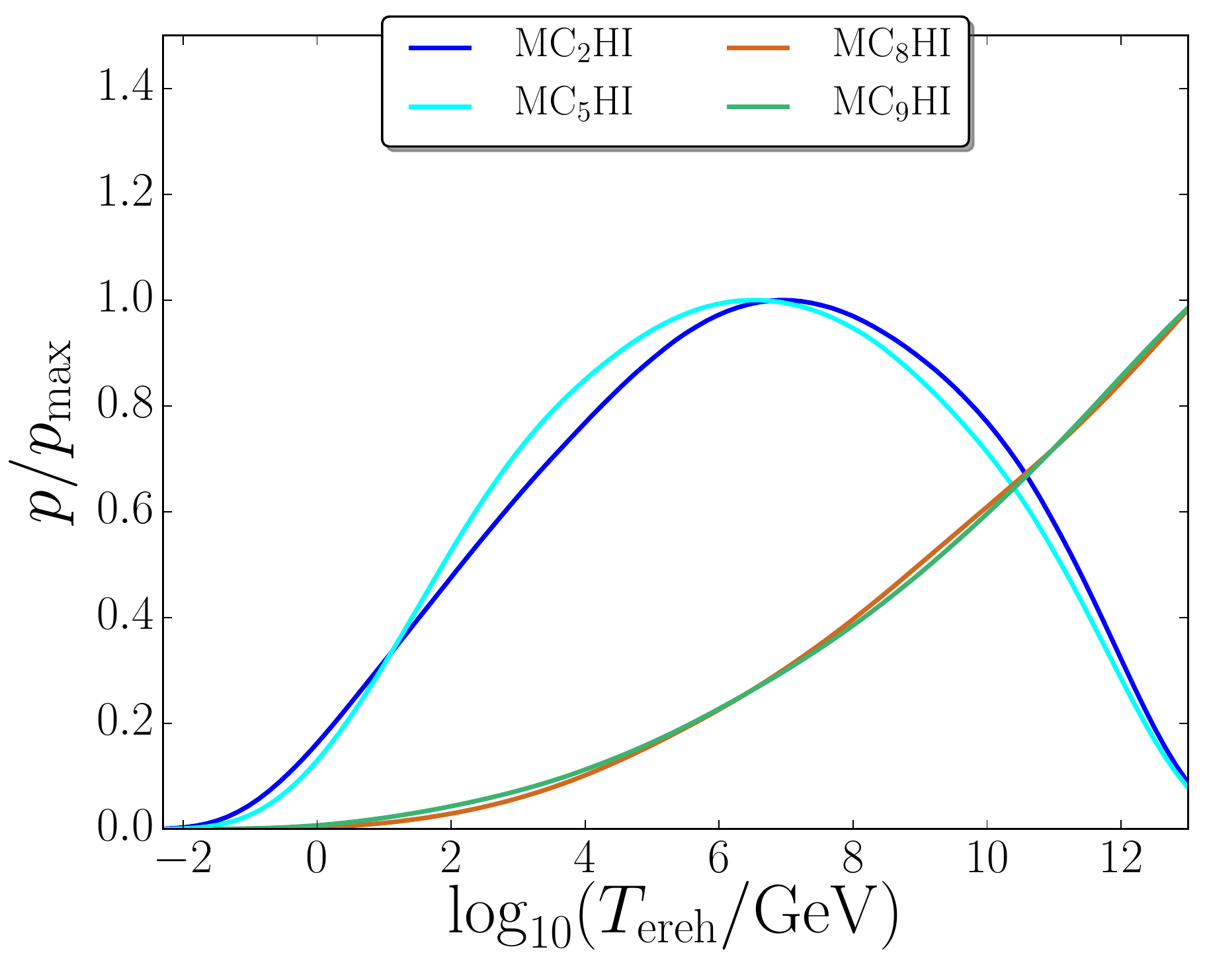}
\includegraphics[width=0.45\textwidth]{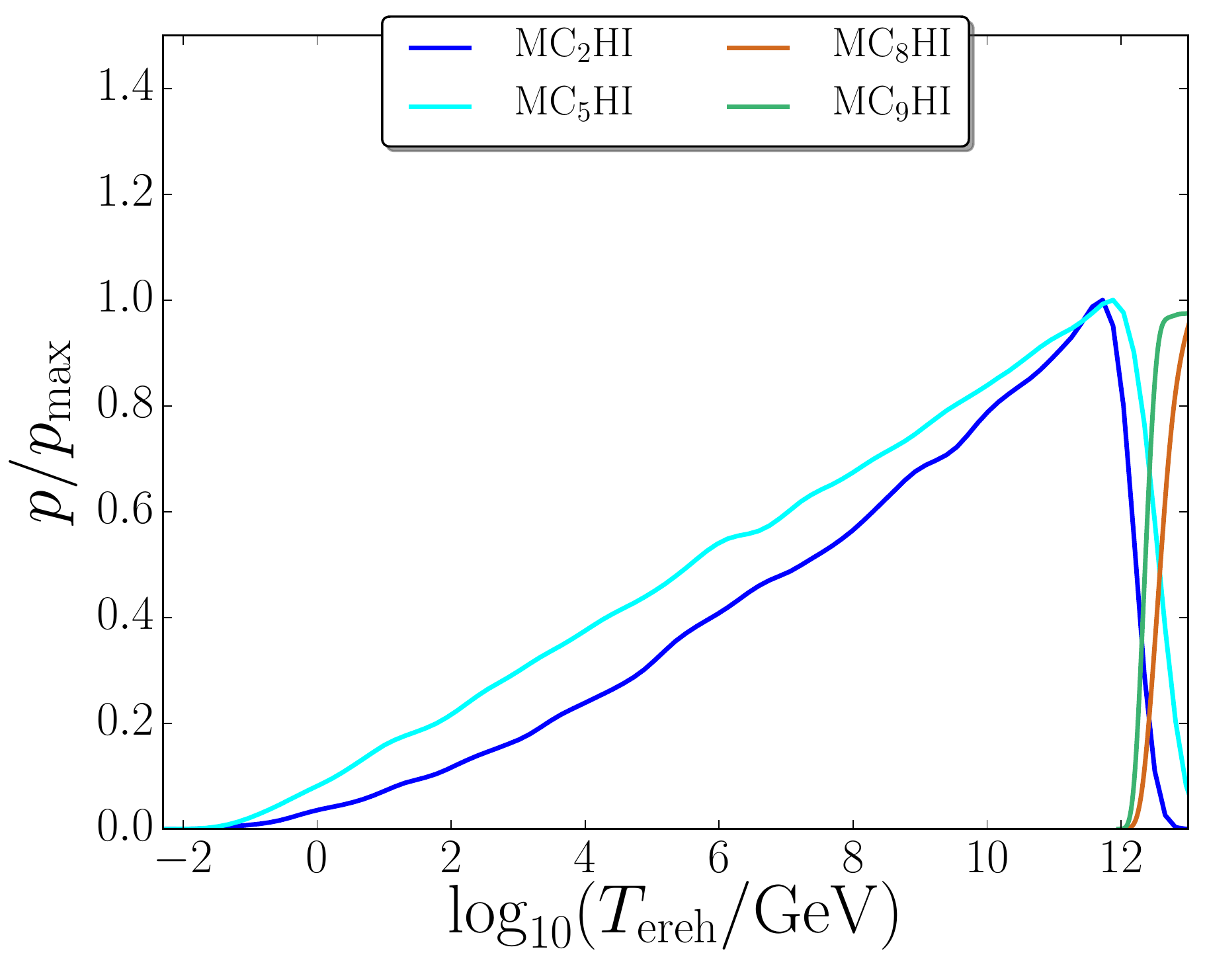}
\includegraphics[width=0.45\textwidth]{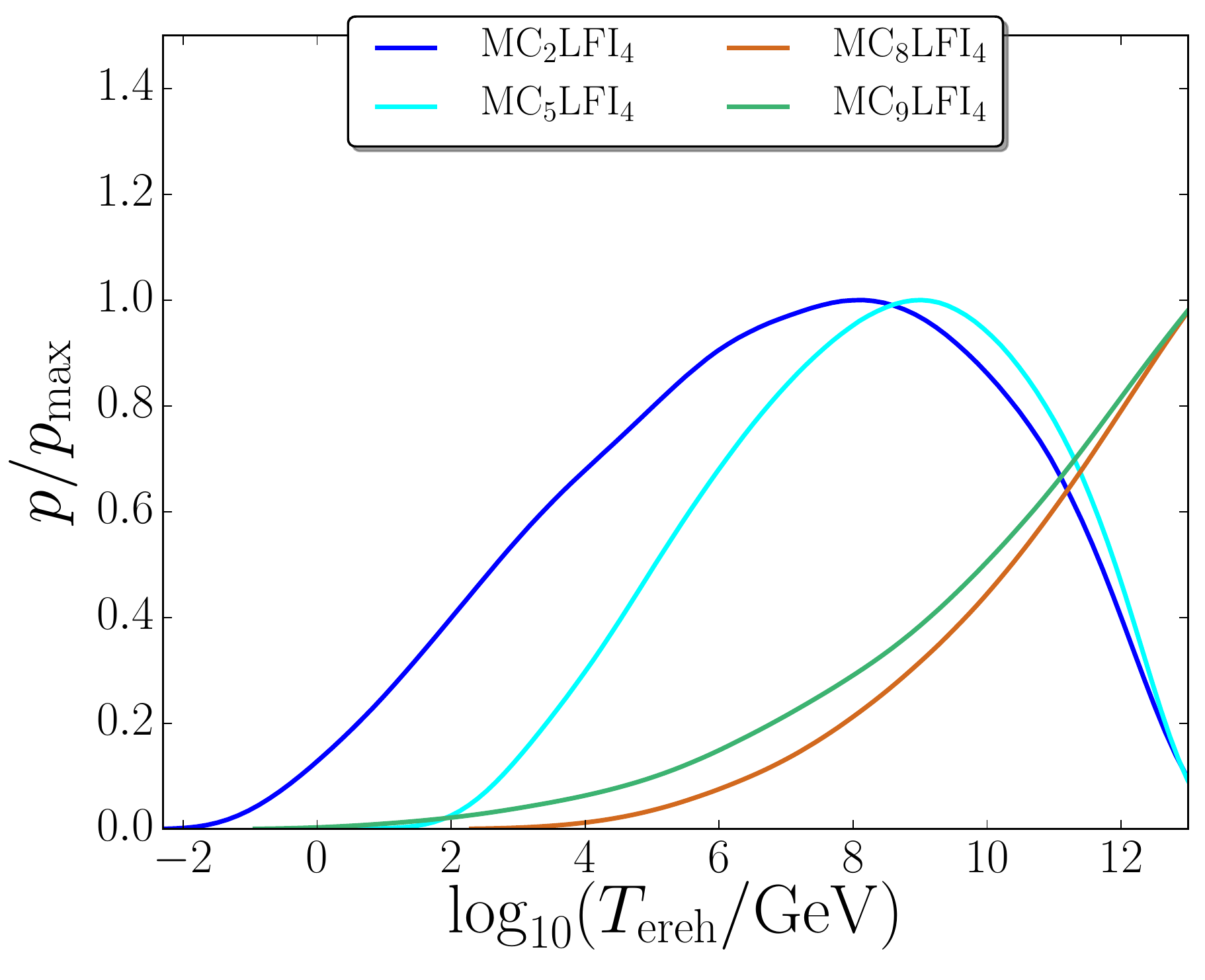}
\includegraphics[width=0.45\textwidth]{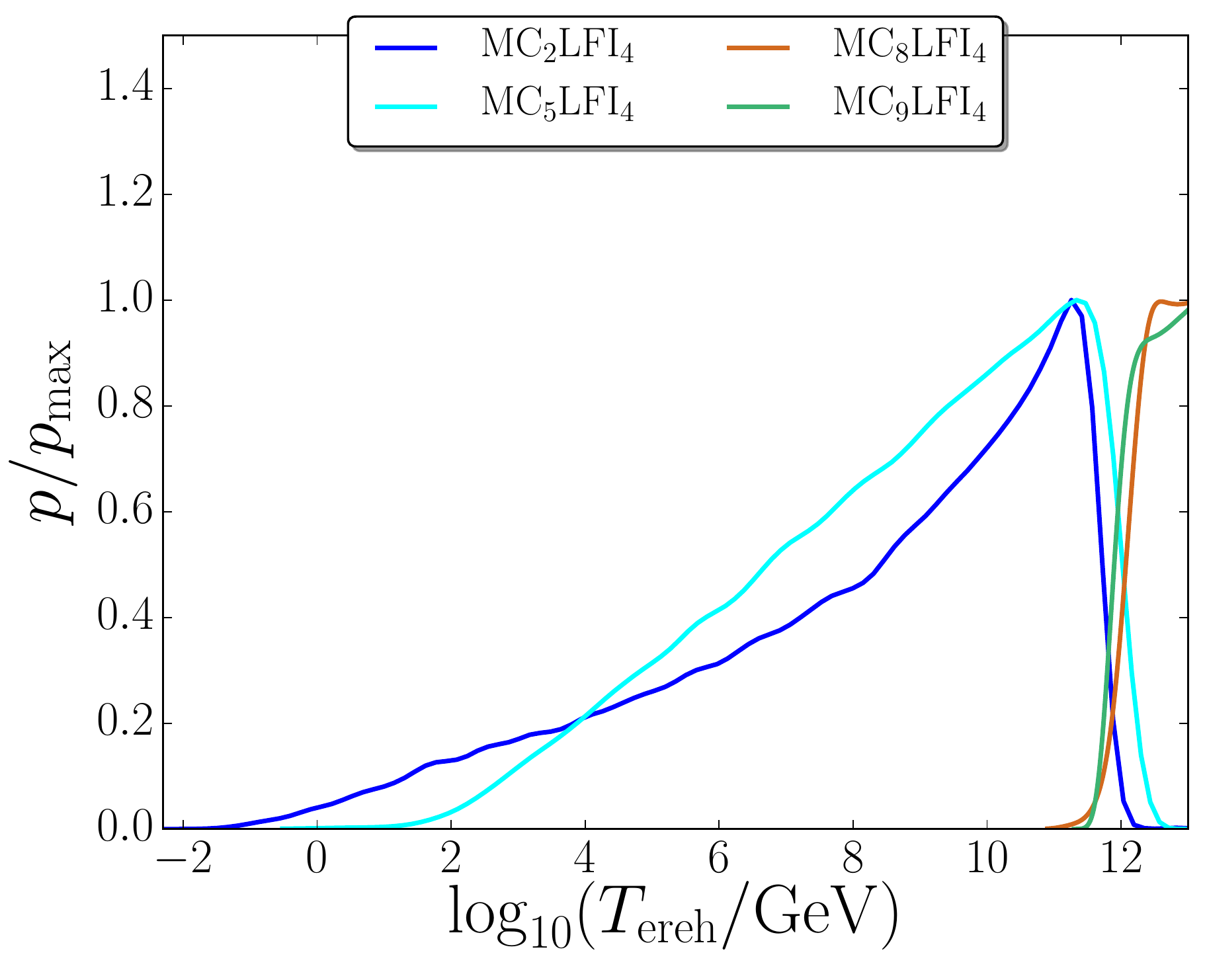}
\includegraphics[width=0.45\textwidth]{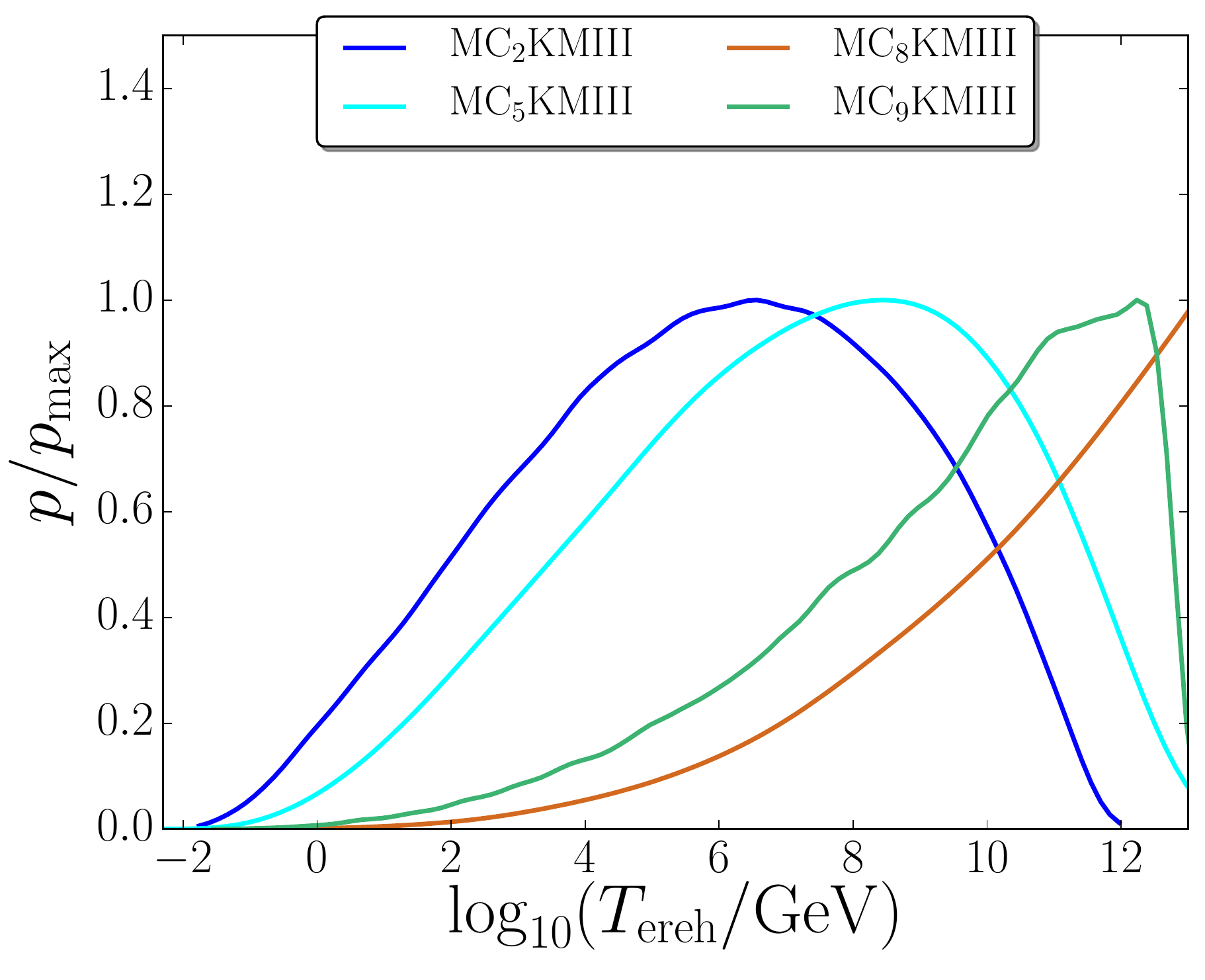}
\includegraphics[width=0.45\textwidth]{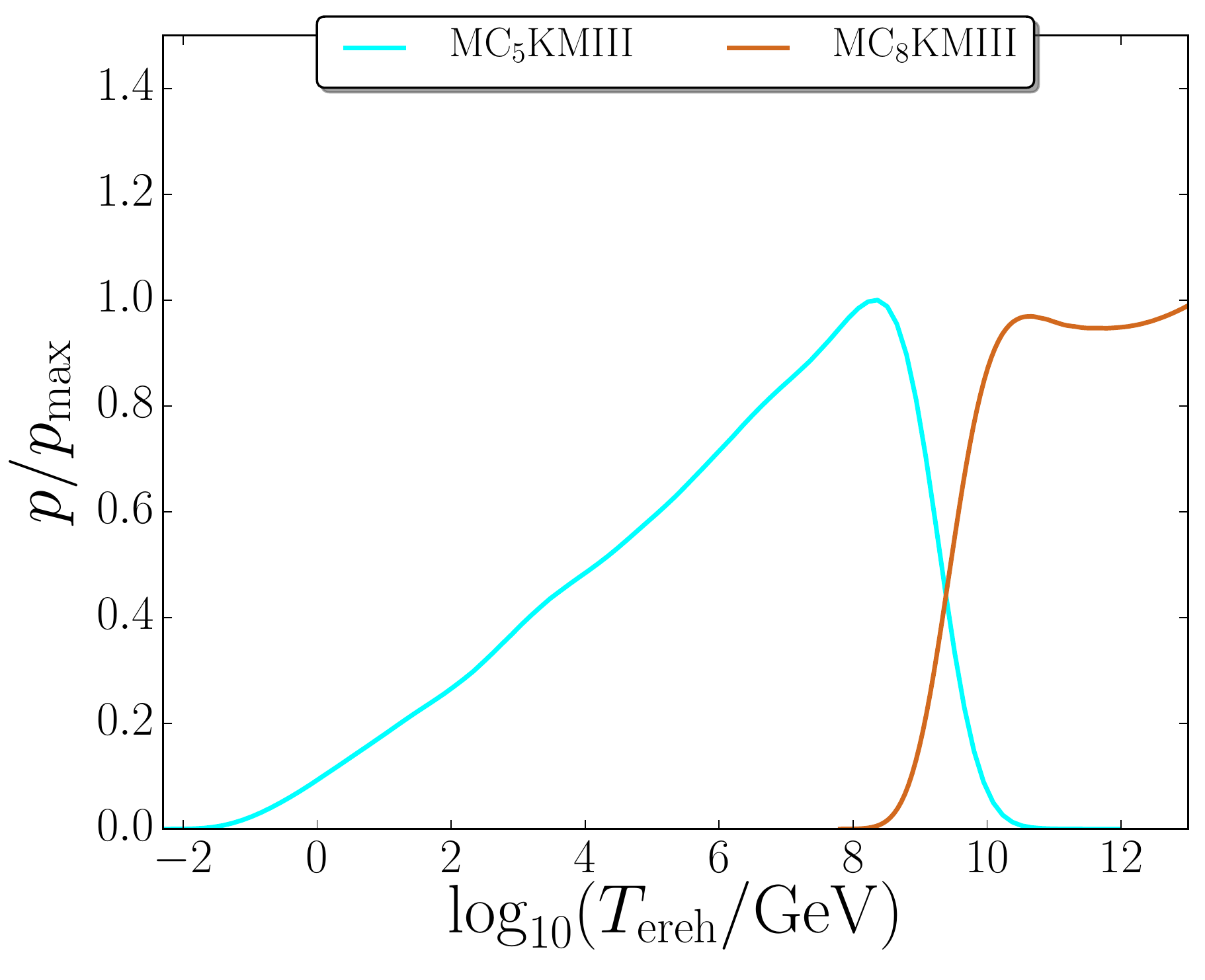}
\caption{Posterior distributions on the early reheating temperature $T_\uereh$ with the plateau potential~(\ref{eq:pot:hi}) of Higgs inflation (top panels), the quartic potential~(\ref{eq:pot:quartic}) (middle panels), and the plateau potential~(\ref{eq:pot:KMIII}) of K\"ahler moduli inflation II (bottom panels). The left panels correspond to the logarithmically flat prior~(\ref{eq:sigmaend:LogPrior}) on $\sigma_\uend$, and the right panels stand for the stochastic prior~(\ref{eq:sigmaend:GaussianPrior}) derived from the equilibrium distribution of a light scalar field in a de Sitter space-time with Hubble scale $H_\uend$. The dashed blue lines correspond to the single-field versions of the models, while the solid coloured lines stand for the $10$ reheating scenarios of \Fig{fig:cases} when an extra light scalar field is present.}
\label{fig:post:Tereh:individual}
\end{center}
\end{figure}
\newpage
\section{Information Density}
\label{Sec:DKLDensity} 
%
%
\subsection{Energy Density at the End of Inflation}
\label{sec:app:DKL:rhoend}
\begin{figure}[!h]
\figpilogsto
\begin{center}
\includegraphics[width=0.45\textwidth]{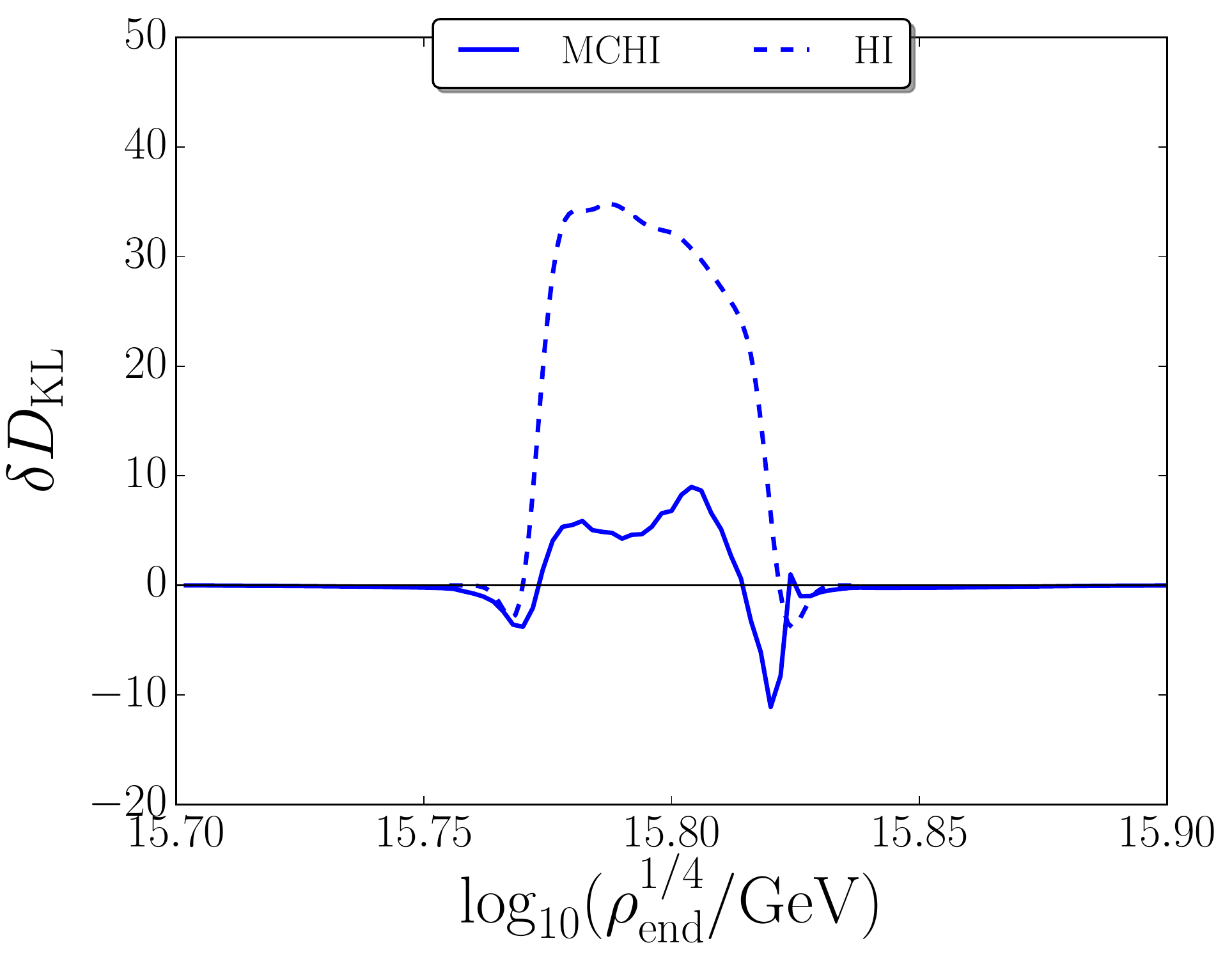}
\includegraphics[width=0.45\textwidth]{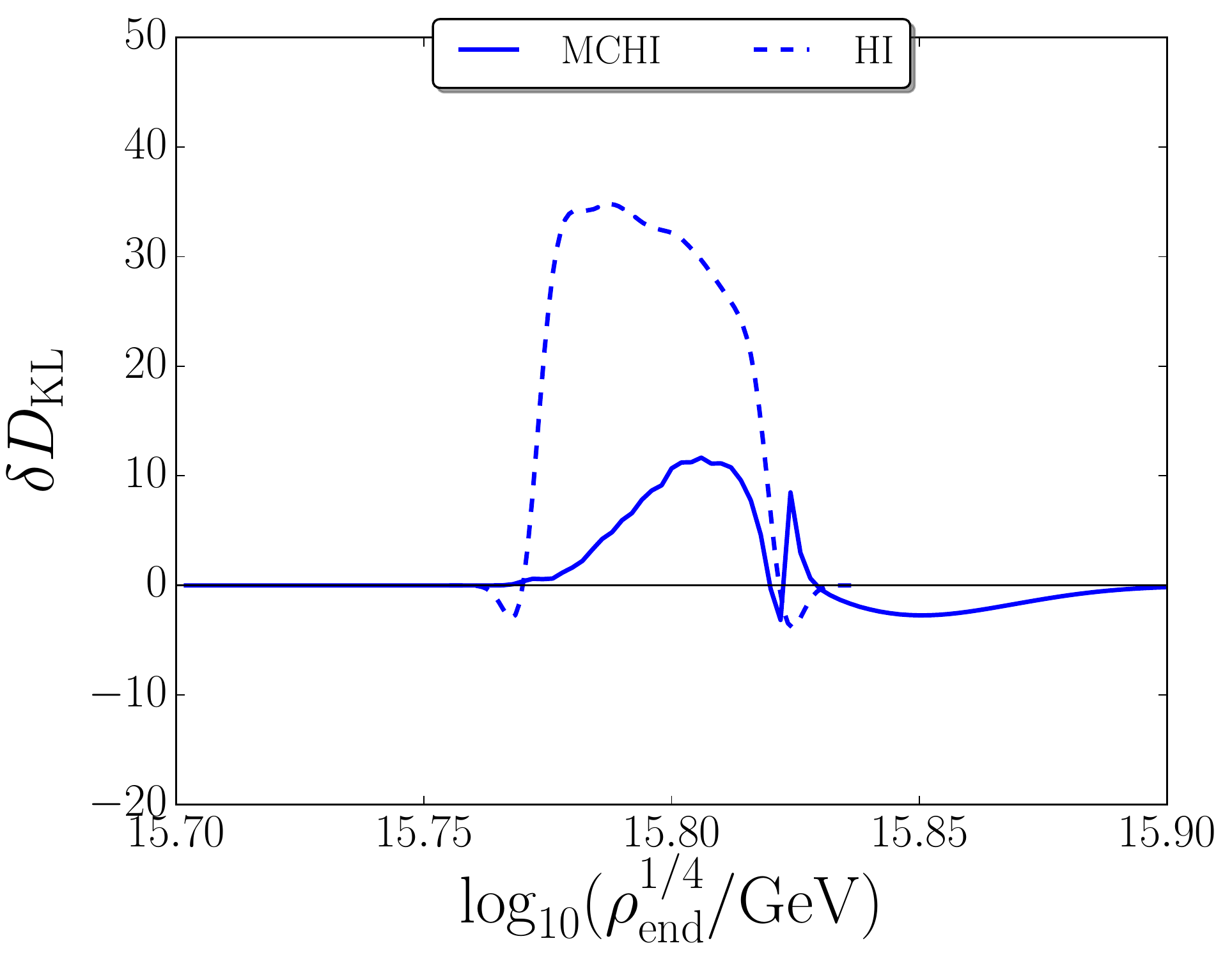}
\includegraphics[width=0.45\textwidth]{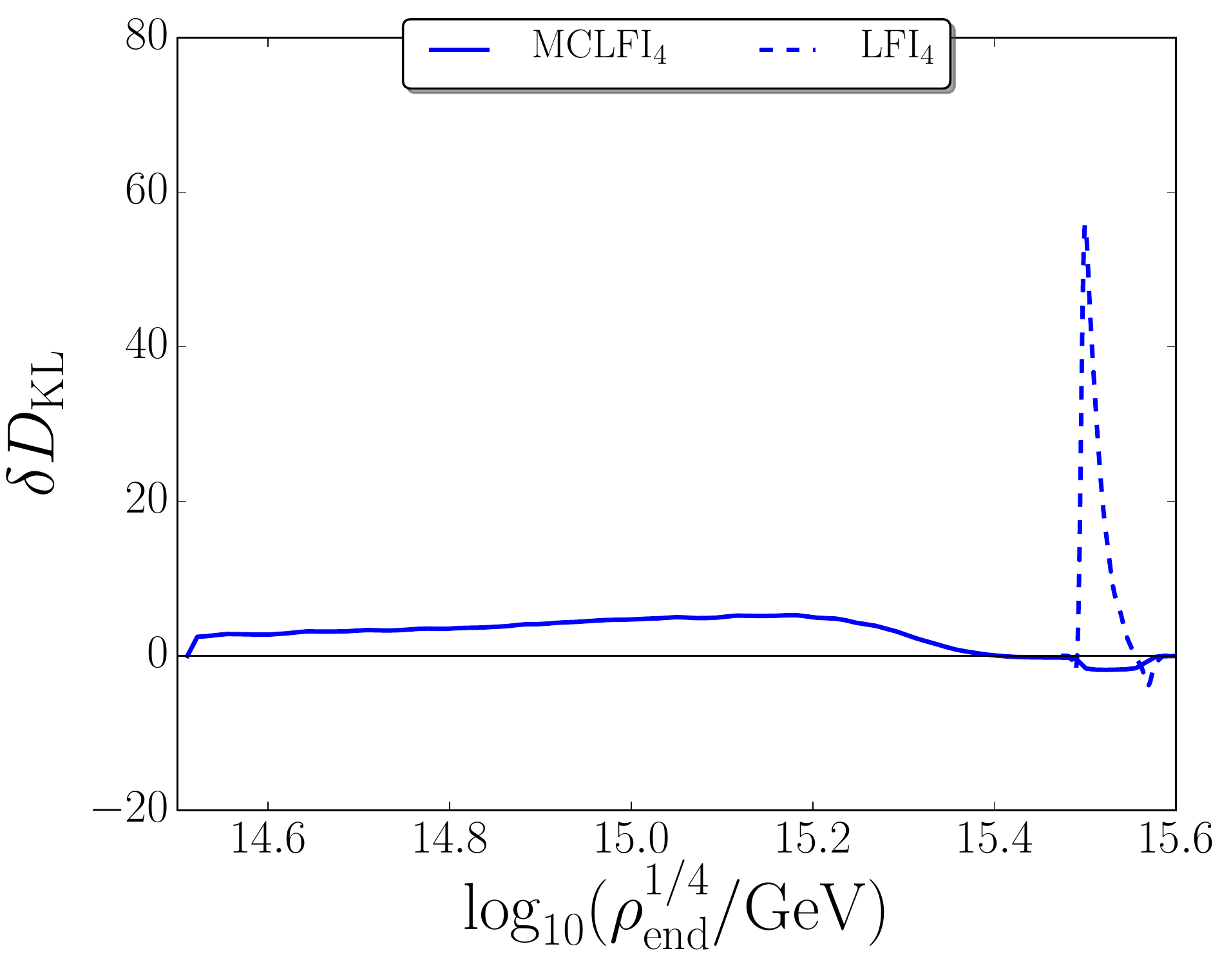}
\includegraphics[width=0.45\textwidth]{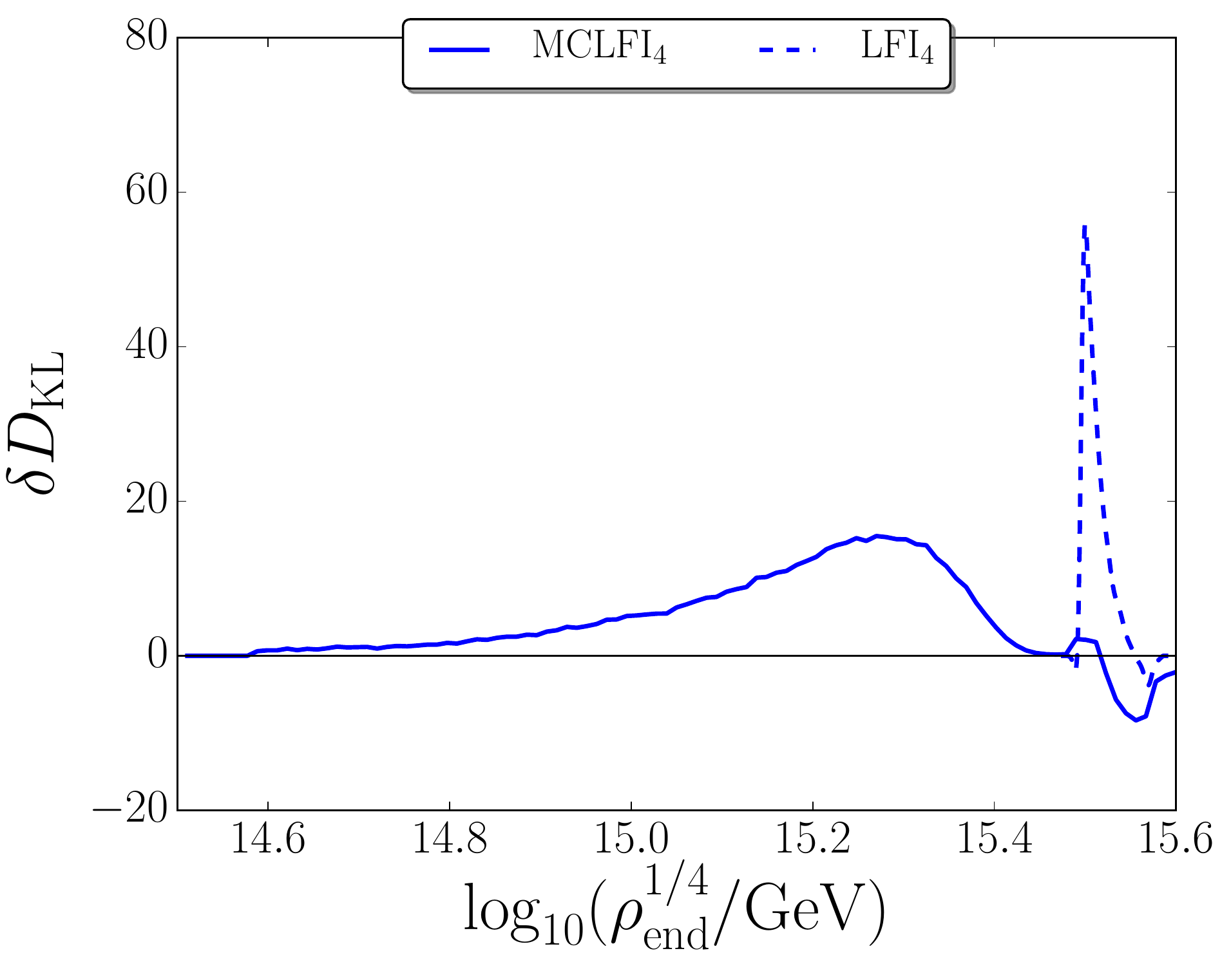}
\includegraphics[width=0.45\textwidth]{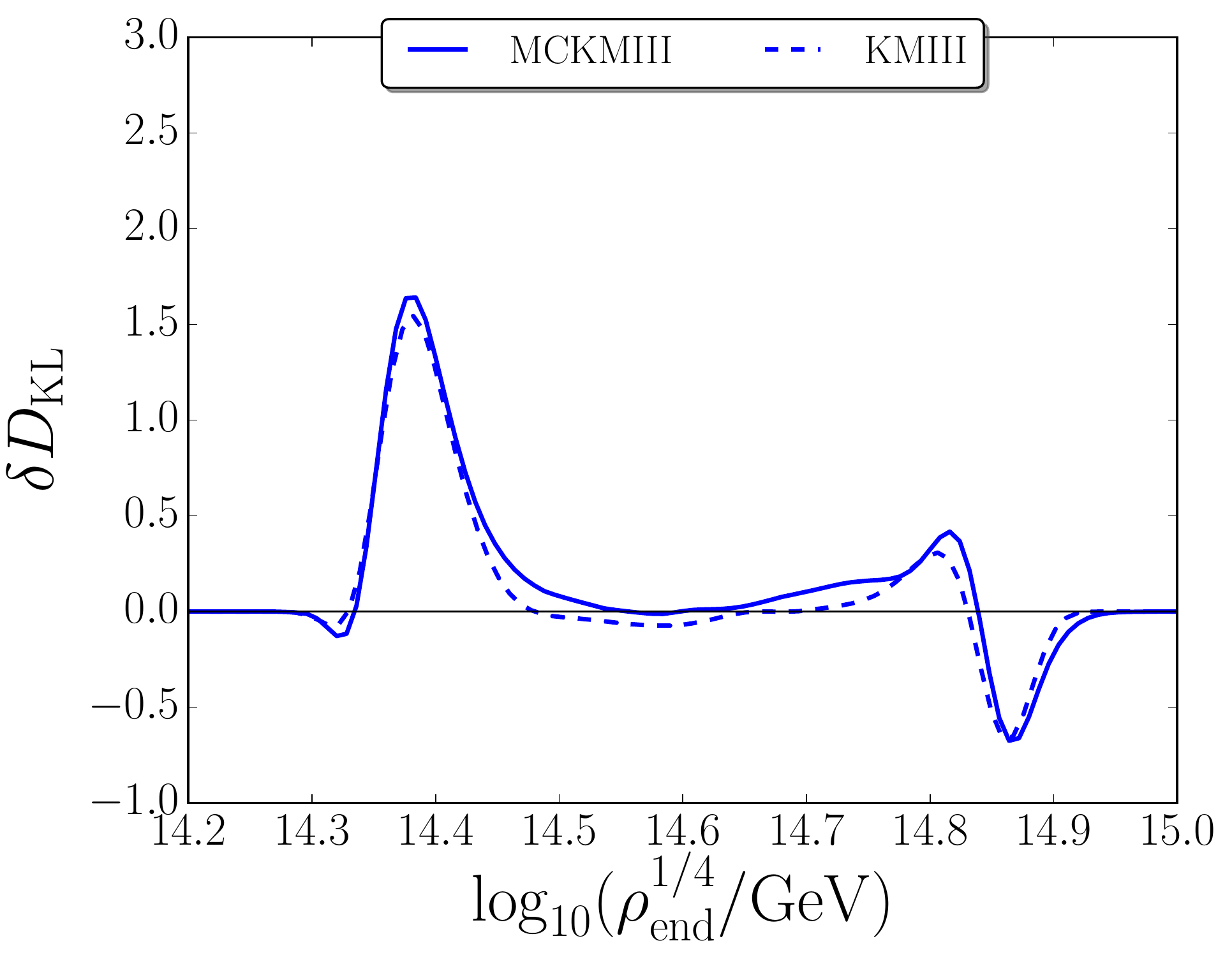}
\includegraphics[width=0.45\textwidth]{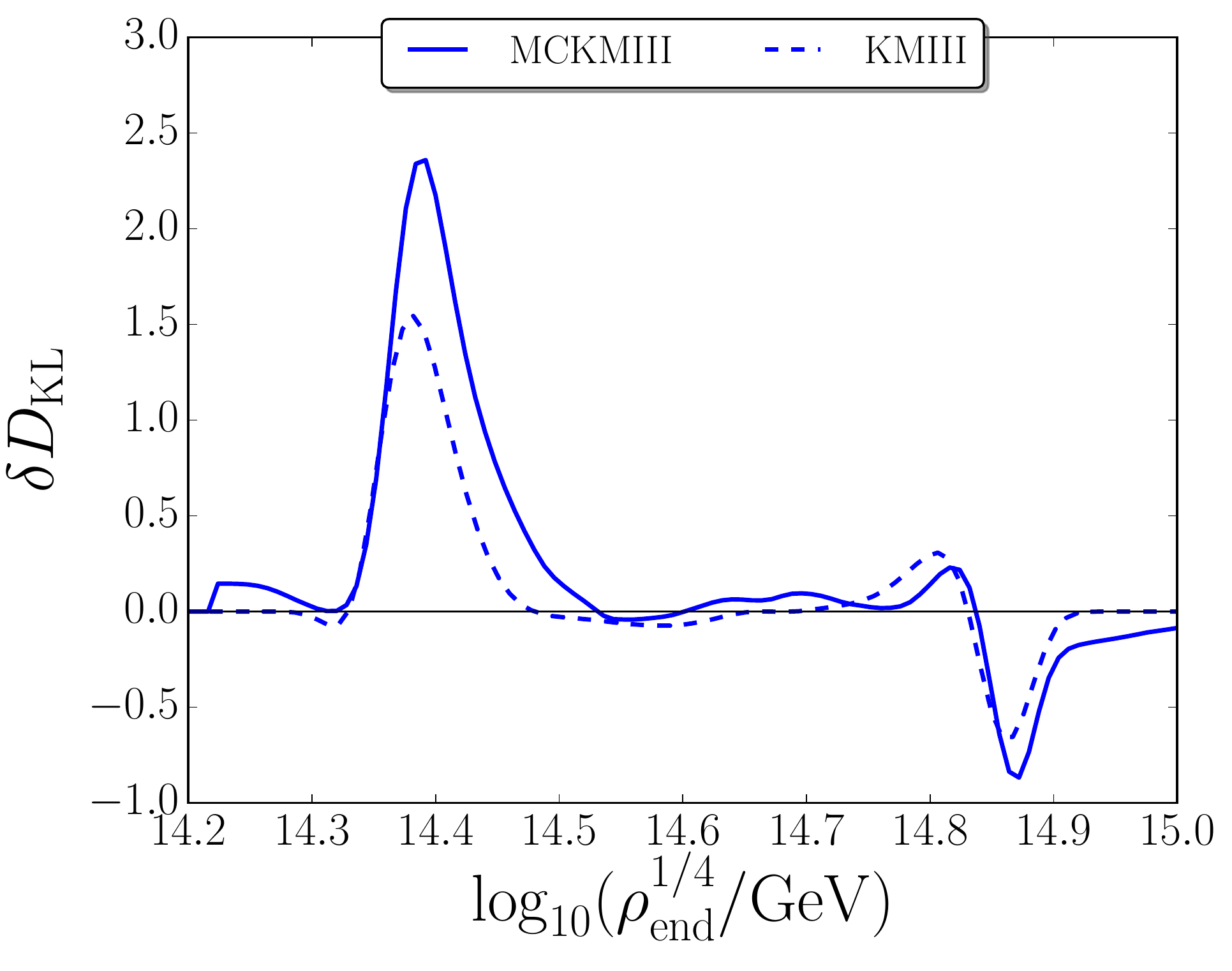}
\caption{Information density on $\rho_\uend$ for Higgs inflation (top panels), quartic inflation (middle panels) and K\"ahler moduli inflation II (bottom panels). The left panels correspond to the logarithmically flat prior~(\ref{eq:sigmaend:LogPrior}) on $\sigma_\uend$, and the right panels stand for the stochastic prior~(\ref{eq:sigmaend:GaussianPrior}). The dashed lines correspond to the single-field versions of the models, while the solid lines are derived from the averaged distributions over all $10$ reheating scenarios. 
}
\label{fig:DKL:rhoend:averaged}
\end{center}
\end{figure}
\newpage
\begin{figure}[!h]
\figpilogsto
\begin{center}
\includegraphics[width=0.45\textwidth]{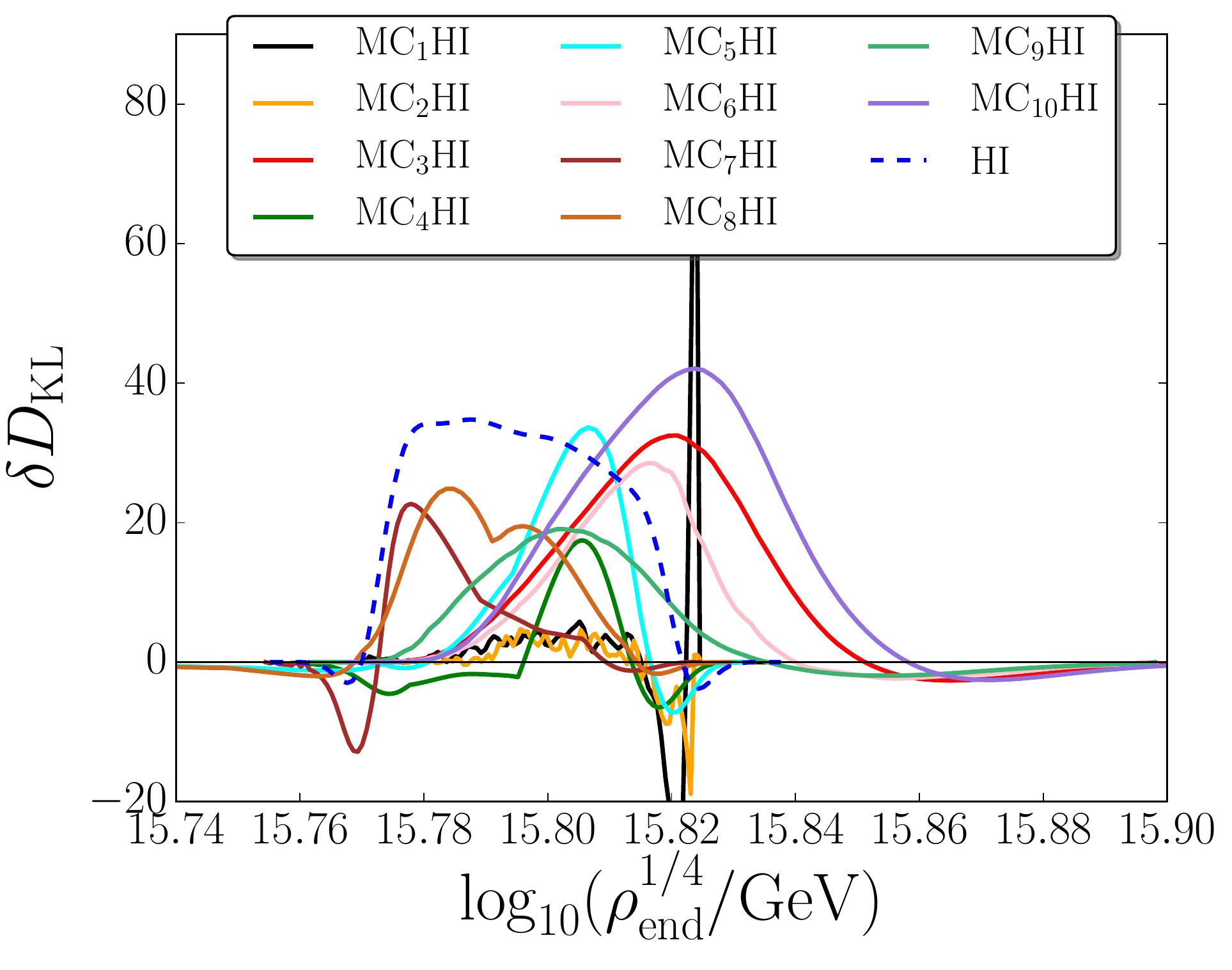}
\includegraphics[width=0.45\textwidth]{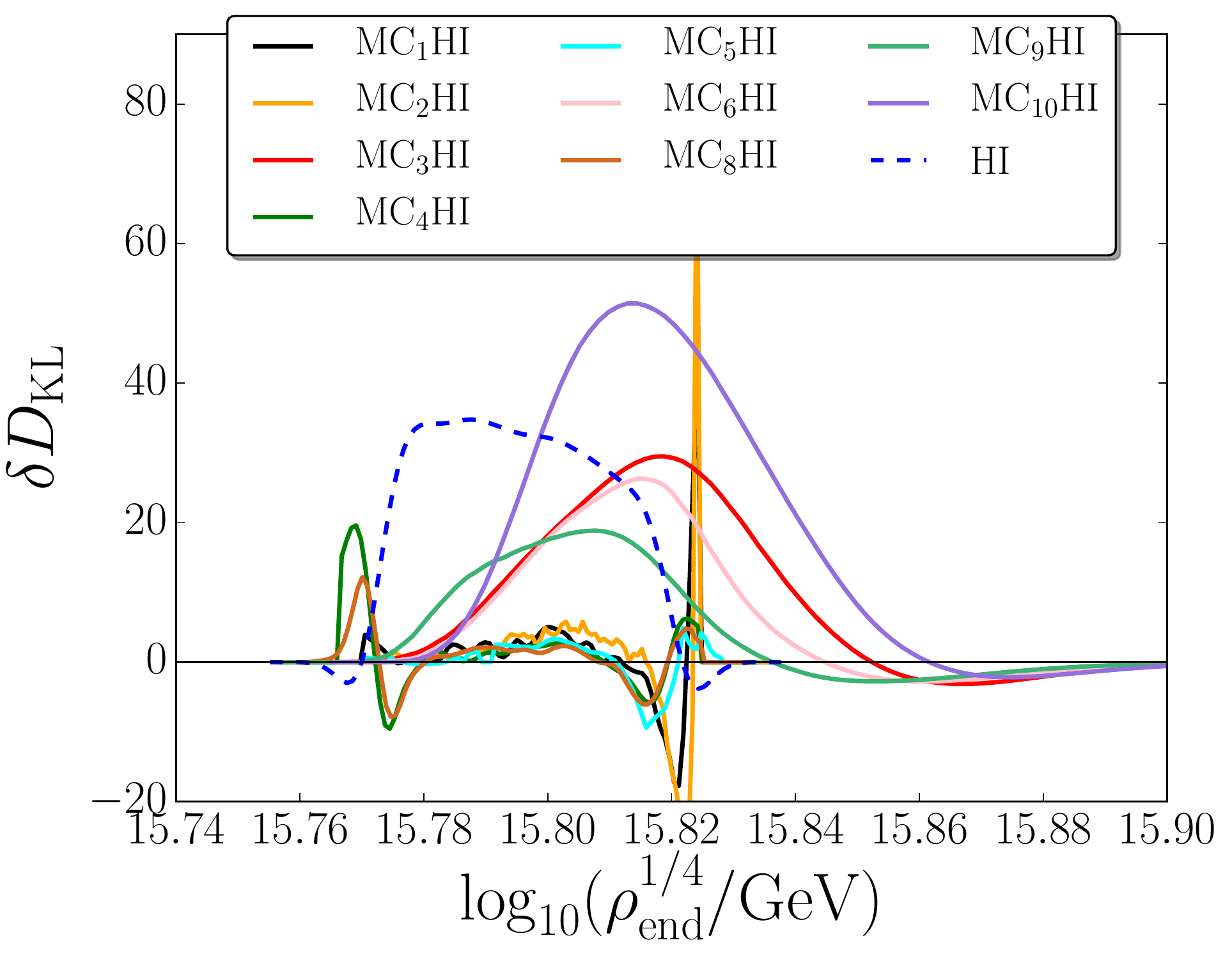}
\includegraphics[width=0.45\textwidth]{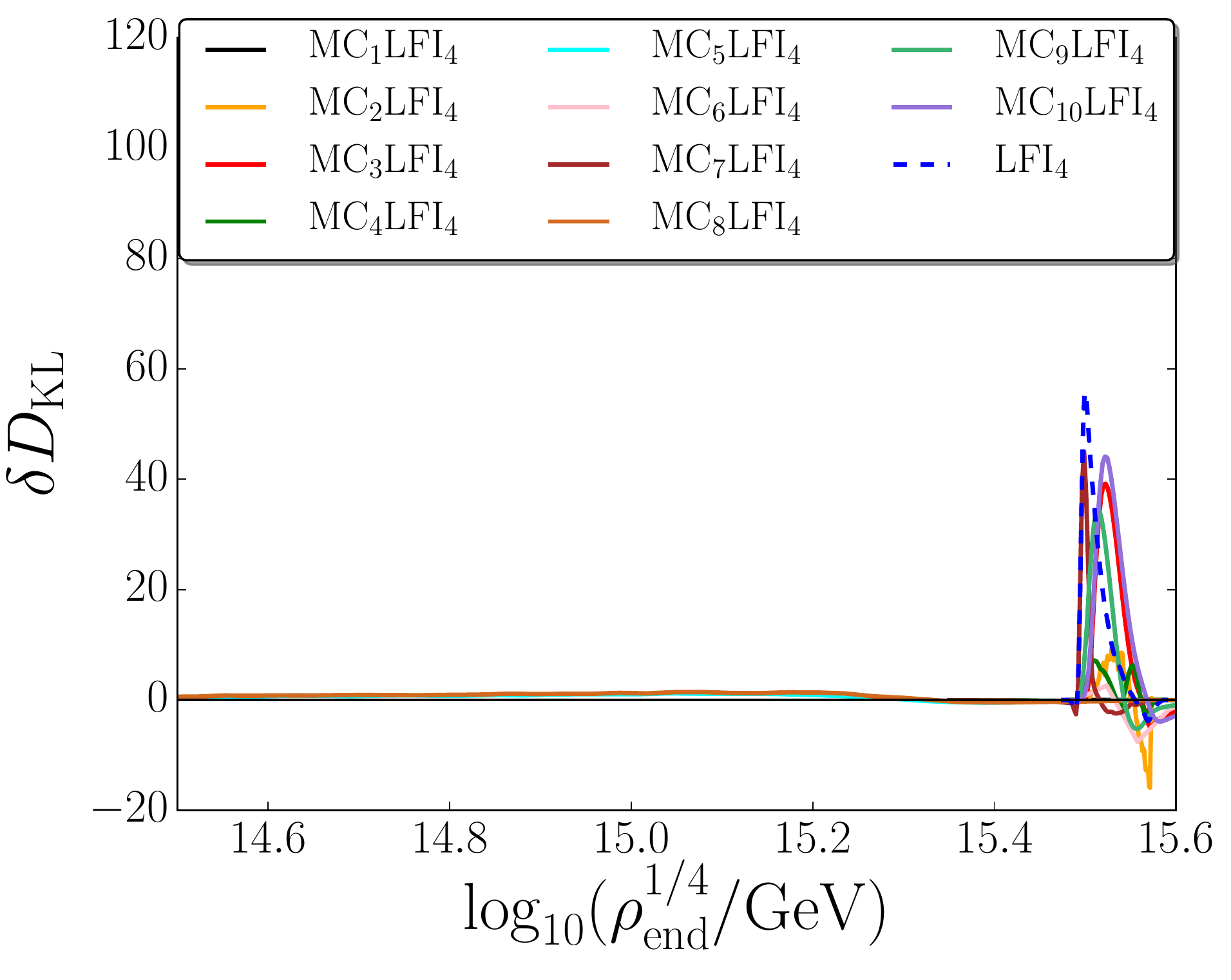}
\includegraphics[width=0.45\textwidth]{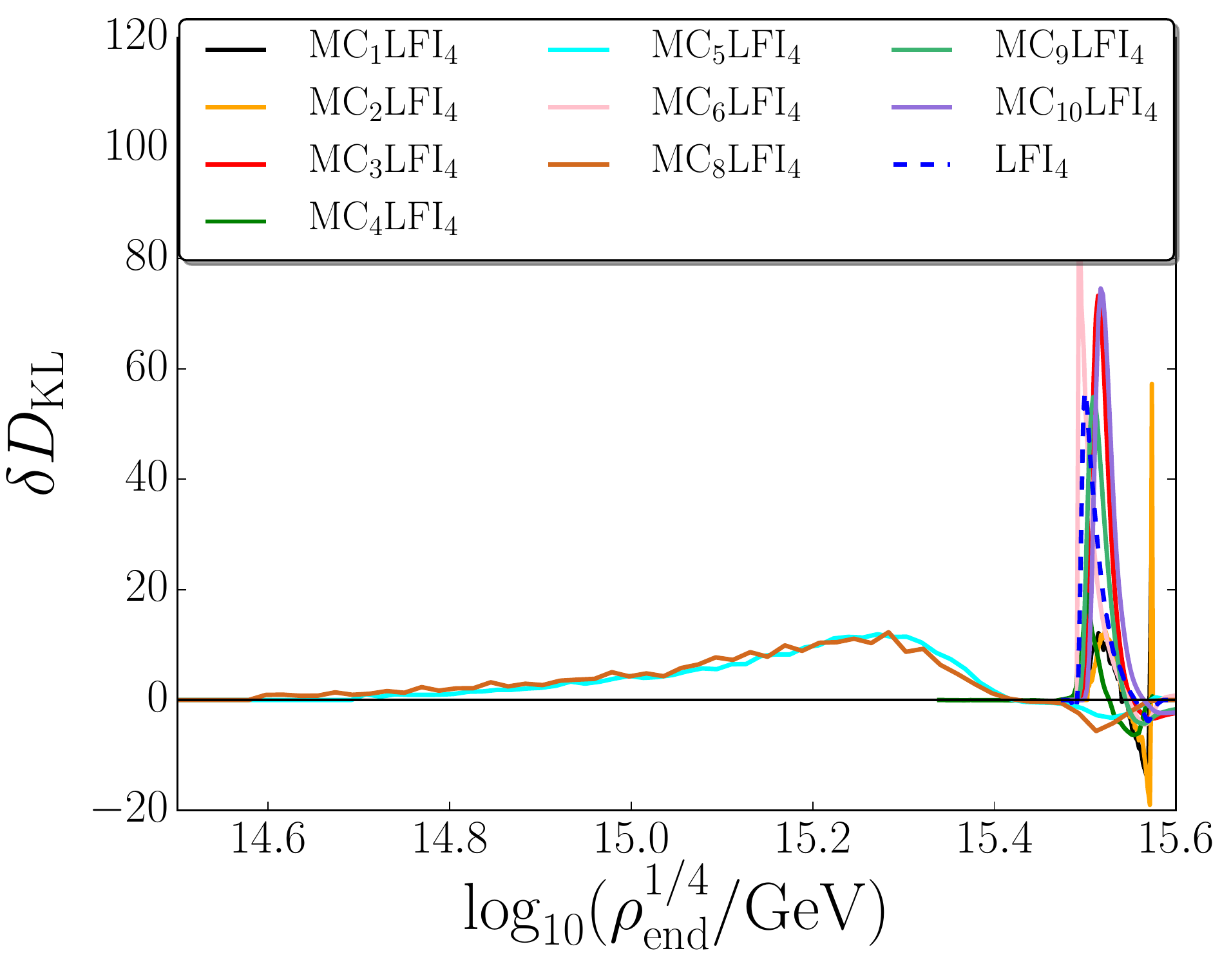}
\includegraphics[width=0.45\textwidth]{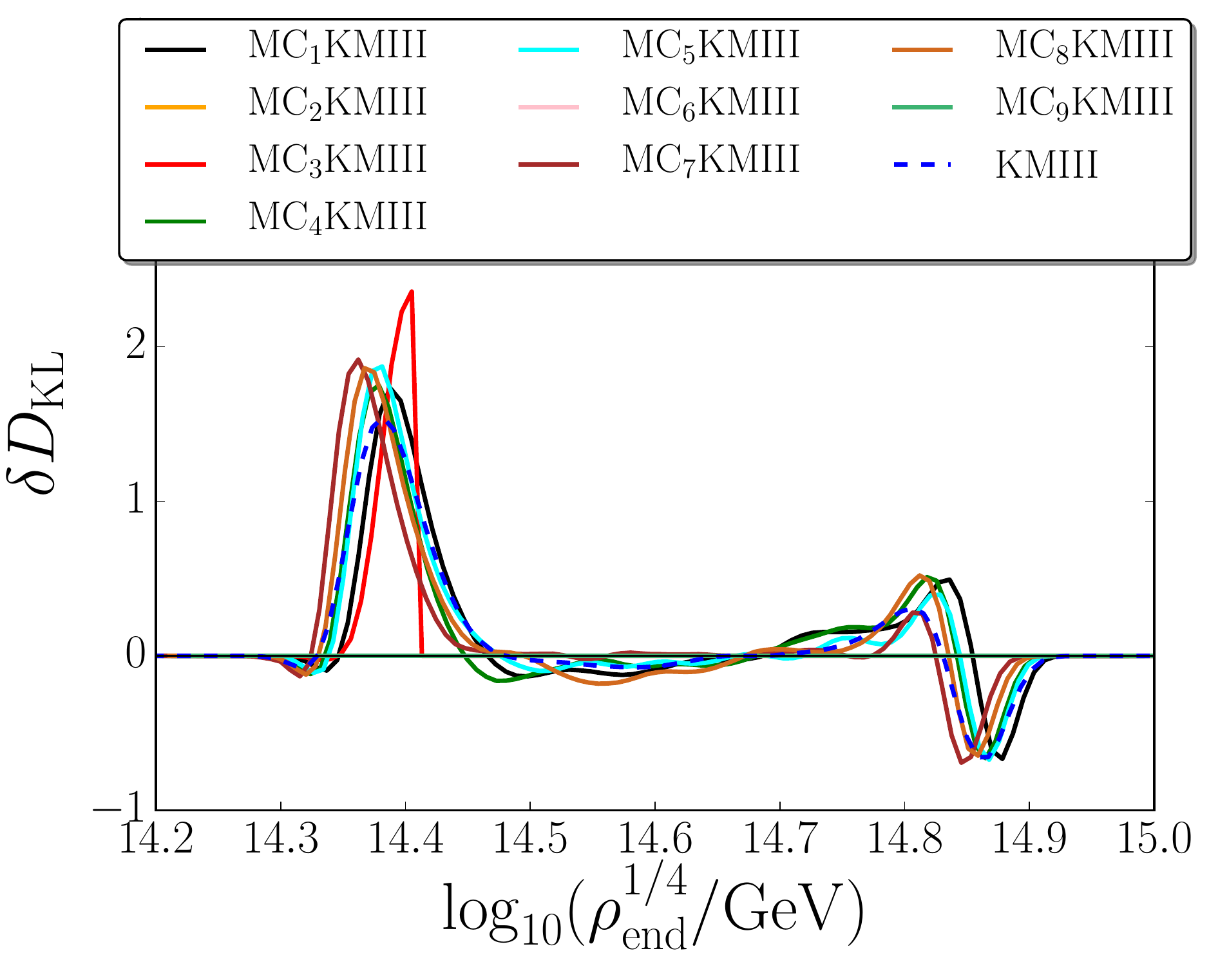}
\includegraphics[width=0.45\textwidth]{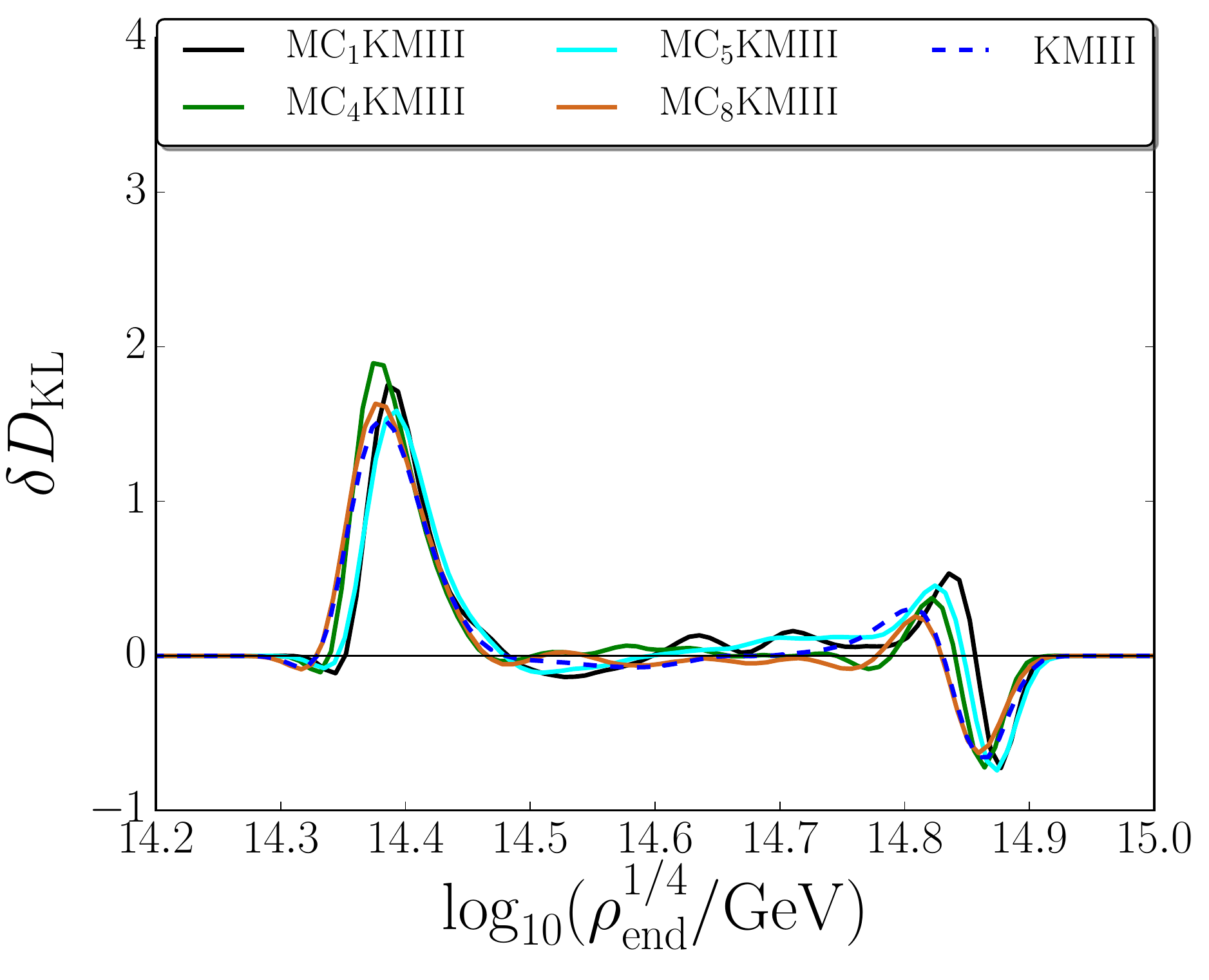}
\caption{Information density on $\rho_\uend$ for Higgs inflation (top panels), quartic inflation (middle panels) and K\"ahler moduli II inflation (bottom panels). The left panels correspond to the logarithmically flat prior~(\ref{eq:sigmaend:LogPrior}) on $\sigma_\uend$, and the right panels stand for the stochastic prior~(\ref{eq:sigmaend:GaussianPrior}) derived from the equilibrium distribution of a light scalar field in a de Sitter space-time with Hubble scale $H_\uend$. The dashed blue lines correspond to the single-field versions of the models, while the solid coloured lines stand for the $10$ reheating scenarios. 
}
\label{fig:DKL:rhoend:individual}
\end{center}
\end{figure}
\newpage
\subsection{Reheating Temperature}
\label{sec:app:DKL:Treh}
\begin{figure}[!h]
\figpilogsto
\begin{center}
\includegraphics[width=0.46\textwidth]{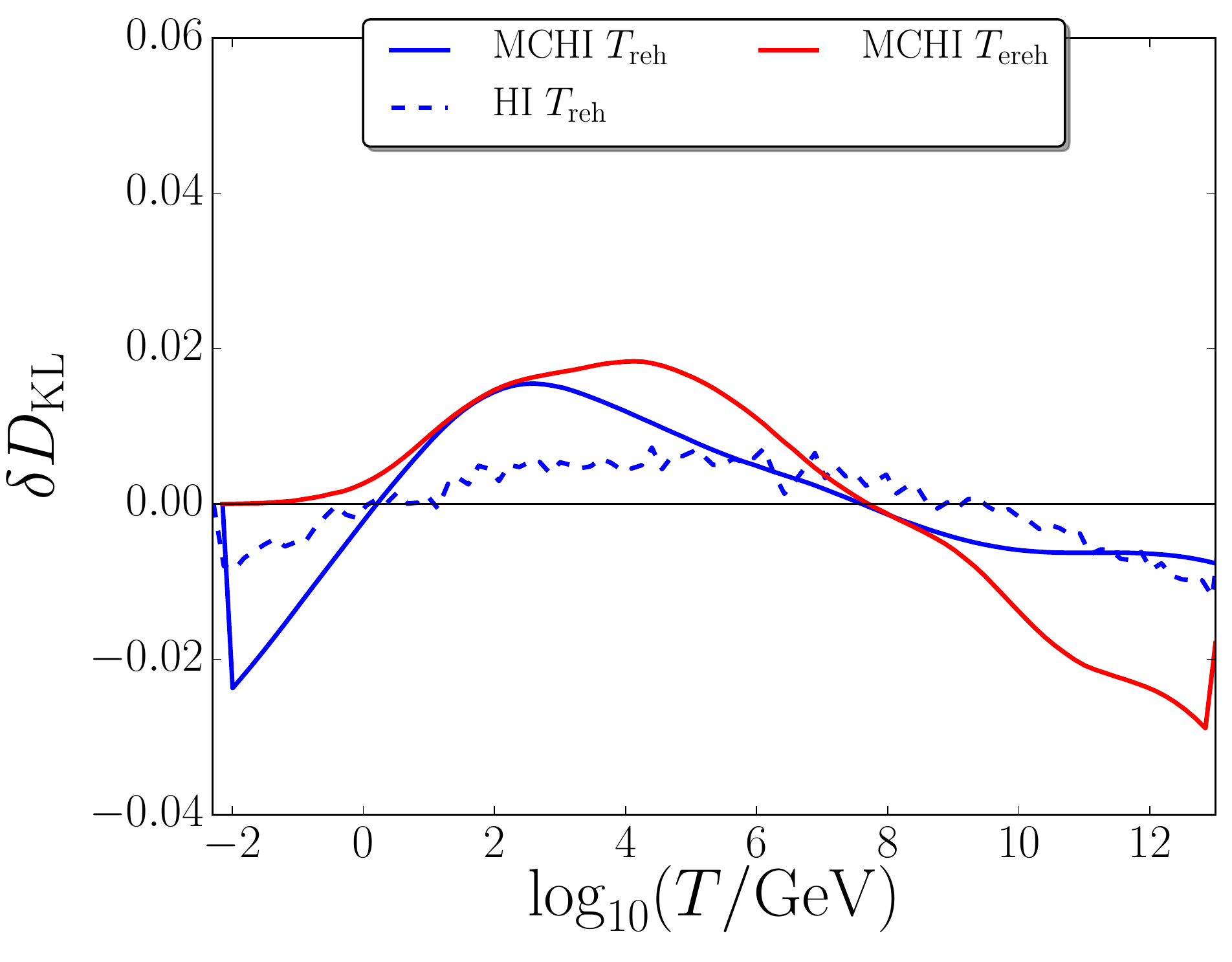}
\includegraphics[width=0.45\textwidth]{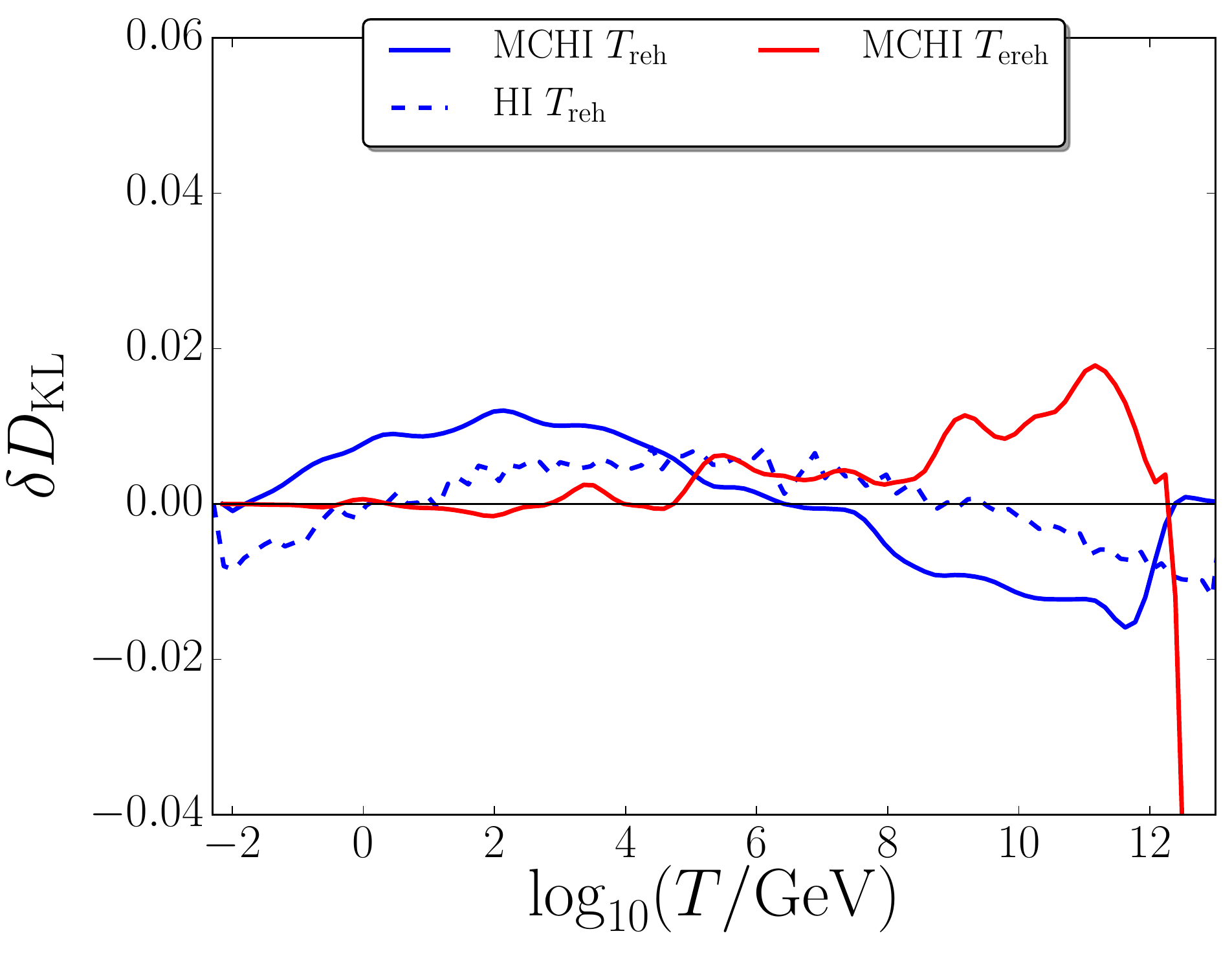}
\includegraphics[width=0.44\textwidth]{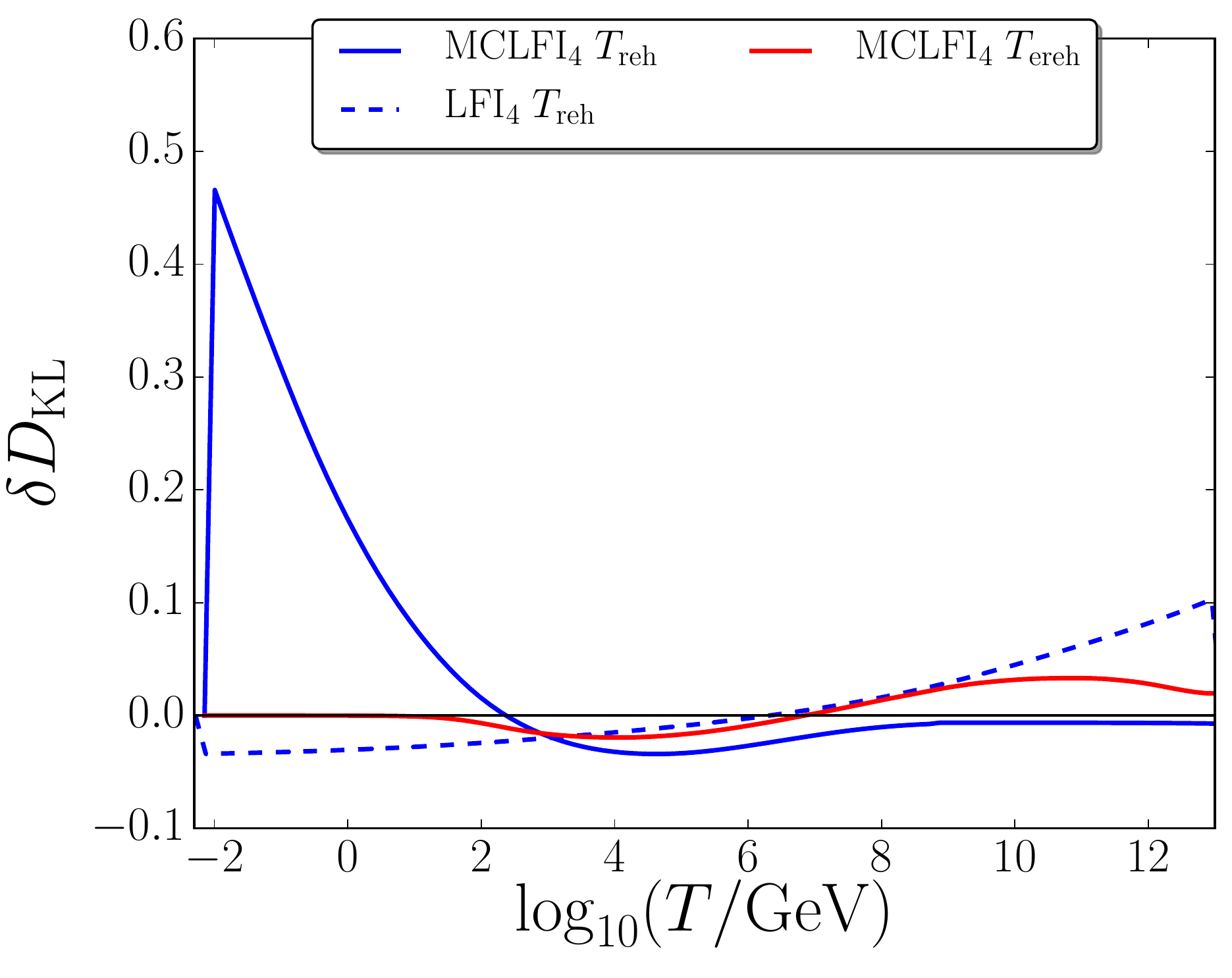}
\includegraphics[width=0.45\textwidth]{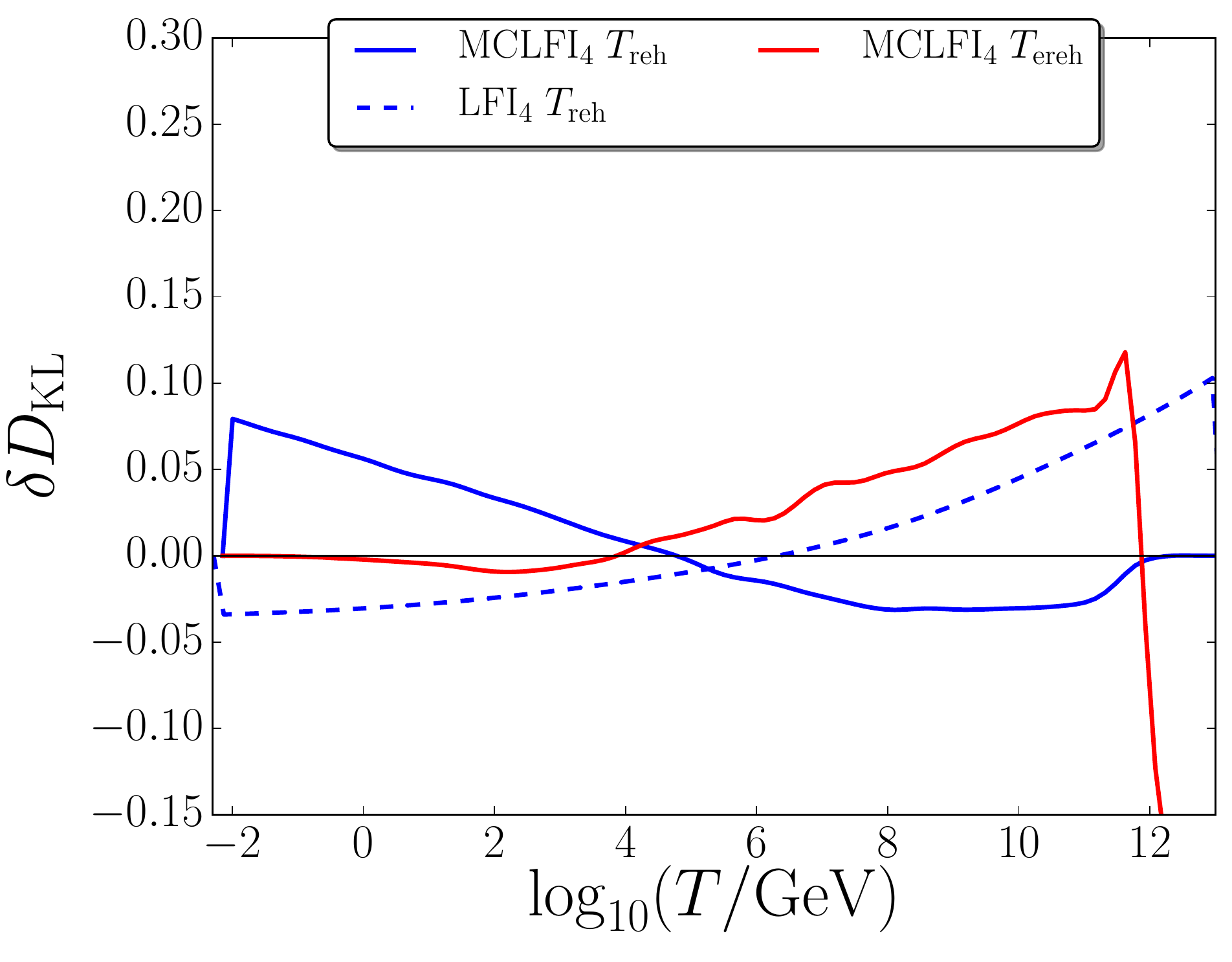}
\includegraphics[width=0.44\textwidth]{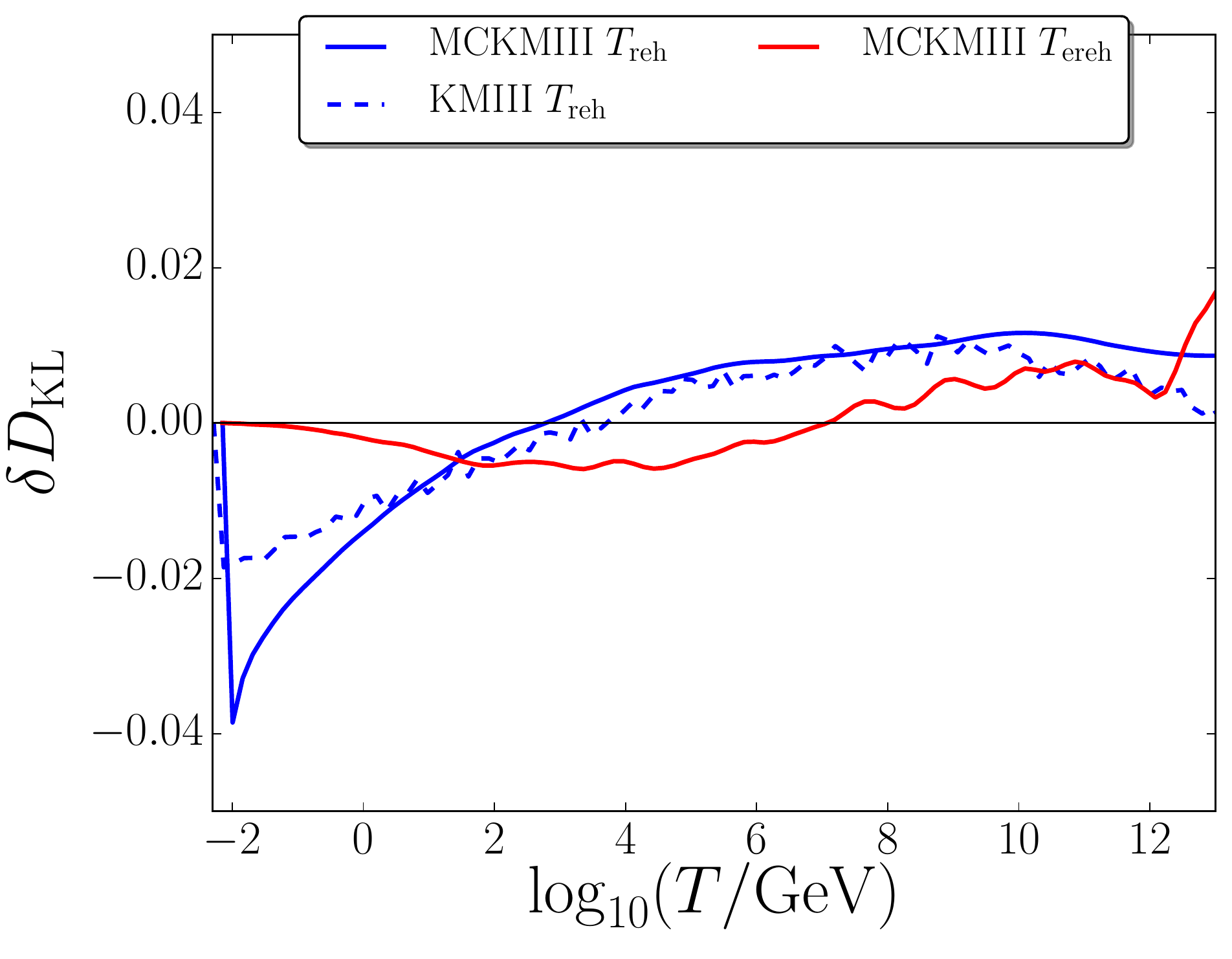}
\includegraphics[width=0.45\textwidth]{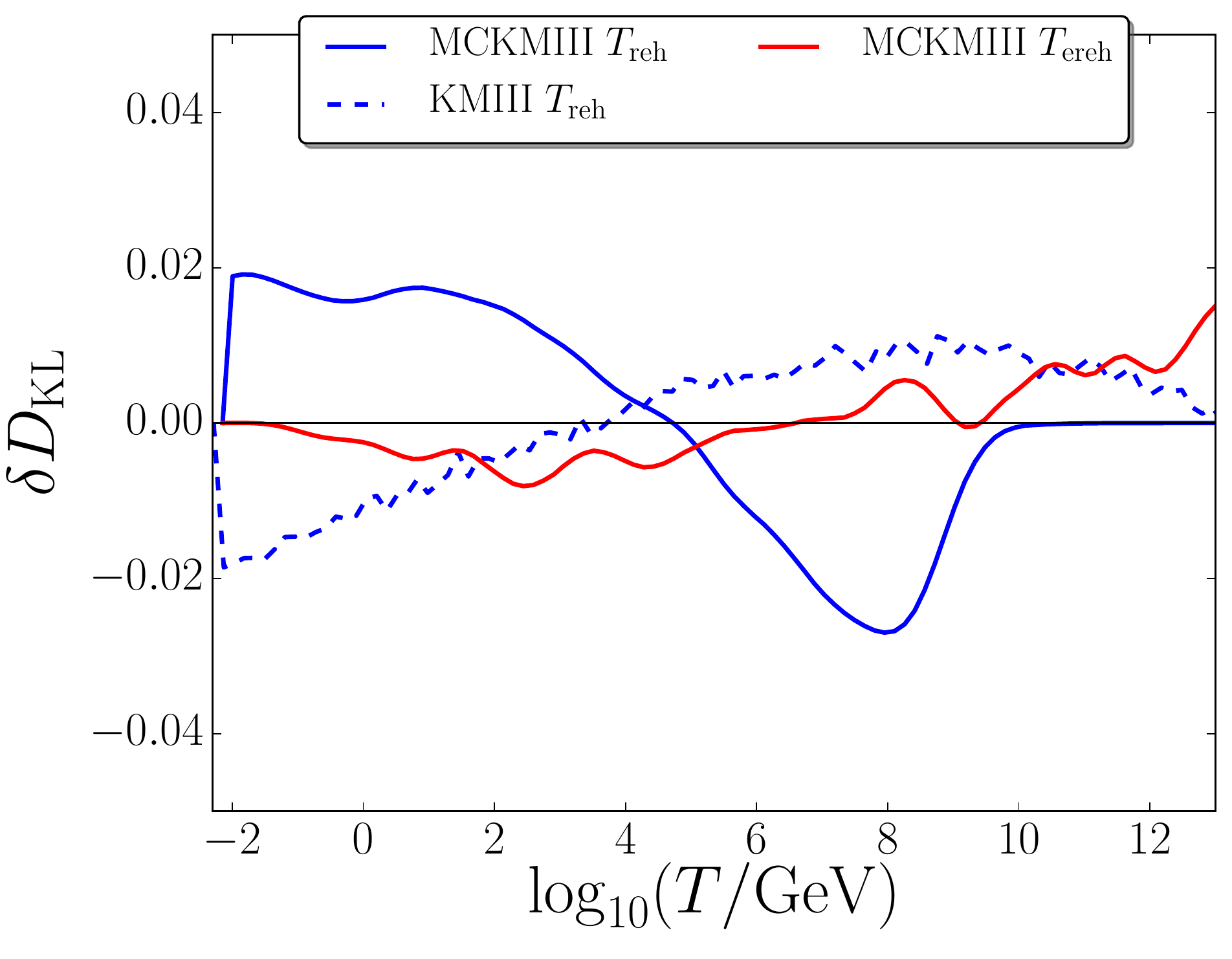}
\caption{Information density on $T_\ureh$ and $T_\uereh$ for Higgs inflation (top panels), quartic inflation (middle panels) and K\"ahler moduli II inflation (bottom panels). The left panels correspond to the logarithmically flat prior~(\ref{eq:sigmaend:LogPrior}) on $\sigma_\uend$, and the right panels stand for the stochastic prior~(\ref{eq:sigmaend:GaussianPrior}) derived from the equilibrium distribution of a light scalar field in a de Sitter space-time with Hubble scale $H_\uend$. The dashed blue lines correspond to the single-field versions of the models, while the solid lines are derived from the averaged distributions on $T_\ureh$ (blue) and $T_\uereh$ (red), when an extra light scalar field is added. 
}
\label{fig:DKL:Treh:averaged}
\end{center}
\end{figure}
\newpage
\begin{figure}[!h]
\figpilogsto
\begin{center}
\includegraphics[width=0.45\textwidth]{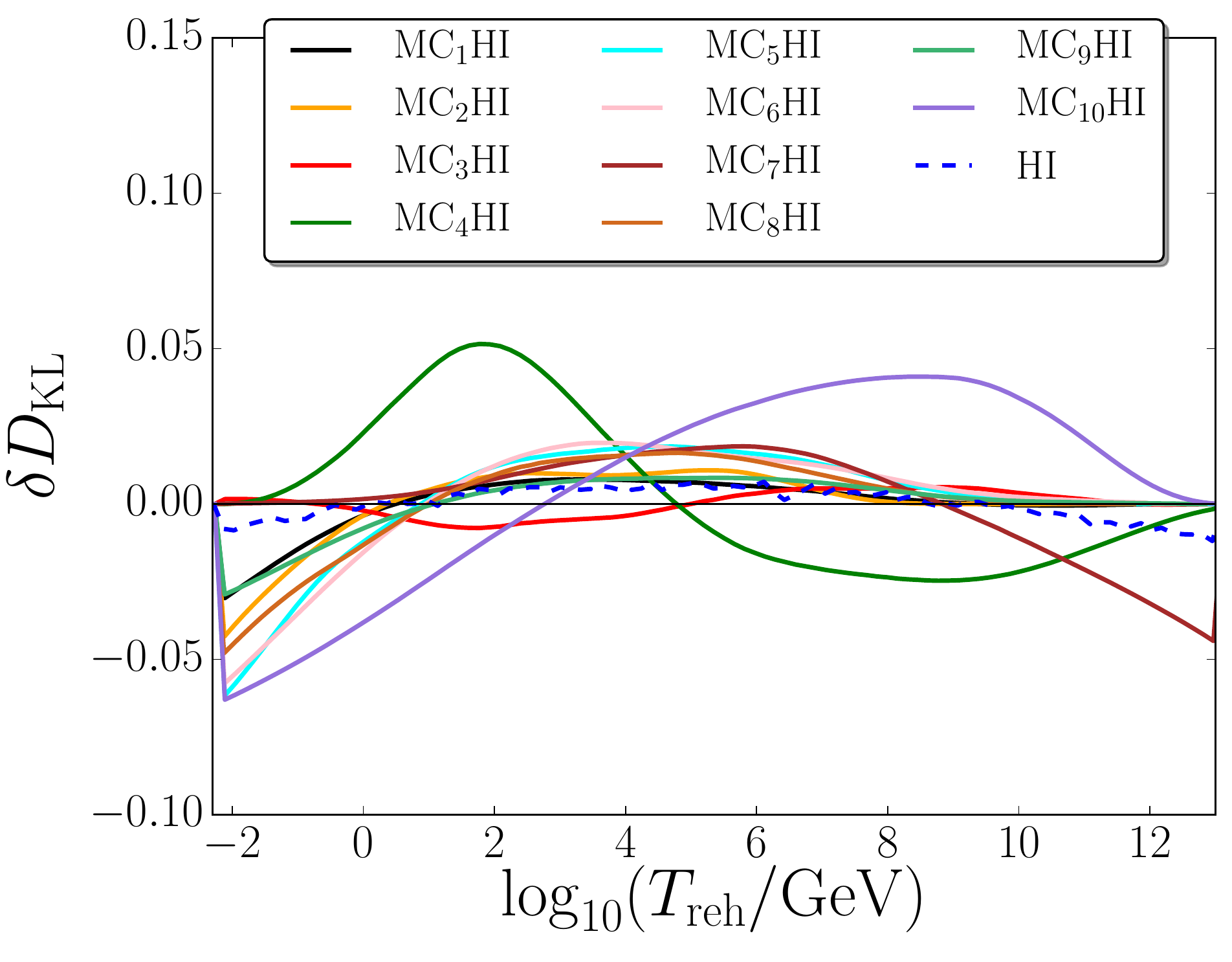}
\includegraphics[width=0.45\textwidth]{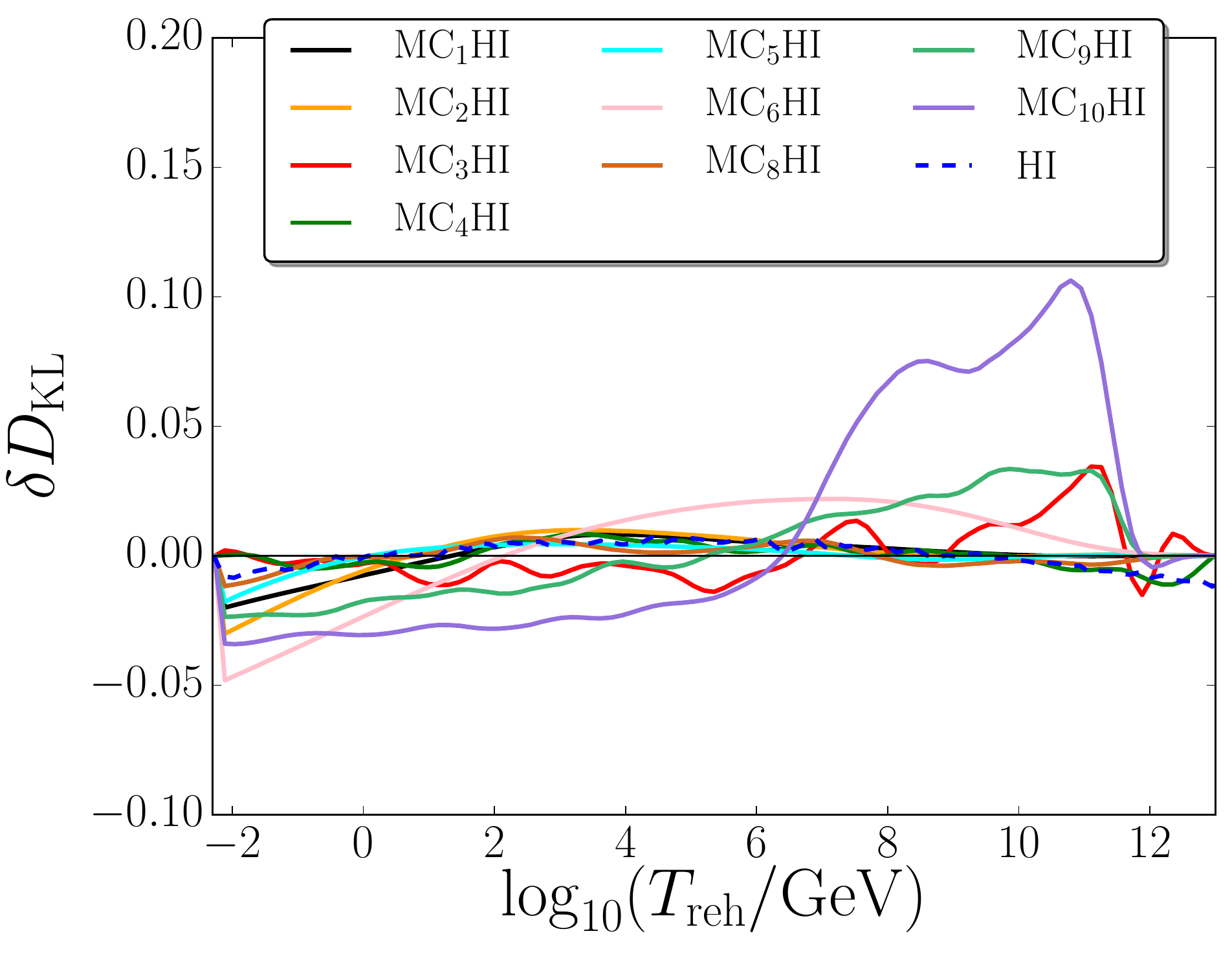}
\includegraphics[width=0.45\textwidth]{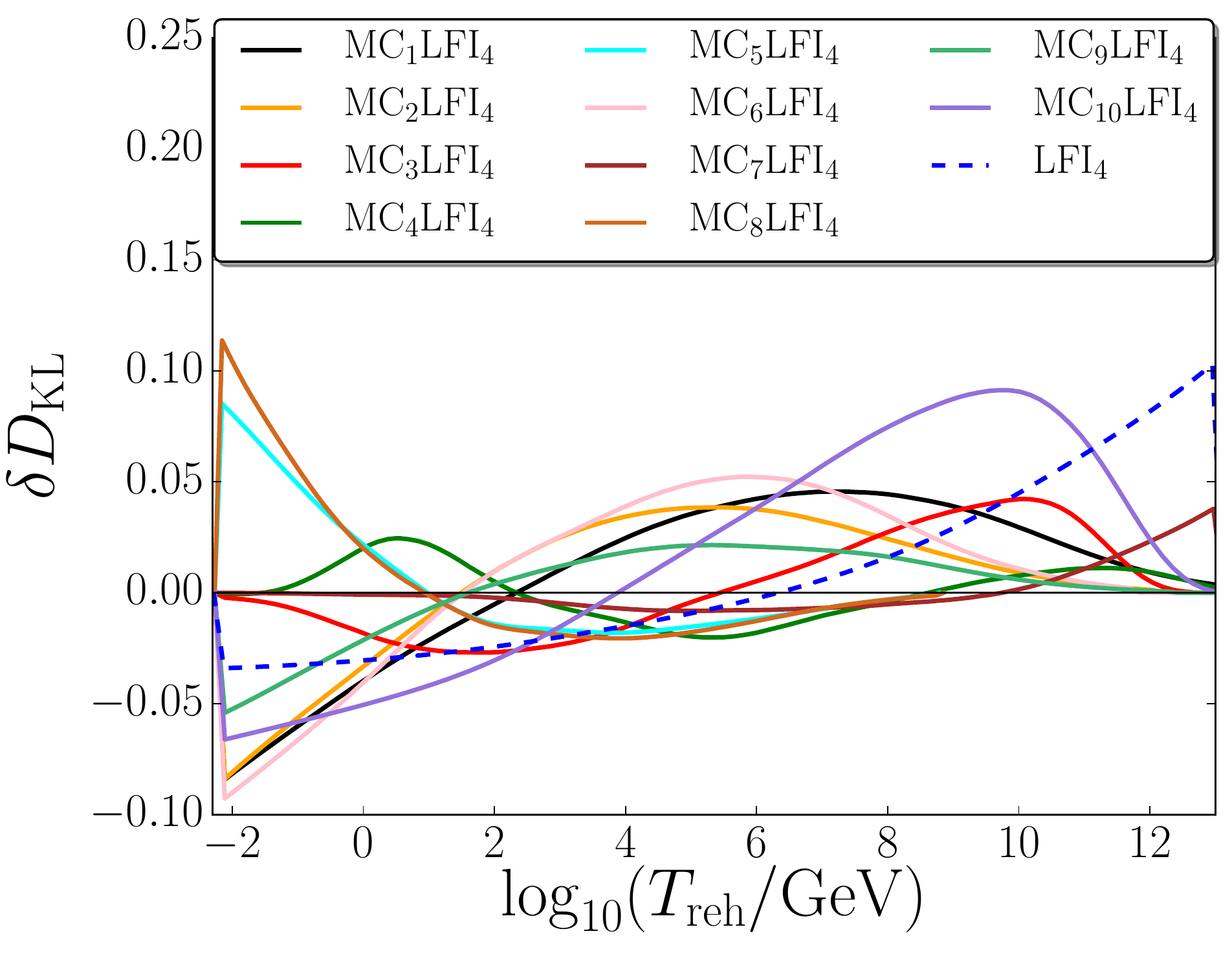}
\includegraphics[width=0.45\textwidth]{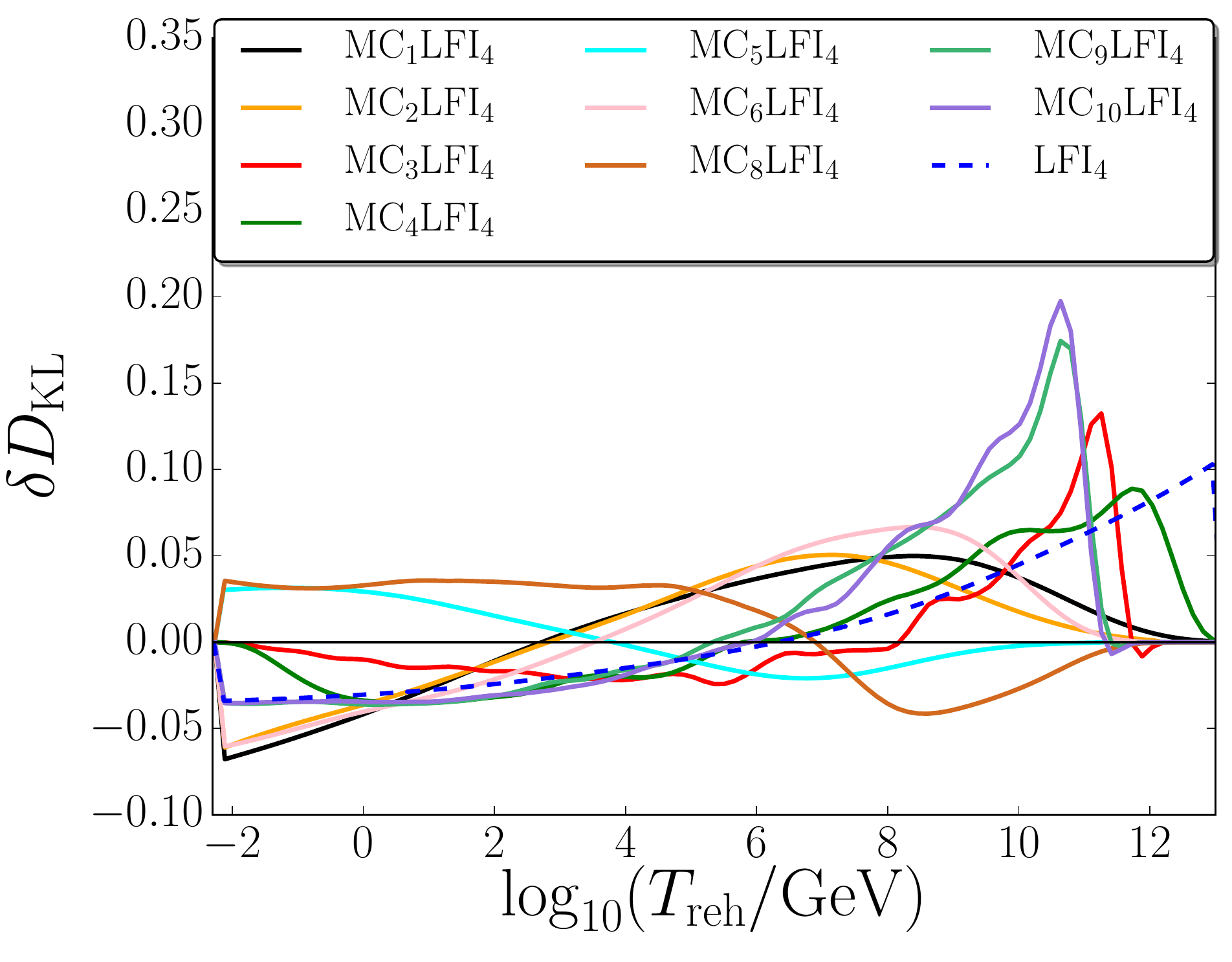}
\includegraphics[width=0.45\textwidth]{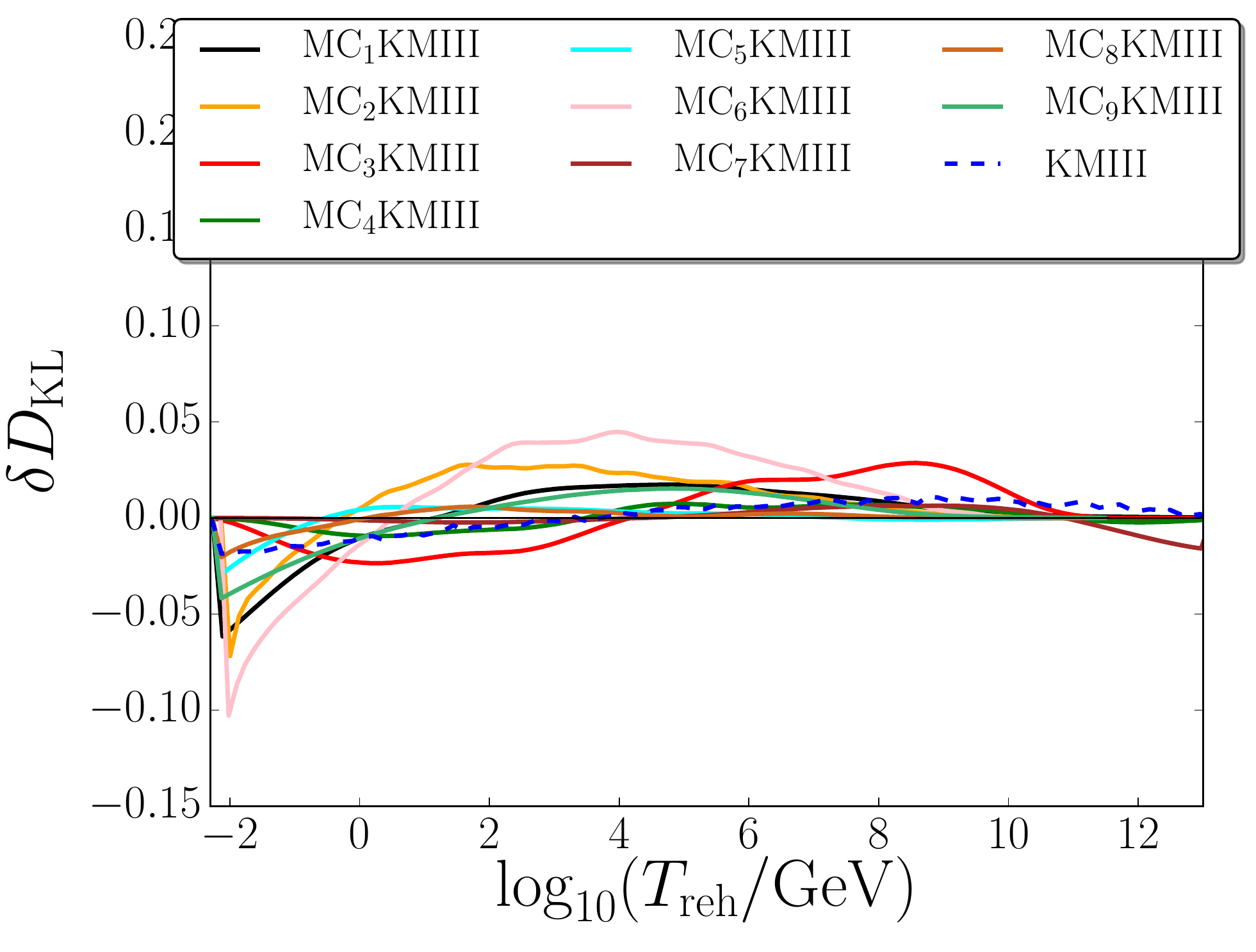}
\includegraphics[width=0.45\textwidth]{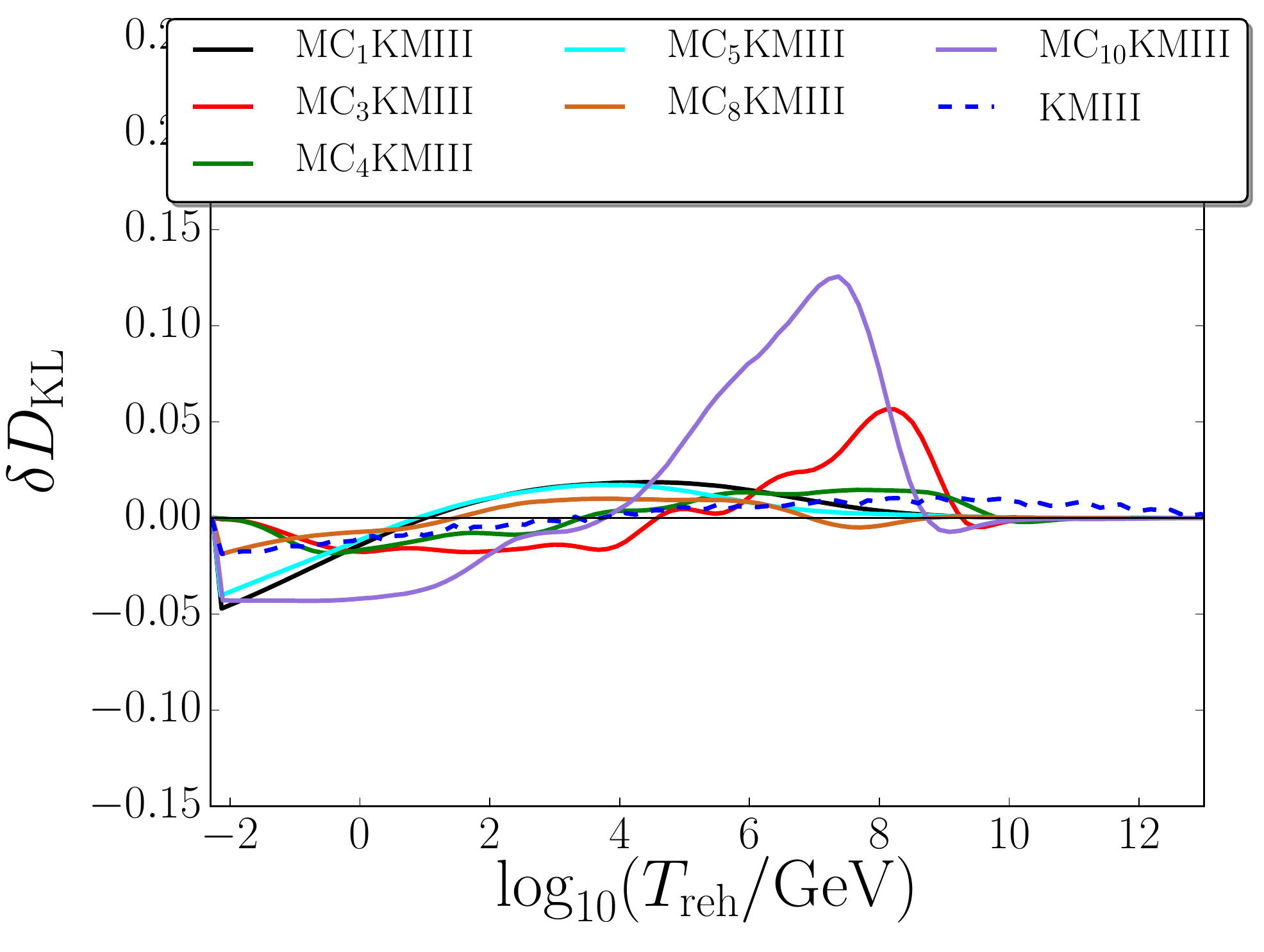}
\caption{Information density on $T_\ureh$ for Higgs inflation (top panels), quartic inflation (middle panels) and K\"ahler moduli II inflation (bottom panels). The left panels correspond to the logarithmically flat prior~(\ref{eq:sigmaend:LogPrior}) on $\sigma_\uend$, and the right panels stand for the stochastic prior~(\ref{eq:sigmaend:GaussianPrior}) derived from the equilibrium distribution of a light scalar field in a de Sitter space-time with Hubble scale $H_\uend$. The dashed blue lines correspond to the single-field versions of the models, while the solid coloured lines stand for the 10 reheating scenarios. 
}
\label{fig:DKL:Treh:individual}
\end{center}
\end{figure}
\newpage
\subsection{Early Reheating Temperature}
\label{sec:app:DKL:Tereh}
\begin{figure}[!h]
\figpilogsto
\begin{center}
\includegraphics[width=0.45\textwidth]{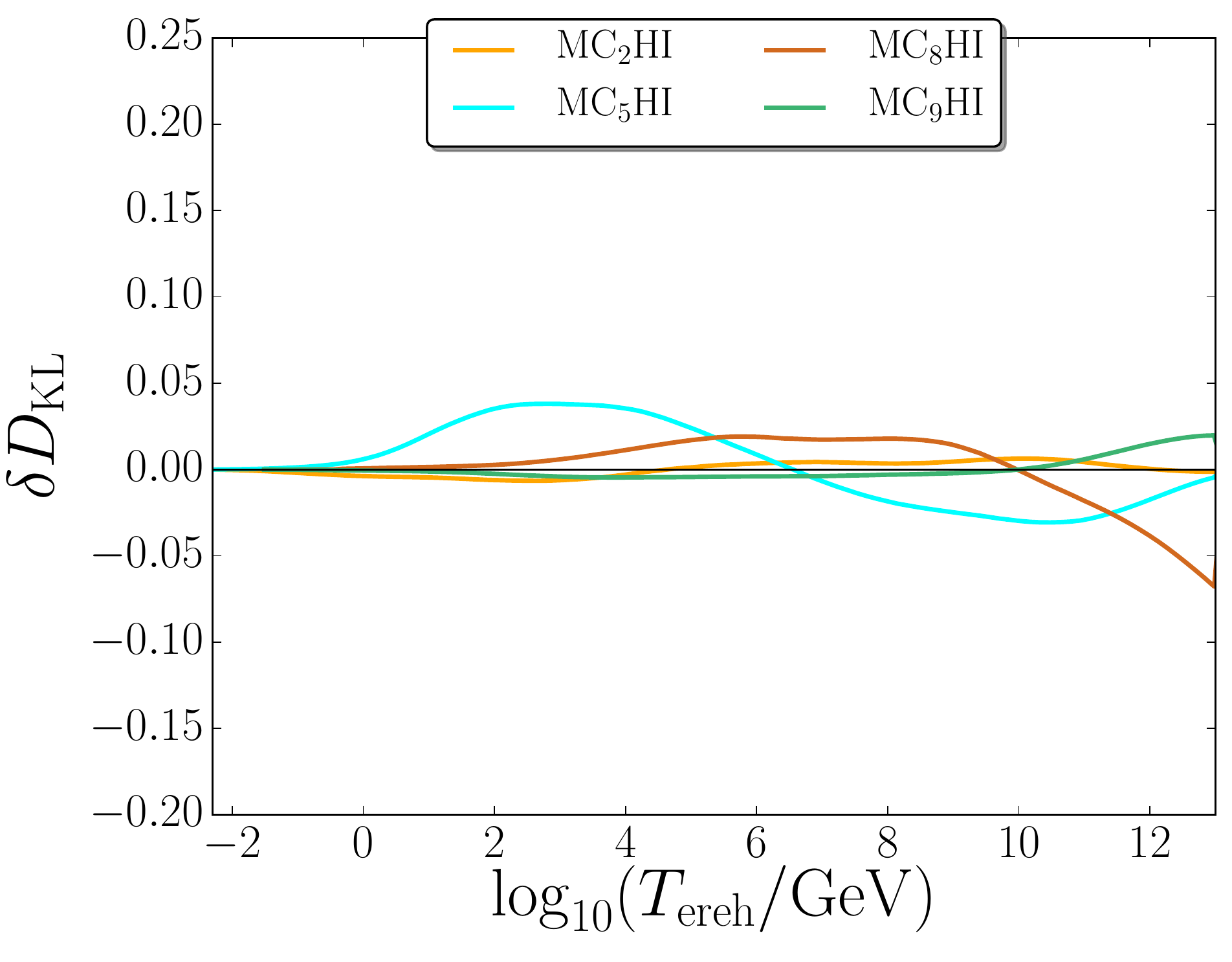}
\includegraphics[width=0.45\textwidth]{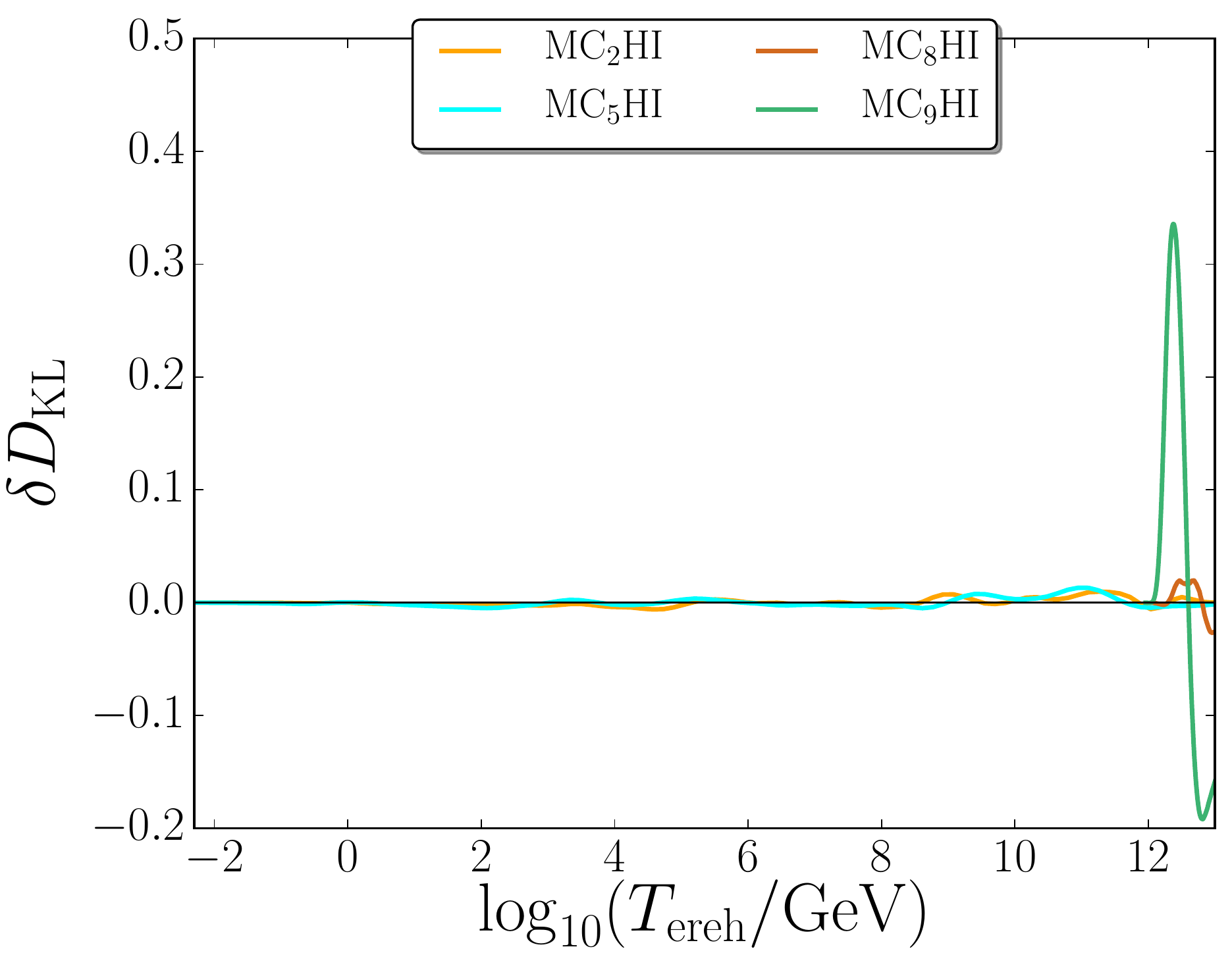}
\includegraphics[width=0.45\textwidth]{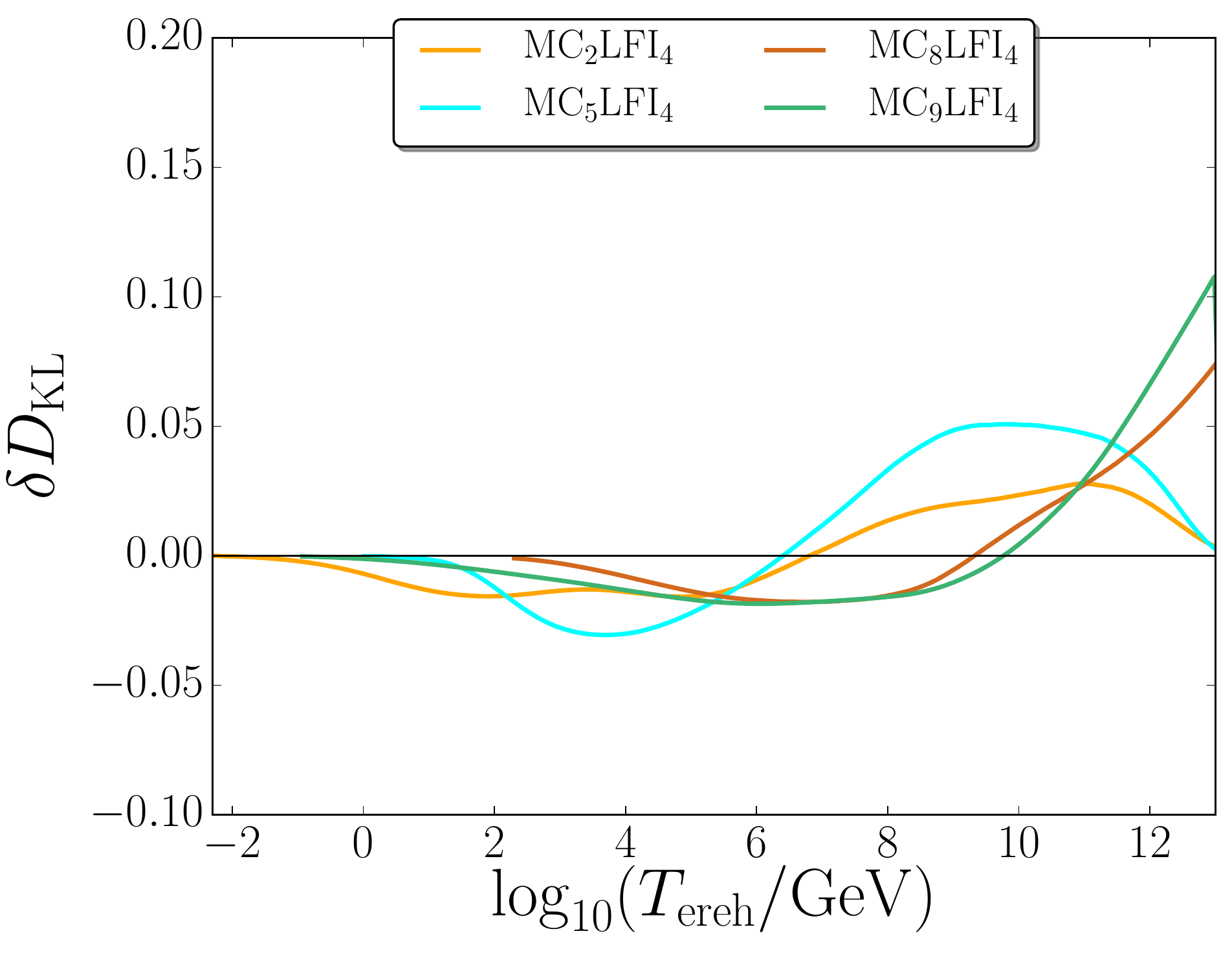}
\includegraphics[width=0.45\textwidth]{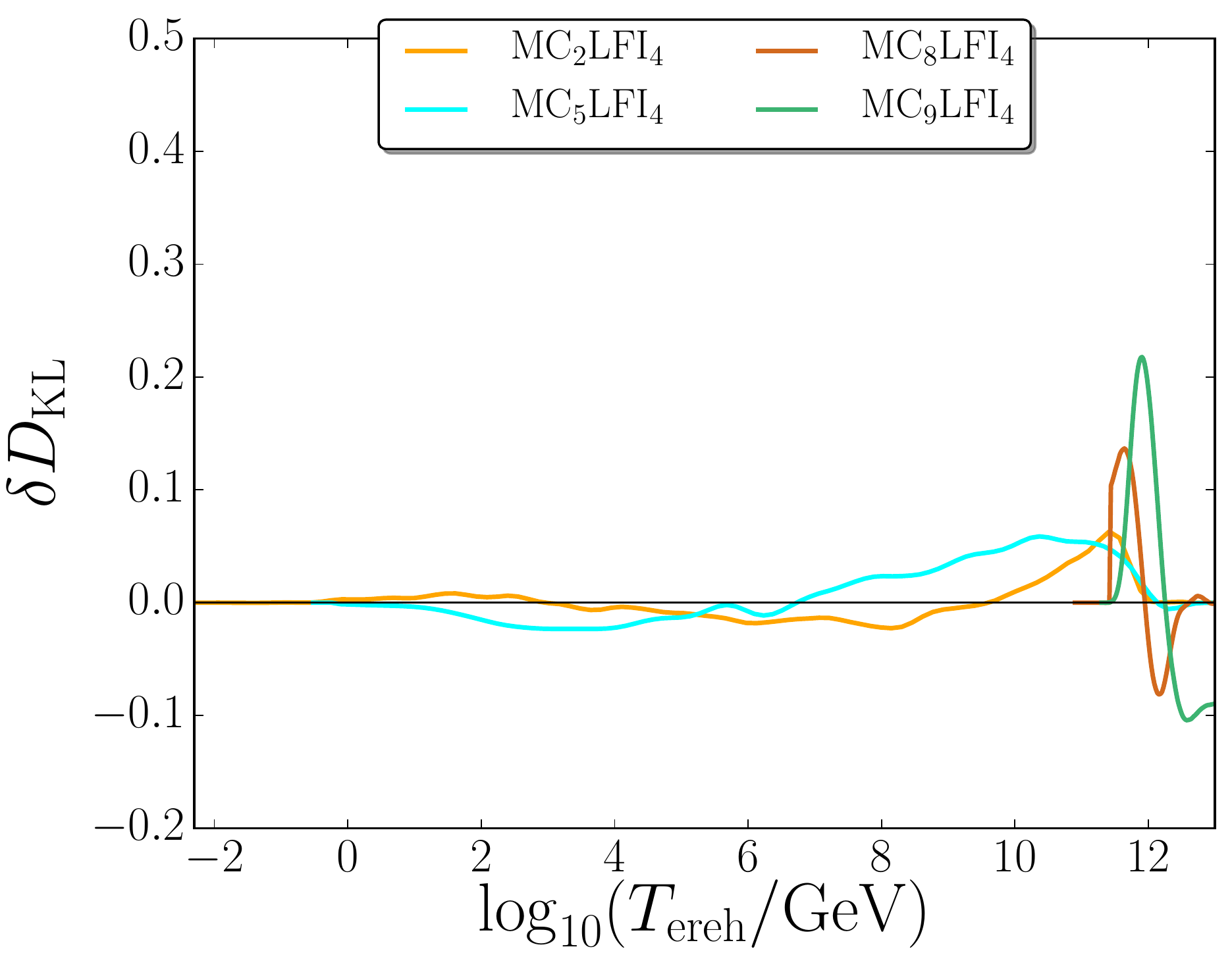}
\includegraphics[width=0.45\textwidth]{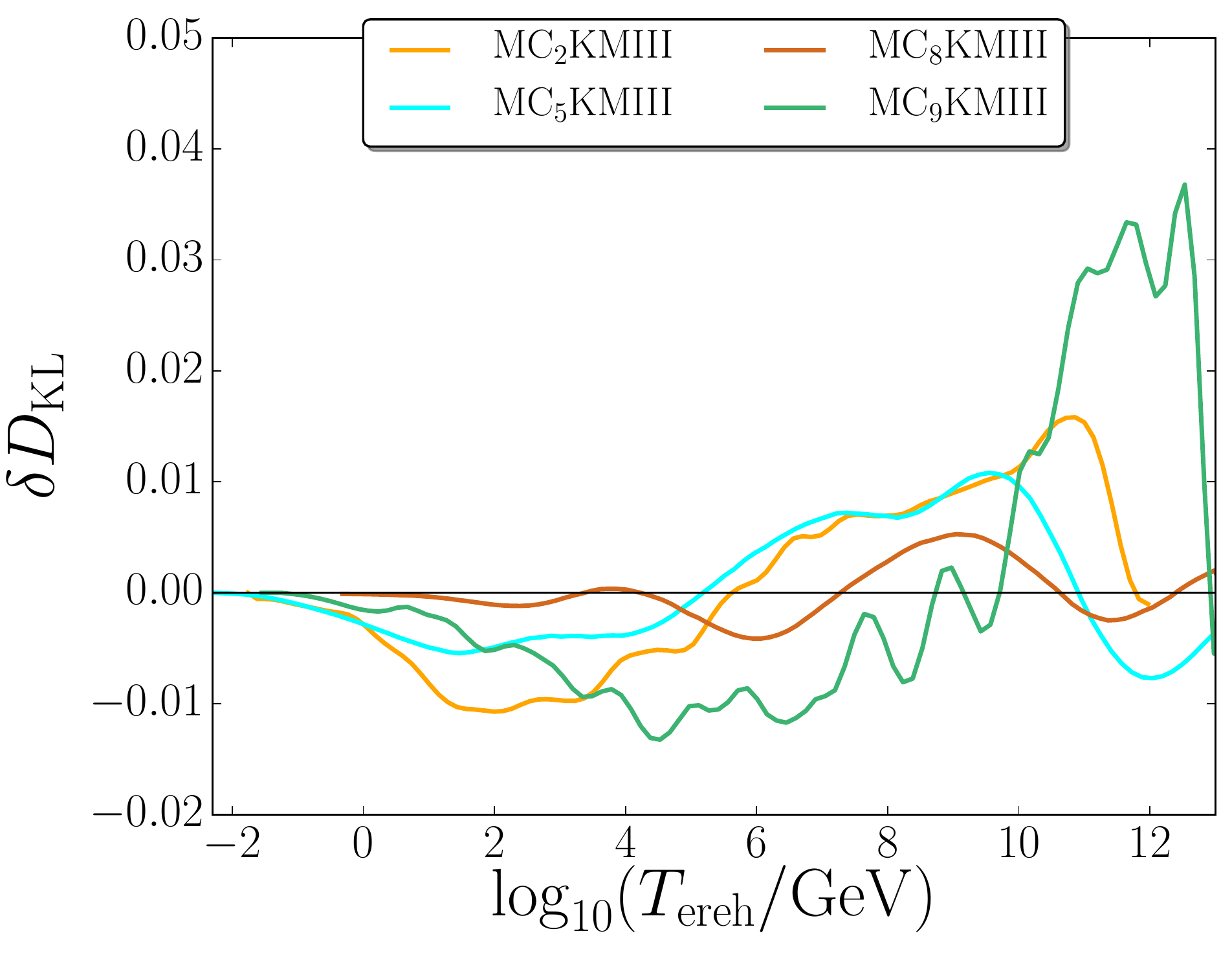}
\includegraphics[width=0.45\textwidth]{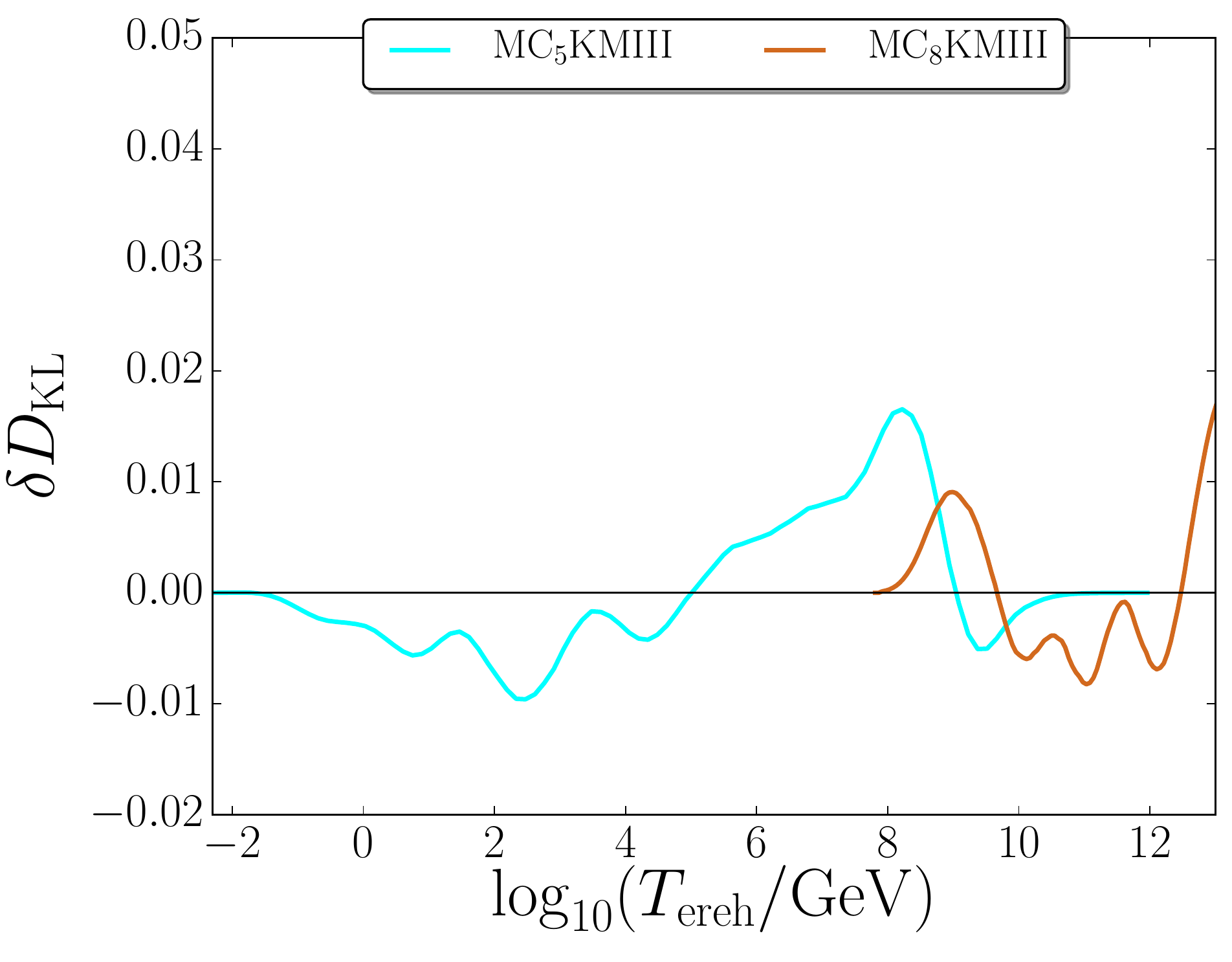}
\caption{Information density on $T_\uereh$ for Higgs inflation (top panels), quartic inflation (middle panels) and K\"ahler moduli II inflation (bottom panels). The left panels correspond to the logarithmically flat prior~(\ref{eq:sigmaend:LogPrior}) on $\sigma_\uend$, and the right panels stand for the stochastic prior~(\ref{eq:sigmaend:GaussianPrior}) derived from the equilibrium distribution of a light scalar field in a de Sitter space-time with Hubble scale $H_\uend$. 
}
\label{fig:DKL:Tereh:averaged}
\end{center}
\end{figure}
\end{appendix}
\clearpage
\bibliographystyle{JHEP}
\bibliography{curvreheat}
\end{document}